\documentclass[superscriptaddress,secnumarabic,
amssymb,amsmath,nobibnotes,aps,prd,showkeys,showpacs,nofootinbib]{revtex4}%
\usepackage{graphicx}
\usepackage{epsf}
\usepackage{bm}
\usepackage{amsmath}
\usepackage{amsfonts}
\usepackage{amssymb}
\usepackage{epstopdf}
\usepackage{natbib}
\usepackage{color}
\usepackage{float}
\providecommand{\U}[1]{\protect\rule{.1in}{.1in}}

\newcommand{\be}{\begin{equation}}
\newcommand{\ee}{\end{equation}}

\newcommand{\mincir}{\raise
-3.truept\hbox{\rlap{\hbox{$\sim$}}\raise4.truept\hbox{$<$}\ }}
\newcommand{\magcir}{\raise
-3.truept\hbox{\rlap{\hbox{$\sim$}}\raise4.truept\hbox{$>$}\ }}

\begin{document}

\title{A review of Quintessential Inflation}


\author{Jaume  de Haro}
\email{jaime.haro@upc.edu}
\affiliation{Departament de Matem\`atiques, Universitat Polit\`ecnica de Catalunya, Diagonal 647, 08028 Barcelona, Spain}

\author{Llibert Arest\'e Sal\'o}
\email{l.arestesalo@qmul.ac.uk} 
\affiliation{School of Mathematical Sciences, Queen Mary University of London, Mile End Road, London, E1 4NS, United Kingdom}

\thispagestyle{empty}

\begin{abstract}
We compute numerically the reheating temperature due to the gravitational production of conformally coupled superheavy  particles during the  phase transition from the end of inflation to the beginning of  kination in two different  Quintessential Inflation (QI) scenarios, namely Lorentzian Quintessential Inflation (LQI) and $\alpha$-attractors in the context of Quintessential Inflation ($\alpha$-QI). Once these superheavy  particles 
have been created, they must decay into lighter ones to form a relativistic plasma, whose energy density will eventually dominate the one of the inflaton field in order to reheat after inflation our universe
with a  very high temperature, in both cases greater than $10^7$ GeV,  contrary to the usual belief that heavy masses suppress the particle production and, thus,  lead  to an inefficient reheating temperature. Finally, we will show that the over-production of Gravitational Waves (GWs) during this phase transition, when one deals with our models, 
does not disturb the Big Bang Nucleosynthesis (BBN) success.
\end{abstract}

\vspace{0.5cm}

\maketitle

Understanding the universe's evolution 
is one of the greatest mysteries in the history of humanity. It is always the primary question: ``where we come from and where we are going".
In particular, its  early and late expansions have been studied a great deal at present time. Looking at the scientific  literature, one can find two popular and well accepted theories -though they are not observationally proved-,
namely the inflation (the early surprisingly fast accelerated  expansion of our universe) and the quintessence as a form of dark energy (the current cosmic acceleration). The inflationary paradigm \cite{guth, linde, {albrecht}} is actually a very fast accelerating phase of the early universe  that lasted for an extremely tiny time and  became able to solve a number of shortcomings associated with the standard Big Bang cosmology, such as the horizon problem, flatness problem or the primordial monopole problem. The  predictive power  of inflation was soon recognized due to its ability to explain the  origin of inhomogeneities in the universe as quantum fluctuations during this epoch \cite{chibisov, starobinsky, pi, bardeen, Linde:1982uu}, because such an explanation greatly matches with the  recent observational data from Planck's team \cite{Planck}. Thus, it is remarkable to note that inflation,  which appeared at the beginning of the 80's,  is still considered
the best way to explain the recent observational data, because it is nowadays the simplest viable theory that describes almost correctly the early universe in agreement with the recent observations \cite{Planck}. 

\

On the other hand, one of the most accepted  explanations for the current cosmic acceleration comes through the introduction of some quintessence field \cite{Copeland:2006wr}. 
In fact, soon after the discovery of the current cosmic acceleration at the end of the last century \cite{riess, perlmutter}, a~class of pioneering cosmological models attempting to unify the early- and late-time accelerating expansions were introduced. By construction, unlike the standard quintessence models (see \cite{Tsujikawa} for a review), these models -named as  {\it Quintessential Inflation} (QI) models \cite{pv, Spokoiny, pr}-
 only contain one classical scalar field, also named inflaton as in standard inflation \cite{guth, linde, starobinsky, albrecht}, and it is shown that they succeed 
 in reproducing the two accelerated epochs of the universe expansion (see also \cite{deHaro:2016hpl,hap,deHaro:2016hsh,deHaro:2016ftq,Geng:2017mic,AresteSalo:2017lkv, Haro:2015ljc, hossain1,hossain2,hossain3,hossain4,guendelman1} for other interesting QI~models). 
This idea to unify Inflation with Quintessence was indeed a novel attempt by Peebles and Vilenkin in their seminal paper \cite{pv}, and the novelty of their proposal comes through the introduction of a single potential that at early times allows inflation while at late times provides quintessence. Thus, a unified picture of the universe was effectively proposed connecting the distant early phase to the present one. 
Thanks to this proposal, the origin of the scalar responsible for the current inflation of the universe can be determined and fine-tunning problems are reduced \cite{dimopoulos01}.
In addition, the majority of models we deal with only depend on two parameters, which are determined by observational data. Hence, because of the behavior of the slow-roll regime as an attractor,  the dynamics of the model are obtained with the value of the scalar field and its derivative -initial conditions- at some moment during inflation. This shows the simplicity
of Quintessential Inflation, which from our viewpoint is  simpler than standard quintessence, where a minimum of two fields are needed to depict the evolution of the universe, the inflaton and a quintessence field. Thus, one needs two different potentials and two different initial conditions: one for the inflaton, which has to be fixed during inflation, and another one for the quintessence field, whose initial conditions normally have to be fixed at the beginning of the radiation era.

\

This enhanced more investigation in order to connect Quintessential Inflation with the observational data \cite{dimopoulos1,Giovannini:2003jw,hossain1,hossain3, deHaro:2016hpl,deHaro:2016hsh,deHaro:2016ftq,hap,Geng:2017mic,AresteSalo:2017lkv,Haro:2015ljc,hyp} and, consequently, this particular topic has become a popular area of research at the present time.    
However, an important difference occurs with respect to the standard inflationary paradigm, where 
the potential of the inflaton field has a local minimum (a deep well) and, thus, the inflaton field  releases its energy while it oscillates, producing enough particles \cite{kls, kls1, gkls, stb, Basset} to reheat our universe. 
In contrast, for the  "non-oscillating"  models, i.e., in Quintessential Inflation,  where the inflaton field survives to be able to reproduce the current cosmic acceleration, 
the mechanism of reheating  is completely different: once the inflationary phase is completed, a reheating mechanism keeping  "alive" the inflaton field is needed to match inflation with the Hot Big Bang universe \cite{guth} because the particles existing before the beginning of this period were completely diluted at the end of inflation resulting in a very cold universe. 

\

On this way, 
the most accepted idea to reheat the universe in the context of QI comes through a phase transition of the universe from inflation to kination (a regime where all the energy density of the inflation turns into kinetic \cite{Joyce}) where the adiabatic regime is broken, which allows to create particles. The mechanism to produce particles is not unique in this context since a number of distinct ones are available and can be used.  The first one is the well-known {\it Gravitational Particle Production} studied long time ago  in \cite{Parker,fmm,glm,gmm,ford,Zeldovich}, at the end of the 90's in \cite{Damour, Giovannini} and more recently applied to Quintessential Inflation in \cite{Spokoiny, pv, dimopoulos0, vardayan} for massless particles. A second important  mechanism is the so-called {\it Instant Preheating} introduced in \cite{fkl0} and applied for the first time to  inflation in \cite{fkl}  and recently in \cite{dimopoulos, vardayan} in the context of $\alpha$-attractors in supergravity.  Other less popular mechanisms  are the {\it Curvaton Reheating} applied  to Quintessence Inflation in \cite{FL, ABM},
the production of massive particles self-interacting and coupled to gravity \cite{tommi}
and the reheating via production of heavy massive particles conformally coupled to gravity \cite{kolb, kolb1,Birrell1, hashiba, hyp}.

\

The production of superheavy massive particles conformally coupled to gravity is one of the most important concerns of this  review. These particles are created 
during the phase transition from the end of inflation to the beginning of kination and  should  decay into lighter ones to form a thermal relativistic plasma, which eventually dominates to match this early period with the Hot Big Bang universe. In fact, 
the  main motivation for using a conformally-coupled scalar field is its simplicity, which allows  to employ the well-known Hamiltonian diagonalization method (see \cite{gmmbook} for a review) to calculate the energy density of the gravitationally produced particles, showing that before the beginning of kination the vacuum polarization effects, which are geometric objects
associated to the creation and annihilation of the so-called {\it quasi-particles} \cite{gmmbook},  are sub-dominant and have no relevant effect in the Friedmann equation. On the contrary, after the abrupt phase transition to kination, heavy massive particles are produced
and, since their energy density decreases as $a^{-3}$ before decaying in lighter particles and as $a^{-4}$ after that,
they will eventually dominate the energy density of the inflation whose decrease is as $a^{-6}$. And, thus, the universe will become reheated. 

\

Coming back to 
the original Peebles-Vilenkin model \cite{pv}, the inflationary part is described by a quartic potential and, according to the recent observations, this does not suit well. To be explicit, for the quartic potential in the inflationary part of this potential the two-dimensional contour of ($n_s$, $r$), where $n_s$ is the 
scalar spectral index and $r$ is the ratio of tensor to scalar perturbations, 
does not enter into the 95\% confidence-level of the Planck results \cite{Planck}. However, a simple change in the inflationary piece $-$quartic potential to a plateau one$-$ can solve this issue (see \cite{hap} for a detailed discussion and also see \cite{hyp}). On the other hand, 
the reheating mechanism followed in \cite{pv} is the
gravitational production of massless particles that results in a reheating temperature of  the  order  of  $1$ TeV. This reheating temperature is not
sufficient to solve the overproduction of the  Gravitational Waves (GWs). As a result, the Big Bang Nucleosynthesis (BBN) process could be hampered. 
 
 \

Dealing once again with the mechanisms to reheat the universe, the question related to the bounds of the reheating temperature arises.
Some works already considered the constraints for reheating in Quintessential Inflation models, both on {\it Instant Preheating} \cite{sami} and on {\it Gravitational Particle Production}~ \cite{figueroa}.
A lower bound is obtained recalling  that the
radiation dominated era is prior to the Big Bang  Nucleosynthesis (BBN) epoch which occurs in the  $1$ MeV  regime \cite{gkr}. 
As a consequence,  the reheating temperature has to be greater
than $1$ MeV (see also \cite{hasegawa}   where the authors obtain lower limits on the reheating temperature in the MeV  regime assuming both radiative and hadronic decays of relic particles only gravitationally interacting and taking into account effects of neutrino self-interactions and oscillations   in the neutrino thermalization calculations).  The upper bound may depend on the theory we are dealing with; for instance, many supergravity and superstring theories contain particles such as the gravitino or a modulus field with only gravitational interactions and, thus,
the late-time decay of these relic products may disturb  
the success of the standard BBN \cite{lindley}, but
this problem can be successfully removed if the  reheating temperature is
of the order of $10^9$ GeV (see for instance \cite{eln}).  This is the reason why  we will restrict  the reheating temperature to remain, more or less,  between $1$ MeV and $10^9$ GeV.

\

Finally, one has to take into account that a viable reheating mechanism has to deal with the effect of the Gravitational Waves (GWs), which are also produced during the phase transition,  in the BBN success by satisfying the observational bounds coming from the overproduction of the GWs \cite{pv} or related to the logarithmic spectrum of their energy density \cite{maggiore}.
As we will see throughout this review, the overproduction of GWs 
does not disturb the BBN success in the models studied (Lorentzian Quintessential Inflation and $\alpha$-attractors in the context of QI), when the reheating is due to the production of superheavy particles, which shows their viability. 


\

In summary, due to the lack of works that explain all the outstanding contributions made in Quintessential Inflation since the appearance of the seminal Peebles-Vilenkin article until now,
the main goal of this review is to show the reader the most relevant aspects
obtained on this topic during this time period. To do it we have chosen the most prominent papers on the subject from our viewpoint, and we have used the results provided by them to write this review.

\

 The review is organized as follows:
In Section \ref{secii}
we review, with great detail, the seminal Peebles-Vilenkin model, calculating the value of the two parameters on which the model depends, the evolution of the inflaton field  throughout history,  the number of e-folds as a function of the reheating temperature, and  additionally  introducing models that improve the original one. Section \ref{seciii} is devoted to the study of a class of Exponential Quintessential Inflation models. We show that, with initial conditions when the pivot scale leaves the Hubble radius (recall that this happens in the slow-roll regime which is an attractor, so one can safely take initial conditions close to the slow-roll solution) during the radiation phase, the solution is never in the basin of attraction of the scaling solution. Therefore, in order to match with the current observational data the potential should be modified introducing a pure exponential term.
In Section \ref{seciv} we review the so-called {\it Lorentzian Quintessential Inflation} (LQI), which is based on the assumption that the main slow-roll parameter as a function of the number of e-folds evolves as the Lorentzian distribution. The $\alpha$-attractors in the context of Quintessential Inflation are revisited in Section \ref{secv}, showing that we obtain the same results as in the case of LQI. In Section \ref{secvi} we review 
other Quintessential Inflation models and in particular 
some aspects of the work of K. Dimopoulos in QI. 
Next, in Section \ref{secvii} we study some  reheating mechanisms in Quintessential Inflation, namely via {\it Gravitational Particle Production}
of massless and superheavy particles, via {\it Instant Preheating}
and via {\it Curvature Reheating},
obtaining 
bounds for the reheating temperature. In Section \ref{secviii}  we deal with the constraints to preserve the BBN success coming from the logarithmic spectrum of GWs and also from its overproduction during the phase transition from the end of inflation to the beginning of kination, showing 
that for the LQI model, when the reheating is via gravitational production of superheavy particles,  all these bounds are easily overpassed.
The last section is devoted to the conclusions of the review.

\

The units used throughout the paper are $\hbar=c=1$, and   the reduced Planck's mass has been denoted by
$M_{pl}\equiv \frac{1}{\sqrt{8\pi G}}\cong 2.44\times 10^{18}$ GeV.

\

\section{The Peebles-Vilenkin model}\label{secii}

The first  Quintessential Inflation  scenario was proposed by Peebles and Vilenkin in their seminal paper \cite{pv} at the end of the last century, soon after the discovery of the current cosmic acceleration \cite{riess, perlmutter}. The corresponding potential is given by 
\begin{eqnarray}\label{pv}
V(\varphi)=\left\{\begin{array}{ccc}
\lambda (\varphi^4+M^4)& \mbox{for}& \varphi\leq 0\\
\lambda \frac{M^8}{\varphi^4+M^4} &\mbox{for}& \varphi\geq 0,
\end{array}\right.
\end{eqnarray}
where $\lambda$ is a dimensionless parameter and $M\ll M_{pl}$ is a very small mass compared with the Planck's one.  At this point, note that the quartic potential is the responsible for inflation and the inverse power law leads to dark energy (in that case quintessence) at late times.

\

One can see that the model is very simple and only depends on two parameters which can be calculated as follows:
First of all, we calculate the main slow-roll parameters
\begin{eqnarray}\label{slowroll}
\epsilon\equiv\frac{M_{pl}^2}{2}\left(\frac{V_{\varphi}}{V}\right)^2\cong \frac{8M_{pl}^2}{\varphi^2},  \qquad 
\eta\equiv{M_{pl}^2}\frac{V_{\varphi\varphi}}{V}=\frac{12M_{pl}^2}{\varphi^2}. \end{eqnarray}
So, the spectral index and the ratio of tensor to scalar perturbations are given by \cite{btw}
\begin{eqnarray}\label{spectral}
n_s\cong 1-6\epsilon_*+2\eta_*\cong 1-\frac{24M_{pl}^2}{\varphi^2_*},
\qquad r\cong 16\epsilon_*\cong \frac{128M_{pl}^2}{\varphi^2_*},\end{eqnarray}
where the star denotes that the quantities are evaluated when the pivot scale leaves the Hubble radius (at the horizon crossing). Thus, one has the important relation
$16(1-n_s)=3r$ between both quantities.

\

On the other hand, using the formula of the power spectrum of scalar perturbations  (see for instance \cite{btw})
\begin{eqnarray}
P_{\zeta}=\frac{H_*^2}{8\pi^2\epsilon_*M_{pl}^2}\sim 2\times 10^{-9},
\end{eqnarray}
for the Peebles-Vilenkin model it yields that 
\begin{eqnarray}
P_{\zeta}\cong \frac{\lambda \varphi_*^4}{24\pi^2\epsilon_*M_{pl}^4}\cong \frac{8\lambda}{3\pi^2\epsilon_*^3}\cong \frac{72\lambda}{\pi^2(1-n_s)^3}\sim 2\times 10^{-9},
\end{eqnarray}
where we have used the formulas (\ref{slowroll}) and (\ref{spectral}). And, taking for example $n_s\cong 0.96$, which is approximately its observational value \cite{planck18, planck18a}, we get
\begin{eqnarray}
\lambda\sim \frac{\pi^2}{36}(1-n_s)^3\times 10^{-9}\sim10^{-14}.
\end{eqnarray}

\

The calculation of the other parameter is more involved and for that one needs to know the evolution of the inflaton field up to the present time. As we will see, the value of the inflaton at the present time is of the order of $40 M_{pl}$ (see Fig. \ref{fig:phi_pv}). 
So, taking into account that in order to reproduce the current cosmic acceleration the kinetic energy has to be negligible compared with the potential one at the present time, 
in order to match with the present energy density we will have
{\begin{eqnarray}
\lambda \frac{M^8}{40^4M_{pl}^4+M^4}\cong \lambda \frac{M^8}{40^4M_{pl}^4}\sim H_0^2M_{pl}^2, 
\end{eqnarray}}
where $H_0\sim 10^{-61} M_{pl}$ is the present value of the Hubble rate. Finally, we get
\begin{eqnarray}
M\sim 40^{1/2}\lambda^{-1/8} H_0^{1/4}M_{pl}^{3/4}\sim 10^{-13} M_{pl}\sim 10^5 \mbox{ GeV}.
\end{eqnarray}

\

\begin{figure}[H]
    \centering
    \includegraphics[width=0.4\textwidth]{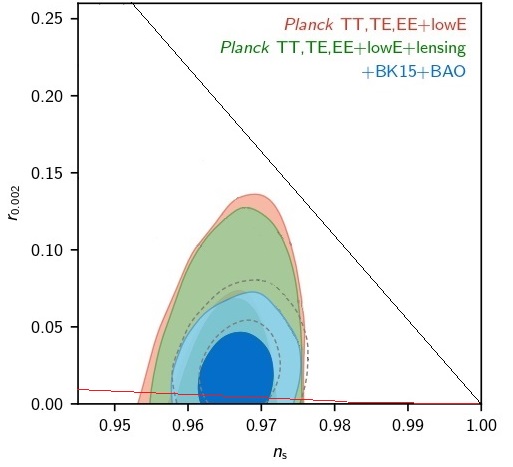}
    \caption{Clearly the line $16(1-n_s)=3r$ (black line) does not enter in the $2\sigma$ Confidence Level, in contrast to the parabola $(1-n_s)^2=\frac{\alpha^2}{2}r$ (red line) with $\alpha=1$ from the improved model studied in next subsection} (Figure courtesy of the Planck 2018 Collaboration \cite{planck18a})
    \label{fig:ns_r_pv}
\end{figure}

Unfortunately, the original Peebles-Vilenkin model provides a spectral index and a tensor/scalar ratio that do not enter in the  two dimensional marginalized joint confidence contour at $2\sigma$ CL for the Planck TT, TE, EE + low E+ lensing + BK14 + BAO likelihood (see Fig. \ref{fig:ns_r_pv}).

\

For this reason, we need to improve the part of the potential which provides inflation, that is, the quartic potential.

\subsection{Improvements}

 For that purpose, one would need to consider plateau potentials \cite{Geng:2017mic} or $\alpha$-attractors \cite{attractor1,vardayan}, such as an Exponential SUSY Inflation type potential

\begin{eqnarray} \label{susy}
V^1_{\alpha}(\varphi)=\left\{\begin{array}{ccc}
\lambda M_{pl}^4\left( 1-e^{\alpha\varphi/M_{pl}} + \left(\frac{M}{M_{pl}}\right)^4\right) & \mbox{for} & \varphi\leq 0\\
\lambda\frac{M^8}{\varphi^{4}+M^4} &\mbox{for} & \varphi\geq 0,\end{array}
\right.
\end{eqnarray}
or a Higgs Inflation-type potential
\begin{eqnarray} \label{higgs}
V^2_{\alpha}(\varphi)=\left\{\begin{array}{ccc}
\lambda M_{pl}^4\left( 1-e^{\alpha\varphi/M_{pl}} + \left(\frac{M}{M_{pl}}\right)^2\right)^2 & \mbox{for} & \varphi\leq 0\\
\lambda\frac{M^8}{\varphi^{4}+M^4} &\mbox{for} & \varphi\geq 0.\end{array}
\right.
\end{eqnarray}

\

For the first model, the slow-roll parameters are given by
\begin{eqnarray}
\epsilon\cong \frac{\alpha^2}{2}\frac{e^{2\alpha\varphi/M_{pl}}}{\left( 1-e^{\alpha\varphi/M_{pl}} \right)^2}, \qquad 
\eta\cong -{\alpha^2}\frac{e^{\alpha\varphi/M_{pl}}}{1-e^{\alpha\varphi/M_{pl}}},\end{eqnarray}
and, taking into account that when the pivot scale leaves the Hubble radius $e^{\alpha\varphi_*/M_{pl}}\ll 1$, one gets
\begin{eqnarray}
\epsilon_*\cong \frac{\alpha^2}{2}{e^{2\alpha\varphi_*/M_{pl}}}, \qquad 
\eta_*\cong -{\alpha^2}{e^{\alpha\varphi_*/M_{pl}}},\end{eqnarray}
and thus, the spectral index and the tensor/scalar ratio are equal to
\begin{eqnarray}
n_s\cong 1-6\epsilon_*+2\eta_*\cong 1-2{\alpha^2}{e^{\alpha\varphi_*/M_{pl}}}, \qquad
r\cong 8{\alpha^2}{e^{2\alpha\varphi_*/M_{pl}}},\end{eqnarray}
obtaining the relation $(1-n_s)^2=\frac{\alpha^2}{2}r$.

\

We can also calculate the number of e-folds from the horizon crossing to the end of inflation, namely ${\mathcal N}=\int_{t_*}^{t_{END}} Hdt$. In the slow-roll regime, where 
\begin{eqnarray}
3H\dot{\varphi}+V_{\varphi}\cong 0,\qquad H^2\cong \frac{V}{3M_{pl}^2},
\end{eqnarray}
it can be approximated by
\begin{eqnarray}
{\mathcal N}\cong \frac{1}{M_{pl}}\int_{\varphi_*}^{\varphi_{END}}\frac{1}{\sqrt{2\epsilon}}d\varphi,
\end{eqnarray}
which for the first potential leads to
\begin{eqnarray}
{\mathcal N}\cong \frac{1}{\alpha M_{pl}}\int_{\varphi_*}^{\varphi_{END}}\left(e^{-\alpha \varphi/M_{pl}}-1
\right)d\varphi=\frac{1}{\alpha^2}\left(e^{-\alpha \varphi_*/M_{pl}}-e^{-\alpha \varphi_{END}/M_{pl}} \right)\nonumber\\-
\frac{1}{\alpha M_{pl}}(\varphi_{END}-\varphi_*)
\cong \frac{1}{\alpha^2}e^{-\alpha \varphi_*/M_{pl}}\cong \frac{2}{1-n_s}.
\end{eqnarray}

\

In the same way, for the second potential, the slow-roll parameters are given by 
\begin{eqnarray}
\epsilon\cong 2{\alpha^2}\frac{e^{2\alpha\varphi/M_{pl}}}{\left( 1-e^{\alpha\varphi/M_{pl}} \right)^2}, \qquad 
\eta\cong -2{\alpha^2}\frac{e^{\alpha\varphi/M_{pl}}}{(1-e^{\alpha\varphi/M_{pl}})^2},\end{eqnarray}
and, taking into account again that when the pivot scale leaves the Hubble radius $e^{\alpha\varphi_*/M_{pl}}\ll 1$, we get
\begin{eqnarray}
\epsilon_*\cong 2{\alpha^2}{e^{2\alpha\varphi_*/M_{pl}}}, \qquad 
\eta_*\cong -2{\alpha^2}{e^{\alpha\varphi_*/M_{pl}}},\end{eqnarray}
and thus, the spectral index and the tensor/scalar ratio are equal to
\begin{eqnarray}
n_s\cong 1-6\epsilon_*+2\eta_*\cong 1-4{\alpha^2}{e^{\alpha\varphi_*/M_{pl}}}, \qquad
r\cong 32{\alpha^2}{e^{2\alpha\varphi_*/M_{pl}}},\end{eqnarray}
obtaining the same relation as for the other potential, namely  $(1-n_s)^2=\frac{\alpha^2}{2}r$.

\

In addition, for the number of e-folds, one also obtains the same result
\begin{eqnarray}
{\mathcal N}
\cong \frac{1}{2\alpha^2}e^{-\alpha \varphi_*/M_{pl}}\cong \frac{2}{1-n_s}, 
\end{eqnarray}
and what is important to note is that
for both potentials  the spectral index and the ratio of tensor to scalar perturbations have the same relation with the number of e-folds
\begin{eqnarray}
n_s\cong 1-\frac{2}{{\mathcal N}},\qquad r\cong \frac{8}{\alpha^2 {\mathcal N}^2},
\end{eqnarray}
which implies that for $\alpha\sim {\mathcal O}(1)$ and for a number of $e$-folds greater than $60$, which is typical in quintessential inflation due to the kination phase,  the ratio of tensor to scalar perturbations is less than $0.003$. Thus, the spectral index and the tensor/scalar ratio 
enter perfectly in the  two dimensional marginalized joint confidence contour at $2\sigma$ CL for the Planck TT, TE, EE + low E+ lensing + BK14 + BAO likelihood (see Fig. \ref{fig:ns_r_pv}).

\subsection{Dynamical evolution of the Peebles-Vilenkin model: from the beginning of kination to the matter-radiation equality}

Now we want to understand the evolution of the Quintessential Inflation models after inflation, which only depends on the tail of the potential. We will focus on the Peebles-Vilenkin one, and for
 that case,  inflation ends when $\epsilon=1$, which occurs for $\varphi_{END}=2\sqrt{2}M_{pl}$. Taking into account that an alternative expression of this slow-roll parameter is $\epsilon=-\frac{\dot{H}}{H^2}$, we conclude that at the end of inflation the effective Equation of State (EoS) parameter, namely $w_{eff}$, satisfies
\begin{eqnarray}\label{EoS}
w_{eff}=\frac{P}{\rho}=\frac{\dot{\varphi}^2/2-V}{\dot{\varphi}^2/2+V}=-1-\frac{2\dot{H}}{3H^2}=-1/3,
\end{eqnarray}
where we have used the Friedmann and Raychaudhuri equations
\begin{eqnarray}\label{FR}
H^2=\frac{\rho}{3M_{pl}^2},\qquad \dot{H}=-\frac{1}{2M_{pl}^2}(\rho+P),
\end{eqnarray}
being $\rho$ the energy density and $P$ the corresponding pressure. 

\

Then, from (\ref{EoS}) one gets the following relation at the end of inflation,
$\dot{\varphi}_{END}^2=V(\varphi_{END})$, meaning that the energy density at the end of inflation is 
\begin{eqnarray}
\rho_{END}=\frac{3}{2}V(\varphi_{END})= \frac{3}{2}\lambda(\varphi_{END}^4+M^4)\cong 
\frac{3}{2}\lambda\varphi_{END}^4=96\lambda M_{pl}^4.
\end{eqnarray}

Next, we have to assume,  as usual,  that there is no drop of energy between the end of inflation and the beginning of kination, which for these models occurs when $\varphi_{kin}\cong 0$. So, at the beginning of kination the energy density is approximately given by $\rho_{kin}\cong 96\lambda M_{pl}^4\Longrightarrow H_{kin}\cong 4\sqrt{2\lambda}M_{pl}$ and,
since all the energy is practically kinetic, the effective EoS parameter is very close to
$w_{eff}\cong 1$. For this value, combining the Friedmann and Raychaudhuri equations (\ref{FR}), one gets 
\begin{eqnarray}
\dot{H}=-3H^2,
\end{eqnarray}
whose solution is given by $H(t)=\frac{1}{3t}$. 

\

Then, coming back to the Friedmann equation and taking into account that during kination all the energy density is kinetic,
we get 
\begin{eqnarray}
\frac{1}{9t^2}=\frac{\dot{\varphi}^2}{6M_{pl}^2}\Longrightarrow \dot{\varphi}(t)=\sqrt{\frac{2}{3}}\frac{M_{pl}}{t}\Longrightarrow 
\varphi(t)=\varphi_{kin}+\sqrt{\frac{2}{3}}M_{pl}\ln \left( \frac{t}{t_{kin}} \right),\end{eqnarray}
which in terms of the Hubble rate could be written as follows:
\begin{eqnarray}
\dot{\varphi}(t)=\sqrt{6}H(t){M_{pl}},  \qquad 
\varphi(t)=\sqrt{\frac{2}{3}}M_{pl}\ln \left( \frac{H_{kin}}{H(t)} \right).\end{eqnarray}

\

At this point, we have to take into account that during the phase transition from the end of inflation to the beginning of kination the adiabatic regime is broken and particles are created. Here we will assume that these particles, which we consider very massive and conformally coupled to gravity, are gravitationally produced. So, they must decay into lighter ones in order to form a relativistic plasma which will eventually dominate the energy density of the universe and will become reheated.
And then, since the kination regime ends when the energy density of the inflaton field is of the same order than the one of the produced particles, 
two different scenarios appear:
\begin{enumerate}
\item Decay before the end of the kination period.
\item Decay after the end of the kination period.
\end{enumerate}

\subsubsection{Decay before the end of kination}

In this case, since the thermalization of the decay products is nearly instantaneous and occurs before the end of kination,  the reheating time coincides with the end of the kination. Hence, we will have
\begin{eqnarray}
\dot{\varphi}_{reh}=\sqrt{6}H_{reh}{M_{pl}},  \qquad 
\varphi_{reh}=\sqrt{\frac{2}{3}}M_{pl}\ln \left( \frac{H_{kin}}{H_{reh}} \right).\end{eqnarray}

Taking into account that at the reheating time the energy density of the inflaton field is the same as the one of the radiation, the Friedmann equation will become $H_{reh}^2=\frac{2\rho_{reh}}{3M_{pl}^2}$ and, using the Stefan-Boltzmann law 
$\rho_{reh}=\frac{\pi^2}{30}g_{reh} T_{reh}^4$, where $g_{reh}$ denotes the effective number of degrees of freedom at the reheating time, we get
\begin{eqnarray}\label{varphireh}
\varphi_{reh}=\sqrt{\frac{2}{3}}M_{pl}\ln\left( \frac{12}{\pi}\sqrt{\frac{10\lambda}{g_{reh}}}
\left(\frac{M_{pl}}{T_{reh}} \right)^2\right)
\qquad \mbox{and}  \qquad  { \dot{\varphi}_{reh}=\sqrt{\frac{\pi^2g_{reh}}{15}} T_{reh}^2}.\end{eqnarray} 

The next step is to calculate the value of the field at the beginning of the matter-radiation equality. During the radiation domination the effective EoS parameter is $w_{eff}=1/3$. Thus, combining once again the Friedmann and Raychaudhuri equations we have $\dot{H}=-2H^2$, whose solution is $H(t)=\frac{1}{2t}$. Then, an approximate solution of the conservation equation
\begin{eqnarray}
\ddot{\varphi}+3H\dot{\varphi}+V_{\varphi}=0
\end{eqnarray}
can be obtained disregarding once again the potential, because during the radiation domination the potential energy is irrelevant compared with the kinetic one. Hence, the equation 
\begin{eqnarray}
\ddot{\varphi}+\frac{3}{2t}\dot{\varphi}=0
\end{eqnarray}
has the following solution,
\begin{eqnarray}
\dot{\varphi}(t)=\dot{\varphi}_{reh}\left(\frac{t_{reh}}{t}   \right)^{3/2}=
\sqrt{\frac{\pi^2g_{reh}}{15}} 
T_{reh}^2\left(\frac{2H(t)}{3H_{reh}}   \right)^{3/2},\nonumber \\
 \varphi(t)=\varphi_{reh}+2\dot{\varphi}_{reh}t_{reh} \left(1-\sqrt{\frac{t_{reh}}{t}}   \right)
 =\varphi_{reh}+2\sqrt{\frac{2}{3}}M_{pl}\left(1-\sqrt{\frac{2H(t)}{3H_{reh}}}\right),
\end{eqnarray}
which at the matter-radiation equality leads to

\begin{eqnarray}\label{xeq}
 \varphi_{eq}
 =\sqrt{\frac{2}{3}}M_{pl}\ln\left( \frac{12}{\pi}\sqrt{\frac{10\lambda}{g_{reh}}}
\left(\frac{M_{pl}}{T_{reh}} \right)^2\right)+2\sqrt{\frac{2}{3}}M_{pl}\left(1-\sqrt{\frac{2}{3}}\left( \frac{g_{eq}}{g_{rh}} \right)^{1/4}\frac{T_{eq}}{T_{rh}}\right), \nonumber\\
\dot{\varphi}_{eq}
=\frac{2\pi}{9}\sqrt{\frac{2g_{eq}}{5}}\left(\frac{g_{eq}}{g_{rh}} \right)^{1/4}\frac{T_{eq}^3}{T_{rh}},
\end{eqnarray} 
where we have used that $H_{eq}^2=\frac{2\rho_{eq}}{3M_{pl}^2}$ with 
$\rho_{eq}=\frac{\pi^2}{30}g_{eq} T_{eq}^4$ being $g_{eq}\cong 3.36$ the effective number of degrees of freedom at the matter-radiation epoch.

\

Finally, noticing the adiabatic evolution after the matter-radiation, we will have
$T_{eq}=\frac{a_0}{a_{eq}}T_0=(1+z_{eq})T_0$, where the sub-index $"0"$ denotes the present time and $z_{eq}$ is the red-shift at the matter-radiation equality. So,
\begin{eqnarray}\label{xeq}
 \varphi_{eq}
 =\sqrt{\frac{2}{3}}M_{pl}\ln\left( \frac{12}{\pi}\sqrt{\frac{10\lambda}{g_{reh}}}
\left(\frac{M_{pl}}{T_{reh}} \right)^2\right)+2\sqrt{\frac{2}{3}}M_{pl}\left(1-\sqrt{\frac{2}{3}}\left( \frac{g_{eq}}{g_{rh}} \right)^{1/4}(1+z_{eq})\frac{T_{0}}{T_{reh}}\right), \nonumber\\
\dot{\varphi}_{eq}
=\frac{2\pi}{9}\sqrt{\frac{2g_{eq}}{5}}\left(\frac{g_{eq}}{g_{rh}} \right)^{1/4}(1+z_{eq})^3\frac{T_0}{T_{reh}}T_0^2,
\end{eqnarray} 
that is, we have obtained the values of the field at the matter-radiation equality as a function of the reheating temperature and the observational data $T_0\cong 2.73$ K and $z_{eq}\sim 3\times 10^3$. In fact, since $T_0\ll T_{reh}$, one can safely do the approximation
\begin{eqnarray}\label{xeq}
 \varphi_{eq}
 \cong \sqrt{\frac{2}{3}}M_{pl}\ln\left( \frac{12}{\pi}\sqrt{\frac{10\lambda}{g_{reh}}}
\left(\frac{M_{pl}}{T_{reh}} \right)^2\right),\qquad 
\dot{\varphi}_{eq}\cong 0,
\end{eqnarray} 
which only depends on the reheating temperature.

\subsubsection{Decay after the end of the kination period}

In that scenario, 
at the end of kination, when the energy density of the field is equal to the one of the produced particles,  one has 
\begin{eqnarray}
\varphi_{end}=\sqrt{\frac{2}{3}}M_{pl}\ln\left( \frac{H_{kin}}{H_{end}} \right), \qquad
\dot{\varphi}_{end}=\sqrt{6}M_{pl}H_{end}.
\end{eqnarray}

Denoting by $\rho_{\varphi}(t)$  the energy density of the inflaton field and by $\rho_{\chi}$ the energy density of the produced particles, at the end of kination we will have
\begin{eqnarray}
\rho_{\varphi, end}=\rho_{\varphi, kin}\left(\frac{a_{kin}}{a_{end}}  \right)^6, \qquad
\rho_{\chi, end}=\rho_{\chi, kin}\left(\frac{a_{kin}}{a_{end}}  \right)^3, 
\end{eqnarray}
and, since $\rho_{\varphi, end}=\rho_{\chi, end}$, 
the important parameter defined by $\Theta\equiv\left( \frac{a_{kin}}{a_{end}}  \right)^3$ and named {\it heat efficiency}
 satisfies
\begin{eqnarray}
\Theta=\frac{\rho_{\chi, kin}}{\rho_{\varphi, kin}}=
\frac{\rho_{\chi, kin}}{96\lambda M_{pl}^4}.
\end{eqnarray}

On the other hand, using that $H_{end}^2=\frac{2\rho_{\varphi,end}}{3M_{pl}^2}$
one arrives at
 $H_{end}=\sqrt{2}H_{kin}\Theta=8\sqrt{\lambda}M_{pl} \Theta$, and then 
 \begin{eqnarray}
\varphi_{end}=-\sqrt{\frac{2}{3}}M_{pl}\ln\left( \sqrt{2}\Theta \right), \qquad
\dot{\varphi}_{end}=8\sqrt{6\lambda}M_{pl}^2\Theta.
\end{eqnarray}

During the period between $t_{end}$ and $t_{reh}$ the universe is matter dominated and, thus, the Hubble parameter evolves as $H=\frac{2}{3t}$. Since the gradient of the potential can also be disregarded at this
epoch, the equation of the scalar field becomes $\ddot{\varphi}+\frac{2}{t}\dot{\varphi}=0$, yielding at the reheating time
\begin{eqnarray}
\varphi_{reh}=\varphi_{end}+\sqrt{\frac{2}{3}}M_{pl}\left( 1-\frac{{t}_{end}}{t_{reh}}  \right)
=\varphi_{end}+\sqrt{\frac{2}{3}}M_{pl}\left( 1-\frac{H_{reh}}{2H_{end}}\right)\nonumber\\ =
-\sqrt{\frac{2}{3}}M_{pl}\ln\left( \sqrt{2}\Theta \right)
+\sqrt{\frac{2}{3}}M_{pl}\left( 1-
{\frac{\pi}{48}\sqrt{\frac{g_{reh}}{5\lambda}}\frac{T_{reh}^2}{M_{pl}^2\Theta}}\right) \end{eqnarray}
and
\begin{eqnarray}
\dot{\varphi}_{reh}
=\sqrt{\frac{2}{3}}\frac{M_{pl}{t}_{end}}{t_{reh}^2}
=\frac{\sqrt{3}}{16\sqrt{2\lambda}}\frac{H_{rh}^2}{\Theta}
=\frac{\sqrt{3}\pi^2}{720\sqrt{2\lambda}}\frac{g_{reh}T_{reh}^4}{M_{pl}^2\Theta}.\end{eqnarray}


Finally, 
in all  the radiation period one can continue disregarding the potential,  obtaining
\begin{eqnarray}
\varphi(t)=\varphi_{reh}+2\dot{\varphi}_{reh}t_{ter}\left(1-\sqrt{\frac{t_{reh}}{t}}\right),
\end{eqnarray}
and thus, at the matter-radiation equality one has
\begin{align}\label{eq}
 \varphi_{eq} =\varphi_{rh}+\sqrt{\frac{2}{3}}M_{pl}\frac{H_{reh}}{H_{end}}\left(1-\sqrt{\frac{4H_{eq}}{3H_{rh}}}\right)\nonumber\\
 =\varphi_{rh}+
 \frac{\pi}{24}\sqrt{\frac{g_{reh}}{30\lambda}}\frac{T_{reh}^2}{M_{pl}\Theta}
 \left(1-\sqrt{\frac{4}{3}}\left( \frac{g_{eq}}{g_{reh}} \right)^{1/4}\frac{T_{eq}}{T_{reh}}\right) \nonumber \\
=
-\sqrt{\frac{2}{3}}M_{pl}\ln\left( \sqrt{2}\Theta \right)
+\sqrt{\frac{2}{3}}M_{pl}\left( 1-
{\frac{\pi}{48}\sqrt{\frac{g_{reh}}{5\lambda}}\frac{T_{reh}^2}{M_{pl}^2\Theta}}\right)\\ \nonumber+\frac{\pi}{24}\sqrt{\frac{g_{reh}}{30\lambda}}\frac{T_{reh}^2}{M_{pl}\Theta}
 \left(1-\sqrt{\frac{4}{3}}\left( \frac{g_{eq}}{g_{reh}} \right)^{1/4}\frac{T_{eq}}{T_{reh}}\right)\\ \nonumber
=-\sqrt{\frac{2}{3}}M_{pl}\ln\left( \sqrt{2}\Theta \right)
+\sqrt{\frac{2}{3}}M_{pl}
-\frac{\pi}{36}\sqrt{\frac{g_{reh}}{10\lambda}}
 \left( \frac{g_{eq}}{g_{reh}} \right)^{1/4}(1+z_{eq})\frac{T_{0}T_{reh}}{M_{pl}\Theta}
 ,
  \end{align}
 and in the same way,  
 \begin{align}\label{doteq}
\dot{\varphi}_{eq}=\dot{\varphi}_{reh}\left(\frac{t_{reh}}{t_{eq}}\right)^{\frac{3}{2}}=\left(\frac{16g_{eq}}{9g_{reh}}\right)^{\frac{3}{4}}\left(\frac{T_{eq}}{T_{reh}}\right)^3 \dot{\varphi}_{reh}
=\frac{\pi^2(1+z_{eq})^3}{270\sqrt{2\lambda}}\left(\frac{g_{eq}^3}{g_{reh}} \right)^{\frac{1}{4}}\frac{T_0^3T_{reh}}{M_{pl}^2\Theta}.
\end{align}

\

As we will see, by calculating the value of $\Theta$ for several reheating mechanisms, we can safely do the approximation
\begin{align}\label{eq}
 \varphi_{eq} \cong 
-\sqrt{\frac{2}{3}}M_{pl}\ln\left( \sqrt{2}\Theta \right)
+\sqrt{\frac{2}{3}}M_{pl}
 ,\qquad \dot{\varphi}_{eq}\cong 0,
  \end{align}
which only depends on the value of the parameter $\Theta$. Effectively, using
for example {\it Instant Preheating} (see subsection 7.2), one gets $\Theta=\frac{g^2}{2\pi^2}$, which for the viable value of the coupling $g\sim 10^{-4}$ yields $\Theta\sim 10^{-9}$, and one can easily check that the approximation holds.

\subsection{Dynamical evolution of the Peebles-Vilenkin model: from the matter-radiation equality
up to now}

After the matter-radiation equality the dynamical equations cannot be solved analytically and, thus, one needs to use numerics to compute them. In order to do that, we need to use a "time" variable  that we choose to be the number of $e$-folds up to the present epoch, namely $N\equiv -\ln(1+z)=\ln\left( \frac{a}{a_0}\right)$. Now, using  the variable $N$,  one can recast the  energy density of radiation and matter respectively as
\begin{eqnarray}
\rho_{r}(a)={\rho_{r, eq}}\left(\frac{a_{eq}}{a}  \right)^4\Longrightarrow \rho_{r}(N)= \frac{\rho_{eq}}{2}e^{4(N_{eq}-N)} 
\end{eqnarray}
and
\begin{eqnarray}
\rho_{m}(a)={\rho_{m,eq}}\left(\frac{a_{eq}}{a}  \right)^3\Longrightarrow \rho_{m}(N)=\frac{\rho_{eq}}{2}e^{3(N_{eq}-N)},
\end{eqnarray}
where $\rho_{eq}=\rho_{m,eq}+\rho_{r,eq}$ and  the value of the energy density of the matter or radiation $\rho_{m,eq}=\rho_{r,eq}$ at the matter-radiation equality 
can be calculated as follows:
Using the present value of the ratio of the matter energy density to the critical one $\Omega_{m,0}=0.308$ and the present value of the Hubble rate $H_0\cong 68\; \mbox{Km/sec/Mpc}\sim 10^{-33}$ eV,
one can deduce that  the present value of the matter energy density is $\rho_{m,0}=3H_0^2M_{pl}^2\Omega_{m,0}\cong 3\times 10^{-121} M_{pl}^4$, and at the matter-radiation equality $\rho_{m,eq}=\rho_{m,0}(1+z_{eq})^3$.




\

In order to obtain the dynamical system for this scalar field model, we 
introduce the following dimensionless variables,
 \begin{eqnarray}
 x=\frac{\varphi}{M_{pl}}, \qquad y=\frac{\dot{\varphi}}{H_0 M_{pl}},
 \end{eqnarray}
 where $H_0\sim 1.42 \times 10^{-33}$ eV denotes once again the current value of the Hubble parameter. Now, using the variable
 $N = - \ln (1+z)$ defined above and also using the conservation equation $\ddot{\varphi}+3H\dot{\varphi}+V_{\varphi}=0$, one can construct the following  non-autonomous dynamical system,
 \begin{eqnarray}\label{system}
 \left\{ \begin{array}{ccc}
 x^\prime & =& \frac{y}{\bar H}~,\\
 y^\prime &=& -3y-\frac{\bar{V}_x}{ \bar{H}}~,\end{array}\right.
 \end{eqnarray}
 where the prime represents the derivative with respect to $N$, $\bar{H}=\frac{H}{H_0}$   and $\bar{V}=\frac{V}{H_0^2M_{pl}^2}$. Moreover, the Friedmann equation now looks as  
 \begin{eqnarray}\label{friedmann}
 \bar{H}(N)=\frac{1}{\sqrt{3}}\sqrt{ \frac{y^2}{2}+\bar{V}(x)+ \bar{\rho}_{r}(N)+\bar{\rho}_{m}(N) }~,
 \end{eqnarray}
where we have introduced the dimensionless energy densities
 $\bar{\rho}_{r}=\frac{\rho_{r}}{H_0^2M_{pl}^2}$ and 
 $\bar{\rho}_{m}=\frac{\rho_{m}}{H_0^2M_{pl}^2}$.

Then, we have to integrate the dynamical system, starting at $N_{eq}=-\ln(1+z_{eq})$, with initial conditions $x_{eq}$ and $y_{eq}$ which are obtained analytically in the previous subsection. The value of the parameter $M$ is obtained equaling at $N=0$ the equation (\ref{friedmann}) to $1$, i.e., imposing $\bar{H}(0)=1$.



\

\begin{figure}[H]
\begin{center}
\includegraphics[scale=0.45]{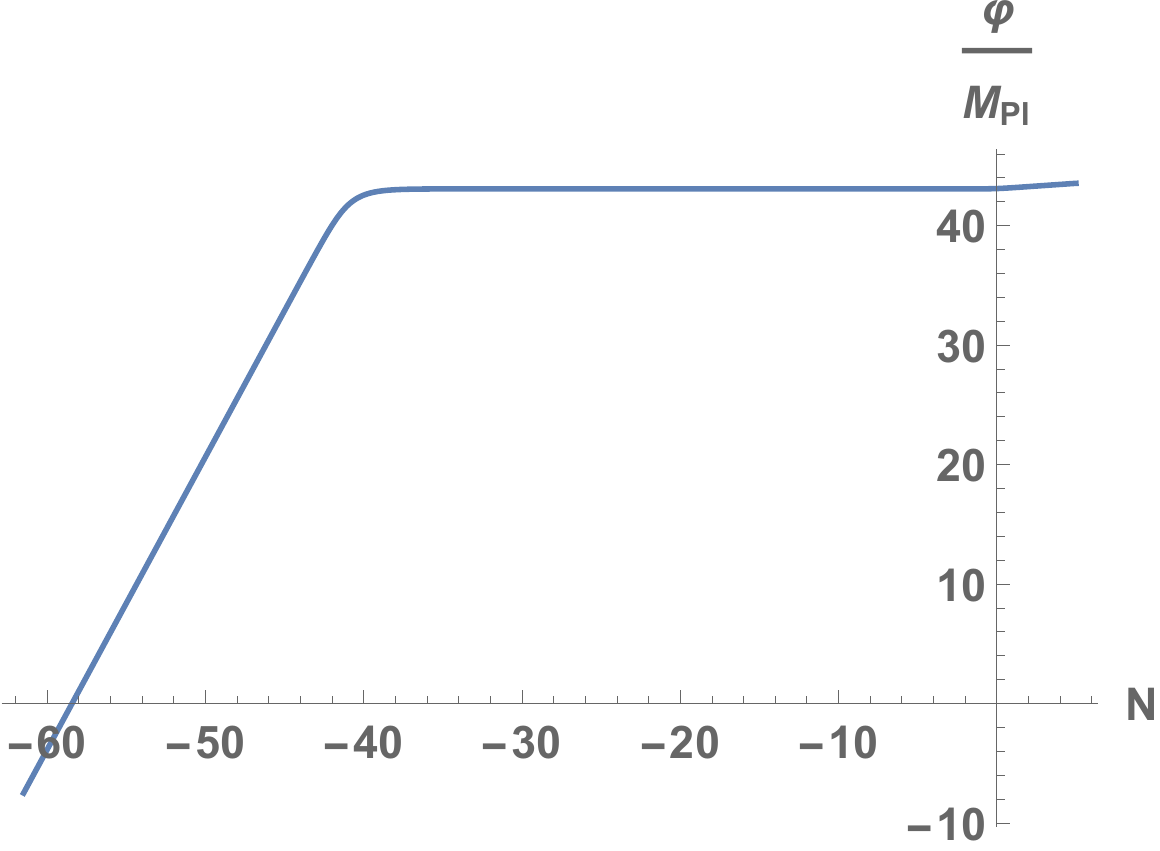}
\end{center}
\caption{Evolution of the scalar field $\varphi$ in the improved Peebles-Vilenkin model.}
\label{fig:phi_pv}
\end{figure}

\begin{figure}[H]
\begin{center}
\includegraphics[scale=0.45]{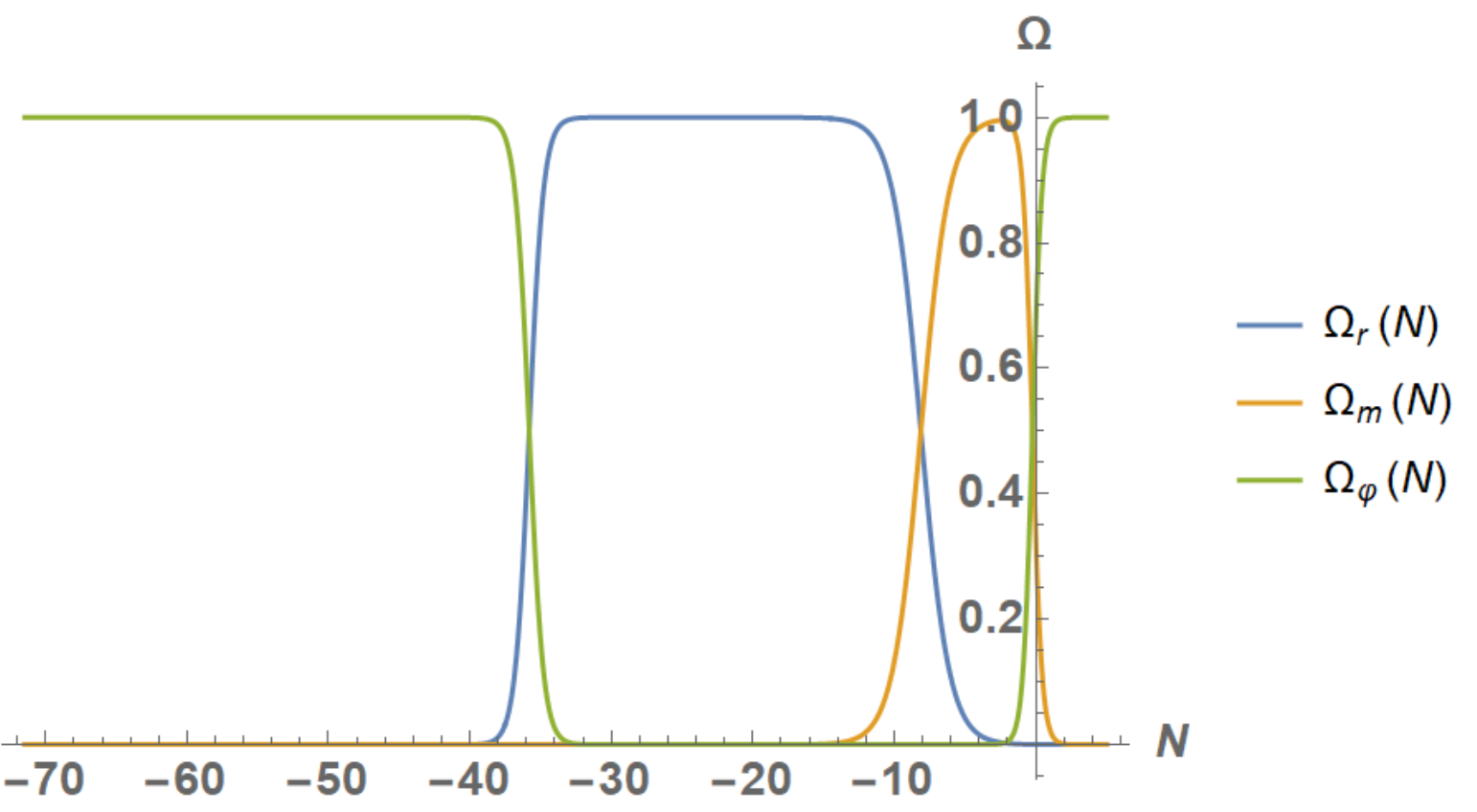}
\includegraphics[scale=0.45]{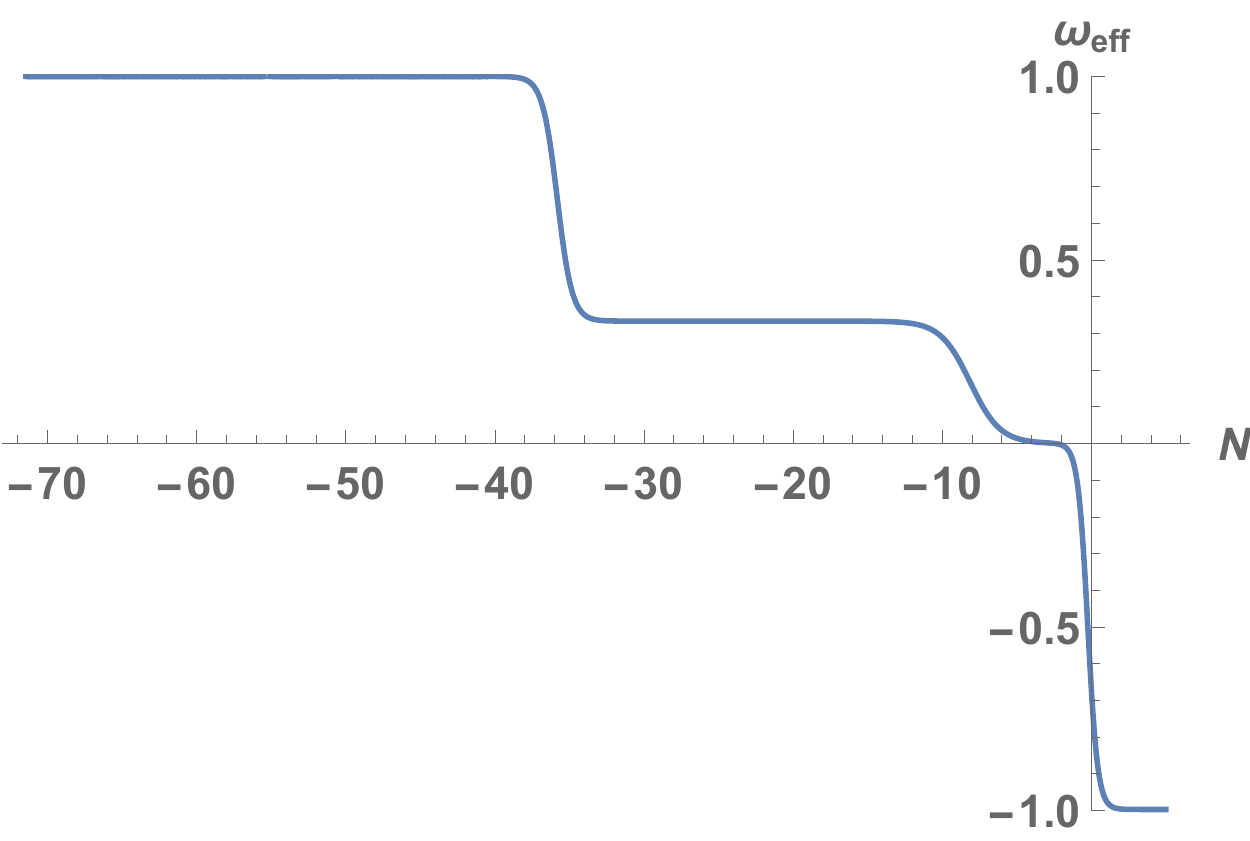}
\end{center}
\caption{\it{\textbf{Left:} The density parameters $\Omega_m=\frac{\rho_m}{3H^2M_{pl}^2}$ (orange curve), $\Omega_r=\frac{\rho_r}{3H^2M_{pl}^2}$ (blue curve) and $\Omega_{\varphi})=\frac{\rho_{\varphi}}{3H^2M_{pl}^2}$, from kination to future times. \textbf{Right:} The effective Equation of State parameter $w_{eff}$, from kination to future times. As one can see in the picture, after kination the Universe enters in a large period of time where radiation dominates. Then, after the matter-radiation equality, the Universe becomes matter-dominated and, finally, near the present, it enters in a new accelerated phase where $w_{eff}$ approaches $-1$.}}
\label{fig:modifiedPV2}
\end{figure}

Numerical calculations show that $M\sim 10^5$ GeV for a reheating temperature of $10^5$ GeV, as in our heuristic reasoning. In addition, it remains of the same order for the other values of the reheating temperature within the allowed range.
In Figure \ref{fig:phi_pv} we display the evolution of the scalar field $\varphi$ and in Figure \ref{fig:modifiedPV2} we plot the graph of the parameter $\Omega_{i}\equiv \frac{\rho_i}{3H^2M_{pl}^2}$ for matter, radiation and dark energy, and also the evolution of the effective EoS parameter. 

\subsection{The number of e-folds}

In this subsection we aim to find the relation between the number of e-folds from the horizon crossing to the end of inflation and the reheating temperature $T_{reh}$.
We start with the well-known formula 
\cite{ll}
\begin{eqnarray}\frac{k_*}{a_0H_0}=e^{-{\mathcal N}}\frac{H_*}{H_0}\frac{a_{END}}{a_{kin}}\frac{a_{kin}}{a_{reh}}
\frac{a_{reh}}{a_{eq}}\frac{a_{eq}}{a_{0}},
\end{eqnarray}
where, as previously, $a$ is the scale factor and $a_{END}$, $a_{kin}$, $a_{reh}$, $a_{eq}$ and $a_0$ denote respectively its value at the end of inflation, at the beginning of kination, radiation and  matter domination,  and finally at the present time.

\

Taking into account the evolution during kination and radiation we have 
\begin{eqnarray}
\left(\frac{a_{kin}}{a_{reh}} \right)^6=\frac{\rho_{reh}}{\rho_{kin}} \quad \mbox{and}\quad 
\left(\frac{a_{reh}}{a_{eq}} \right)^4=\frac{\rho_{eq}}{\rho_{reh}}
\end{eqnarray}
and, noting that $H_0\sim 2\times 10^{-4} \mbox{Mpc}^{-1}\sim 6\times 10^{-61} M_{pl}$ and $k_*=a_0k_{phys}$, where we have chosen as usual $k_{phys}=0.02\mbox{Mpc}^{-1}$,  we have obtained
\begin{eqnarray}
{\mathcal N}\cong -5+\ln\left(\frac{H_*}{H_0}  \right)+ \ln\left(\frac{a_{END}}{a_{kin}}  \right)+\frac{1}{4}\ln\left(\frac{g_{eq}}{g_{reh}}  \right)
+\frac{1}{6}\ln\left(\frac{\rho_{rh}}{\rho_{kin}}  \right)+ \ln\left(\frac{T_0}{T_{rh}}  \right),
\end{eqnarray}
where we have used once again that after the matter-radiation equality the evolution is adiabatic, that is, $a_0T_0=a_{eq} T_{eq}$, as well as the Stefan-Boltzmann law $\rho_{eq}=\frac{\pi^2}{30}g_{eq}T_{eq}^4$ and $\rho_{reh}=\frac{\pi^2}{30}g_{reh}T_{reh}^4$ being $g_{eq}=3.36$ the degrees of freedom at the matter-radiation equality. And we have chosen as degrees of freedom at the reheating time the ones of the Standard Model, i.e., 
$g_{reh}=106.75$.

\

Now, from the formula of the power spectrum  we infer that $H_*\sim 4\pi \times 10^{-4}\sqrt{\frac{\epsilon_*}{10}} M_{pl}
\sim 4\times 10^{-4}\sqrt{\epsilon_*} M_{pl}$, obtaining
\begin{eqnarray}
{\mathcal N}\cong 125 +\frac{1}{2}\ln\epsilon_*+ \ln\left(\frac{a_{END}}{a_{kin}}  \right)
+\frac{1}{6}\ln\left(\frac{\rho_{reh}}{\rho_{kin}}  \right)+ \ln\left(\frac{T_0}{T_{reh}}  \right),
\end{eqnarray}
and introducing the current value of the temperature of the universe $T_0\sim  10^{-31}M_{pl}$ we get 
\begin{eqnarray}
{\mathcal N}\cong 54+\frac{1}{2}\ln\epsilon_*+ \ln\left(\frac{a_{END}}{a_{kin}}  \right)
-\frac{1}{3}\ln\left(\frac{H_{kin}T_{reh}}{M_{pl}^2 } \right).
\end{eqnarray}

\

Taking into account that for the Peebles-Vilenkin model one has $H_{kin}\sim 4\sqrt{2\lambda}M_{pl}\sim{4\sqrt{2}\times 10^{-7} M_{pl}}$ and $\epsilon_*\cong (1-n_s)/3$, we get
\begin{eqnarray}
{\mathcal N}\cong 59+\frac{1}{2}\ln(1-n_s)+ \ln\left(\frac{a_{END}}{a_{kin}}  \right)
-\frac{1}{3}\ln\left(\frac{T_{reh}}{M_{pl} } \right)
\end{eqnarray}
and, using the value of the spectral index $n_s\cong 0.96$ and assuming that $\ln\left(\frac{a_{END}}{a_{kin}}  \right)$ is negligible, we finally obtain
\begin{eqnarray}
{\mathcal N}\cong 57
-\frac{1}{3}\ln\left(\frac{T_{reh}}{M_{pl} } \right).
\end{eqnarray}

\


Since the scale of  nucleosynthesis is $1$ MeV and in order to avoid the late-time decay of gravitational relic products such as moduli fields or gravitinos  which could jeopardise  the nucleosynthesis success, one needs temperatures in the range $1 \mbox{ MeV}\leq T_{reh}\leq 10^9 \mbox{ GeV}$, which leads to constrain the number of e-folds to $64\lesssim \mathcal{N}\lesssim 74$.

\

A final remark is in order: Note that in Quintessential Inflation we have obtained a number of e-folds greater than in standard inflation, which normally ranges between $50$ and $60$ e-folds. This fact is due to the kination regime appearing in QI, which increases significantly  the number of e-folds.

\

\section{Exponential Quintessential Inflation}\label{seciii}

In this section
we will consider more complex models as the Peebles-Vilenkin one: the same  Exponential  Inflation-type potentials  studied for the first time in  \cite{hossain2},
\begin{eqnarray}\label{exp}
V(\varphi)=\
V_0e^{-\gamma{\varphi}^n/{M_{pl}}^n},
\end{eqnarray}
where $\gamma$ is a dimensionless parameter and $n$ is an integer.

\

In this case the power spectrum of scalar perturbations, its spectral index and the ratio of tensor to scalar perturbations 
are given by (see for a detailed derivation \cite{Geng:2017mic})
\begin{eqnarray}\label{power}
{\mathcal P}_{\zeta}=\frac{V_0e^{-\gamma{\varphi}^n/{M_{pl}}^n}}{12\pi^2n^2\gamma^2M_{pl}^{6-2n}\varphi^{2n-2}}\sim 2\times 10^{-9},
\end{eqnarray}
\begin{eqnarray}\label{index}
n_s=1-n\gamma\left( \frac{\varphi}{M_{pl}}\right)^{n-2}\left( 2n-2+\gamma n\left( \frac{\varphi}{M_{pl}}\right)^{n} \right)
\end{eqnarray}
and
\begin{eqnarray}\label{r}
r=8n^2\gamma^2\left( \frac{\varphi}{M_{pl}}\right)^{2n-2}.
\end{eqnarray}

An important relation is obtained combining the equations (\ref{index}) and (\ref{r}),
\begin{eqnarray}\label{relation}
\gamma\left( \frac{\varphi}{M_{pl}}\right)^n=\frac{r(2n-2)}{n(8(1-n_s)-r) },
\end{eqnarray}
which leads to the formula for the power spectrum
\begin{eqnarray}
{\mathcal P}_{\zeta}=\frac{2V_0e^{-\frac{r(2n-2)}{n(8(1-n_s)-r) }}}{3\pi^2M_{pl}^{4}r}\sim 2\times 10^{-9},
\end{eqnarray}
and thus,
\begin{eqnarray}
V_0\sim 3\pi^2re^{\frac{r(2n-2)}{n(8(1-n_s)-r) }}\times 10^{-9} M_{pl}^4,
\end{eqnarray}
which, for the viable values of $n_s\cong 0.96$ (the central value of the spectral index) and $r=0.02$ (whose value, as we will immediately see, guarantees that the model provides a viable a number of e-folds), 
leads to 
\begin{eqnarray}\label{V0}
V_0\sim 6\pi^2\times 10^{-11} M_{pl}^4.
\end{eqnarray}

\

Next, it is important to realize that a way to find theoretically the possible values of the parameter $\gamma$ is to combine the equations (\ref{r}) and (\ref{relation}) to get
\begin{eqnarray}
\gamma=\left( \frac{n(8(1-n_s)-r)}{2r(n-1)} \right)^{n-1}\left(\frac{r}{8n^2}\right)^{n/2}
\end{eqnarray}
and, using the theoretical values $r\leq 0.1$ and $n_s=0.9649\pm 0.0042$ (see for instance \cite{planck18,planck18a}), one can find the candidates of $\gamma$ at $2\sigma$ C.L. These values have to be checked for the joint contour in the plane $(n_s,r)$ at $2\sigma$ C.L., when the number of e-folds is approximately between 60 and 75, which is what usually happens in Quintessential Inflation due to the kination phase \cite{ll,  deHaro:2016ftq} -the energy density of the scalar field is only kinetic \cite{Joyce, Spokoiny}- after the end of the  inflationary period.

\

For this kind of potentials, in order to compute the number of e-folds we need to calculate the main slow-roll parameter 
\begin{eqnarray}
\epsilon\equiv \frac{M_{pl}^2}{2}\left(\frac{V_{\varphi}}{V} \right)^2=\frac{\gamma^2n^2}{2}\left( \frac{\varphi}{M_{pl}} \right)^{2n-2},
\end{eqnarray}
whose value at the end of inflation is $\epsilon_{END}=1$, meaning that at the end of this epoch  the field reaches the value 
$\varphi_{END}=\left( \frac{2}{n^2\gamma^2}\right)^{\frac{1}{2n-2}} M_{pl}$. 

\

Then, the number of e-folds is given by
\begin{eqnarray}\label{N}
{\mathcal N}=\frac{1}{M_{pl}}\int_{\varphi}^{\varphi_{END}}\frac{1}{\sqrt{2\epsilon}}d\varphi=\frac{1}{n\gamma(n-2)}
\left[\left( \frac{\varphi}{M_{pl}} \right)^{2-n}-\left( \frac{2}{n^2\gamma^2}\right)^{\frac{2-n}{2n-2}} \right].
\end{eqnarray}
Thus, combining the equations  (\ref{index}), (\ref{r}) and (\ref{N}), one obtains the spectral index and the tensor/scalar ratio as a function of the number of e-folds and the parameter $\gamma$.

\

On the other hand, since
inflation ends at $\varphi_{END}=\left( \frac{2}{n^2\gamma^2}\right)^{\frac{1}{2n-2}} M_{pl}$, when the effective Equation of State (EoS) parameter is equal to $-1/3$,
meaning that $\dot{\varphi}_{END}^2=V(\varphi_{END})$,
the energy density at the end of inflation is
 \begin{eqnarray}\rho_{\varphi, END}=\frac{3}{2} V(\varphi_{END})\cong 9\pi^2 e^{-\gamma\left( \frac{2}{n^2\gamma^2}\right)^{\frac{n}{2n-2}}}
 \times 10^{-11} M_{pl}^4
 \end{eqnarray}
 and the corresponding value of the Hubble rate is given by
\begin{eqnarray}\label{HEND}
  H_{ END}\cong \sqrt{\frac{3}{10}}
  e^{-\frac{\gamma}{2} \left( \frac{2}{n^2\gamma^2}\right)^{\frac{n}{2n-2}}}
 \times 10^{-5} M_{pl},
 \end{eqnarray}
which will constrain very much the values of the parameter $\gamma$ because in all viable inflationary models at the end of the inflation the value of the Hubble rate is of the order of $10^{-6} M_{pl}$  \cite{linde}.
 In fact, when  (\ref{HEND}) is of the order of $10^{-6} M_{pl}$ one gets
 \begin{eqnarray}
 \gamma\sim \frac{(2/n^2)^{n/2}}{\ln^{n-1}(30\pi^2)} .
 \end{eqnarray}

 \
 
  Therefore, to perform numerical calculations,  we will use the values of
 $n=10$ and $r=0.02$ and, thus, for { $\gamma\sim  10^{-16}$ we obtain  approximately $67$ e-folds, which is a viable value in Quintessential Inflation.}

 \

 After this, we have to find out the time when kination starts, which can be chosen at the moment when the effective EoS parameter is very close to $1$.
 Assuming for instance that kination starts at $w_{eff}\cong 0.99$, we have numerically obtained $\varphi_{kin}\cong 47 M_{pl}$ and $\dot{\varphi}_{kin}\cong 6 \times 10^{-9} M_{pl}^2$, hence $H_{kin}\cong 2\times 10^{-9} M_{pl}$.
 
 \


Next,  we want to calculate the value of the scalar field and its derivative at the reheating time (for example,  in the case that the decay was before the end of kination). Using for our model the formulas (\ref{varphireh})
\begin{eqnarray}
\varphi_{reh}=\varphi_{kin}+\sqrt{\frac{2}{3}}M_{pl}\ln\left( \frac{ H_{kin}}{\sqrt{\frac{\pi^2g_{reh}}{45}} \frac{T_{reh}^2}{M_{pl}}}\right),
\qquad   { \dot{\varphi}_{reh}=\sqrt{\frac{\pi^2g_{reh}}{15}} T_{reh}^2},\end{eqnarray} 
and  assuming that the effective number of degrees of freedom is the same as in the Standard Model, i.e. $g_{reh}=106.75$, 
we get for a viable reheating 
 temperature $T_{reh}\sim  10^{7}$ GeV that at the beginning of the radiation the initial conditions of the scalar field are
 \begin{eqnarray}\label{CI}
 \varphi_{reh}\cong {  70}M_{pl} \qquad {\dot{\varphi}_{reh}\cong  { 5\times 10^{-22}} M_{pl}^2}.
 \end{eqnarray}

\subsection{The scaling and tracker solutions}

The scaling and tracker solutions appear in pure exponential potentials and approximately in more generic exponential potentials as (\ref{exp}) (see for instance \cite{btw,Geng:2017mic}). 

\subsubsection{The scaling solution}

Assuming as usual the flat  Friedmann-Lema\^{i}tre-Robertson-Walker (FLRW) geometry of our universe, we consider a pure exponential potential, $V(\varphi)=V_0e^{-\gamma \varphi/M_{pl}}$, and  a barotropic fluid with Equation of State (EoS) parameter $w=1/3$, i.e., a radiation fluid,  whose energy density  we continue denoting by $\rho_r$. Then, the dynamical system is given by the equations
\begin{eqnarray}\left\{ \begin{array}{ccc}
\ddot{\varphi}+3H\dot{\varphi}+V_{\varphi}&=& 0\\
\dot{\rho}_r+4H\rho_r &=& 0,
\end{array}\right.
\end{eqnarray}
where $H=\frac{1}{\sqrt{3}M_{pl}}\sqrt{\rho_{\varphi}+\rho_r}$.

\

The scaling solution is a solution with the property that the energy density of the scalar field scales as the one of radiation, meaning that the Hubble parameter evolves as $H(t)=\frac{1}{2t}$.  At this point, we look for a solution of the form
\begin{eqnarray}
\varphi_{sc}(t)=\frac{M_{pl}}{\gamma}\ln\left( \frac{t^2}{\bar{t}^2} \right), \qquad \rho_r(t)=\frac{C^2}{t^2},
\end{eqnarray}
where $C$ and $\bar{t}$ are constants.
The equation $\dot{\rho}_r+4H\rho_r = 0$ is satisfied for any value of $C^2$ and the equation $\ddot{\varphi}+3H\dot{\varphi}+V_{\varphi}= 0$ requires that
$\bar{t}^2=\frac{M_{pl}^2}{\gamma^2V_0}$.

\

On the other hand,  equating $H(t)=\frac{1}{2t}$ with $H=\frac{1}{\sqrt{3}M_{pl}}\sqrt{\rho_{\varphi}+\rho_r}$, which yields the relation
\begin{eqnarray}
C^2=\frac{3}{4}\left(1-\frac{4}{\gamma^2} \right)M_{pl}^2,
\end{eqnarray}
shows that the scaling solution only exists for $\gamma>2$, and  the corresponding energy densities evolve as follows,
\begin{eqnarray}
\rho_{\varphi_{sc}}(t)=\frac{3M_{pl}^2}{\gamma^2t^2} \qquad \rho_r(t)= \frac{3}{4}\left(1-\frac{4}{\gamma^2} \right)\frac{M_{pl}^2}{t^2},
\end{eqnarray}
and thus, one could derive the corresponding density parameters,
\begin{eqnarray}
\Omega_{\varphi_{sc}}=\frac{4}{\gamma^2},\qquad \Omega_r=1-\frac{4}{\gamma^2}.\end{eqnarray}

An alternative approach to get this solution goes as follows
(see for instance \cite{copeland}): one can  introduce the dimensionless variables 
\begin{eqnarray}
\tilde{x}\equiv \frac{\dot{\varphi}}{\sqrt{6}M_{pl}H} \quad \mbox{and} \quad \tilde{y}\equiv \frac{\sqrt{V}}{\sqrt{3}M_{pl}H},
\end{eqnarray}
which enable us to write down the following autonomous dynamical system,
 \begin{eqnarray}\label{dynamical}
 \left\{\begin{array}{ccc}
 \tilde{x}'&=& -3\tilde{x}+\sqrt{\frac{3}{2}}\gamma \tilde{y}^2+\frac{3}{2}\tilde{x}\left[(1-w)\tilde{x}^2+(1+w)(1-\tilde{y}^2)\right]\\
  \tilde{y}'&=& -\sqrt{\frac{3}{2}}\gamma\tilde{x}\tilde{y}+\frac{3}{2}\tilde{y}\left[(1-w)\tilde{x}^2+(1+w)(1-\tilde{y}^2)\right]\ \end{array},
 \right.
 \end{eqnarray}
where $w$ is the Equation of State (EoS) parameter, which in our case is equal to $1/3$, 
together with the constraint
\begin{eqnarray}
\tilde{x}^2+\tilde{y}^2+\Omega_r=1.
\end{eqnarray}

One can see that for $w=1/3$ the dynamical system (\ref{dynamical}) has the attractor solution $\tilde{x}=\left(\sqrt{\frac{2}{3}}\; \right)\frac{2}{\gamma},\ \ \tilde{y}=\frac{2}{\sqrt{3}}\frac{1}{\gamma}$, whose energy density scales as radiation, and $\Omega_{\varphi}=\frac{4}{\gamma^2}$.

So, since during radiation $H=\frac{1}{2t}$, thus, from $\tilde{x}=\left(\sqrt{\frac{2}{3}}\; \right)\frac{2}{\gamma}$, we have 
\begin{eqnarray}
\dot{\varphi}_{sc}=\frac{2}{\gamma t}M_{pl} \Longrightarrow \varphi_{sc}(t)=\frac{M_{pl}}{\gamma}\ln\left( \frac{t^2}{\bar{t}^2}\right).
\end{eqnarray}

Finally, in order to obtain $\bar{t}$, we use the equation $\tilde{y}=\frac{2}{\sqrt{3}}\frac{1}{\gamma}$, which leads to $\bar{t}^2=\frac{M_{pl}^2}{\gamma^2V_0}$, recovering the scaling solution 
\begin{eqnarray}
\varphi_{sc}(t)=\frac{M_{pl}}{\gamma}\ln\left( \frac{\gamma^2V_0t^2}{M_{pl}^2}  \right).
\end{eqnarray}

\subsubsection{The tracker solution}
\label{sec-III}

A tracker solution \cite{Steinhardt:1999nw, UrenaLopez:2000aj} is an attractor solution of the field equation which describes the
dark energy domination at late times. In fact, for an exponential potential, it is the solution of $\ddot{\varphi}+3H\dot{\varphi}+V_{\varphi}=0$ when the universe is only filled with the scalar field.
Once again, we look for a solution of the type
\begin{eqnarray}
\varphi_{tr}(t)=\frac{M_{pl}}{\gamma}\ln\left( \frac{t^2}{\bar{t}^2} \right).
\end{eqnarray}

Inserting this equation into $\ddot{\varphi}+3H\dot{\varphi}+V_{\varphi}=0$, one gets
\begin{eqnarray}
\bar{t}^2=\frac{2M_{pl}^2}{V_0\gamma^4}(6-\gamma^2),
\end{eqnarray}
which means that $0<\gamma <\sqrt{6}$ and, thus, the tracker solution is given by 
\begin{eqnarray}\label{tracker}
\varphi_{tr}=\frac{M_{pl}}{\gamma}\ln\left(\frac{V_0\gamma^4}{2(6-\gamma^2)M_{pl}^2}t^2   \right).
\end{eqnarray}

Moreover, for this equation, it is not difficult to show that 
\begin{eqnarray}
\frac{1}{2}\dot{\varphi}^2_{tr}=\frac{2M_{pl}^2}{\gamma^2 t^2},\qquad 
V(\varphi_{tr})=\frac{2M_{pl}^2}{\gamma^2 t^2}(6-\gamma^2),
\end{eqnarray}
meaning that
the corresponding effective EoS parameter is given by
\begin{eqnarray}
w_{eff}=\frac{\gamma^2}{3}-1,
\end{eqnarray}
which proves that, in order to impose the late-time acceleration of the universe, we need to restrict $\gamma$ as 
$0<\gamma<\sqrt{2}$.

\

Note also that the system (\ref{dynamical}) when $w=0$ (in the matter domination era) has the fixed point $(\tilde{x}, \tilde{y})= \left(\gamma/\sqrt{6}, \sqrt{1- \gamma^2/6} \right)$, which of course corresponds to the solution \eqref{tracker}.

\

Finally, in terms of the time $N=-\ln(1+z)$, taking into account that the relation between $N$ and the cosmic time is given by 
$t=\frac{2}{\gamma^2 H_0}e^{\gamma^2 H/2}$, the tracker solution has the following expression (see Section V of \cite{pvharo} for a detailed derivation),
\begin{eqnarray}
\varphi_{tr}(N)=\gamma NM_{pl}+\frac{M_{pl}}{\gamma}\ln\left(\frac{2V_0}{(6-\gamma^2)M_{pl}^2H_0^2} \right).
\end{eqnarray}

\subsection{Numerical simulation during radiation}
In this subsection we will show numerically that in Quintessential Inflation, for exponential types of potentials,  during radiation the scalar field is never in the basin of attraction of the scaling solution and, thus, does not evolve as the scaling solution.

\

To prove it, first of all we  heuristically calculate the value of the red-shift at the beginning of the radiation epoch
\begin{eqnarray}
1+z_{reh}=\frac{a_0}{a_{reh}}=\frac{a_0}{a_{eq}}\frac{a_{eq}}{a_{rh}}=
(1+z_{eq})\left(\frac{\rho_{r,reh}}{\rho_{r,eq}}  \right)^{1/4}= (1+z_{eq})\left(\frac{g_{reh}}{g_{eq}}  \right)^{1/4}\frac{T_{reh}}{T_{eq}}\nonumber\\=\left(\frac{g_{reh}}{g_{eq}}  \right)^{1/4}\frac{T_{reh}}{T_{0}}\sim \frac{T_{reh}}{T_{0}},
\end{eqnarray}
where  we have used that $\rho_{r,eq}=\rho_{r,reh}\left(\frac{a_{reh}}{a_{eq}} \right)^4$ and the adiabatic evolution after the matter-radiation equality.
Then, taking $z_{eq}\sim 3\times 10^3$ and using the current temperature $T_0\sim 10^{-31} M_{pl}$, for the reheating temperature $T_{reh}\sim  10^9$ GeV we get  $z_{reh}\sim  10^{22}$.

\

In addition, at the beginning of radiation the energy density of the matter will be
\begin{eqnarray}
\rho_{m,reh}=\rho_{m,eq}\left( \frac{a_{eq}}{a_{reh}} \right)^3=\rho_{m,eq}\left(\frac{\rho_{r,reh}}{\rho_{r,eq}}  \right)^{3/4}
=\frac{\pi^2}{30}g_{eq}\left( \frac{g_{reh}}{g_{eq}} \right)^{3/4}(1+z_{eq})T^3_{reh}T_{0}\nonumber \\ \sim { 10^{13}} \mbox{ GeV}^4,
\end{eqnarray}
and, for radiation, 
\begin{eqnarray}
\rho_{r,reh}=\frac{\pi^2}{30}g_{reh}T_{reh}^4
\sim 10^{29} \mbox{GeV}^4.
\end{eqnarray}

\

In this way,  
the dynamical equations after the beginning of radiation can be easily obtained using once again as a time variable
$N\equiv -\ln(1+z)=\ln\left( \frac{a}{a_0}\right)$. Recasting the  energy density of radiation and matter respectively as functions of $N$, we get
\begin{eqnarray}
\rho_{r}(a)={\rho_{r,reh}}\left(\frac{a_{reh}}{a}  \right)^4\Longrightarrow \rho_{r}(N)= {\rho_{r,reh}}e^{4(N_{reh}-N)} 
\end{eqnarray}
and
\begin{eqnarray}
\rho_{m}(a)={\rho_{m,reh}}\left(\frac{a_{reh}}{a}  \right)^3\Longrightarrow \rho_{m}(N)={\rho_{m,reh}}e^{3(N_{reh}-N)},
\end{eqnarray}
where  
$N_{reh}=-\ln(1+z_{reh})\sim -50$ denotes the value of the time $N$ at the beginning of radiation for a reheating temperature around $10^9$ GeV.

\



To obtain the dynamical system for this scalar field model, we will
introduce the  dimensionless variables
 \begin{eqnarray}
 x=\frac{\varphi}{M_{pl}} \qquad \mbox{and} \qquad y=\frac{\dot{\varphi}}{K M_{pl}},
 \end{eqnarray}
 where $K$ is a parameter that we will choose accurately in order to facilitate the numerical calculations. So,  taking into account  the conservation equation $\ddot{\varphi}+3H\dot{\varphi}+V_{\varphi}=0$, one arrives at the following   dynamical system,
 \begin{eqnarray}\label{system}
 \left\{ \begin{array}{ccc}
 x^\prime & =& \frac{y}{\bar H}~,\\
 y^\prime &=& -3y-\frac{\bar{V}_x}{ \bar{H}}~,\end{array}\right.
 \end{eqnarray}
 where the prime is the derivative with respect to $N$, $\bar{H}=\frac{H}{K}$   and $\bar{V}=\frac{V}{K^2M_{pl}^2}$.  It is not difficult to see that  one can write  
 \begin{eqnarray}
 \bar{H}=\frac{1}{\sqrt{3}}\sqrt{ \frac{y^2}{2}+\bar{V}(x)+ \bar{\rho}_{r}(N)+\bar{\rho}_{m}(N) }~,
 \end{eqnarray}
where we have defined the dimensionless energy densities as
 $\bar{\rho}_{r}=\frac{\rho_{r}}{K^2M_{pl}^2}$ and 
 $\bar{\rho}_{m}=\frac{\rho_{m}}{K^2M_{pl}^2}$.

\

Next, to integrate from the beginning of radiation up to the matter-radiation equality, i.e.,   from $N_{reh}\sim -50$ to $N_{eq}\sim -8$,
 we will choose $KM_{pl}= 10^{-17} \mbox{GeV}^2$, yielding 
{\begin{eqnarray}
\bar{\rho}_{r}(N)\sim { 4\times10^{64}}e^{4(N_{reh}-N)}, \qquad \bar{\rho}_{m}(N)\sim  { 8 \times10^{47}}e^{3(N_{reh}-N)},
\end{eqnarray}}
and
{\begin{eqnarray}
\bar{V}(x) = \frac{ V_0\times 10^{34}}{\mbox{GeV}^4} e^{-\gamma x^n}.
\end{eqnarray}}

\


Finally, with the initial conditions for the field being $x_{reh}\sim{ 70 }$ and $y_{rh}\sim {  3\times 10^{32}}$  (these initial conditions are obtained in the equations (\ref{CI}))
and integrating numerically the dynamical system, we conclude that, for values of the parameter $n$ greater than $10$,  the value of  the EoS parameter $w_{\varphi}$ remains $1$ during radiation domination, namely between $N_{rh}$ and $N_{eq}$, thus proving that the inflaton field does not belong to the basin of attraction of the scaling solution
for large values of $n$.
In addition, integrating the system up to now, we can see that the model never matches with the current observational data.

\


\begin{figure}[H]
    \centering
    \includegraphics[width=70mm]{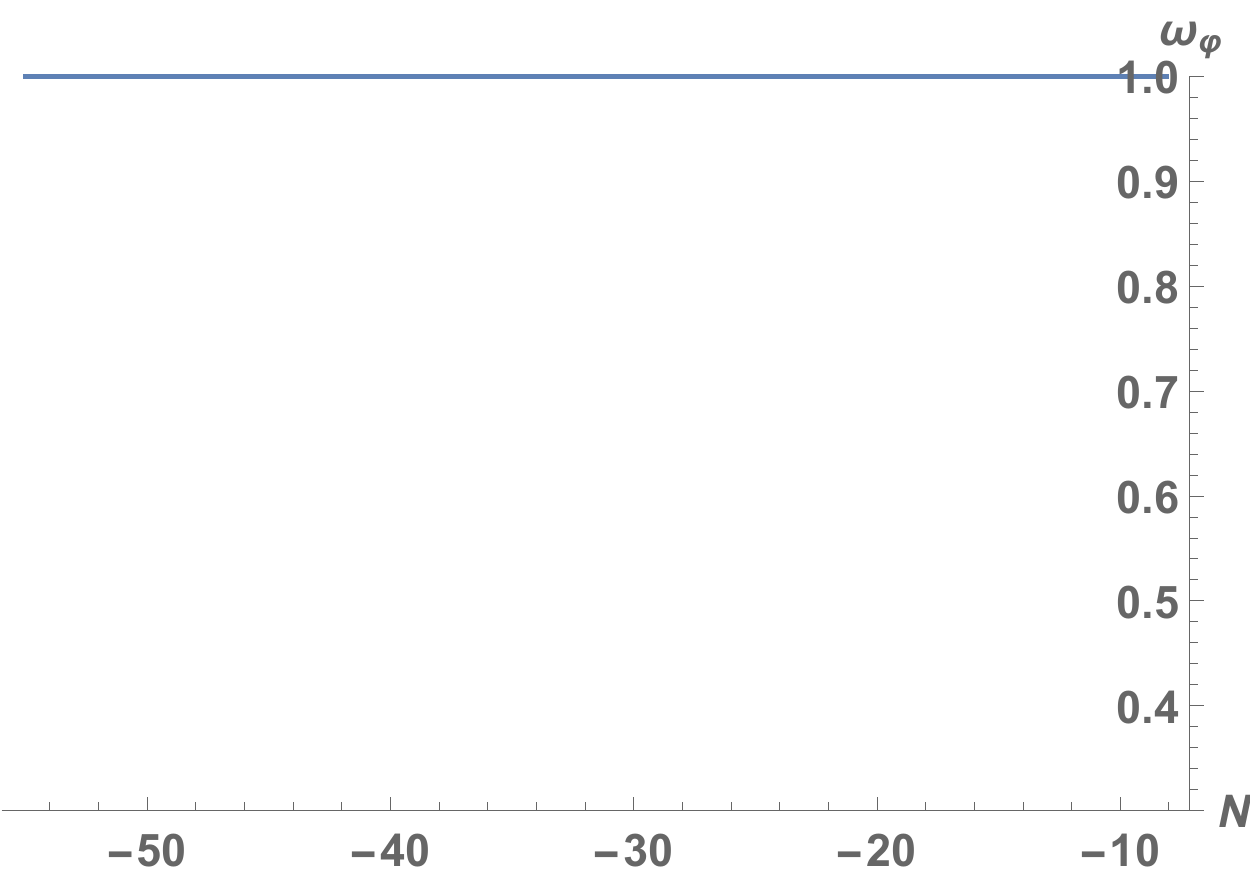}
    \caption{Evolution of $w_{\varphi}$ during radiation domination.}
    \label{fig:wphi_exp}
\end{figure}

\subsection{A viable model}
\label{sec3}

To depict the late-time acceleration, one has to modify the original potential because it does not explain the current observational data
{(for the potential \eqref{exp}, at the present time,  the    density parameter   $\Omega_{\varphi}=\frac{\rho_{\varphi}}{3H^2 M_{pl}^2}$ is far from  its observational value, namely $0.7$)}. For this reason and following the spirit of the Peebles-Vilenkin model \cite{pv}, to match with the current observational data it is mandatory   to introduce a new parameter  $M$  with units of mass which must be calculated numerically. In that case, we will consider the following modification of the potential \eqref{exp} by introducing a new exponential term containing the parameter $M$,
\begin{eqnarray}\label{viable}
V(\varphi)=\
V_0e^{-\bar{\gamma}{\varphi}^n/{M_{pl}}^n}+ M^4e^{-\gamma \varphi/M_{pl}},
\end{eqnarray}
with  $0<\gamma<\sqrt{2}$ in order to obtain, as we have already seen, that at late times the solution is in the basin of attraction of the tracker solution, which evolves as a fluid with effective EoS  parameter $w_{eff}=\frac{\gamma^2}{3}-1$.

\

{\bf Remark.-}
{\it An important remark is in order. Contrary to the potential used in \cite{hossain2, Geng:2017mic}
\begin{eqnarray}
V(\varphi)=\
V_0e^{-\bar{\gamma}{\varphi}^n/{M_{pl}}^n}+ (\bar{\rho}_{\nu}-3\bar{p}_{\nu})e^{\beta \varphi/M_{pl}},
\end{eqnarray}
where $\rho_{\nu}=\bar{\rho}_{\nu}e^{\beta \varphi/M_{pl}}$  is the  neutrino  energy density with constant bare mass $m_{\nu}$ and 
 $\beta>0$ is the non-minimal coupling of neutrinos with the inflaton field (see for details \cite{ Geng:2017mic}), which has a minimum where the inflaton ends its evolution, thus acting as an effective cosmological constant which stands for the current cosmic acceleration, the potential considered here  does not have a minimum and the scalar field continues rolling down the potential following the spirit of the quintessential inflation.} 
 \
 
 On the other hand, in this case one can also justify this choice considering that the inflaton field is non-minimally coupled with neutrinos but with a negative coupling constant, namely $\beta=-\gamma<0$. Therefore, $M^4=\bar{\rho}_{\nu}-3\bar{p}_{\nu}$ and the effective mass of neutrinos is given by  
 $m_{\nu,eff}(\varphi)= m_{\nu}e^{-\gamma \varphi/M_{pl}}$, which tends to zero for large values of the field, so the neutrinos become relativistic, contrary to what happens when $\beta$ is positive, where the neutrinos acquire a heavy mass becoming non-relativistic particles.
 Finally, to match with the current observational data, 
 $M^4=\bar{\rho}_{\nu}-3\bar{p}_{\nu}$ must be very small compared with the Planck's energy density.
 In fact,  for $n=10$ and $\gamma=0.1$ we have numerically obtained $M\sim 10^{-2} \mbox{ eV}$. 

Now, one has   to solve the dynamical system (\ref{system}) with initial conditions at the beginning of the radiation era. Choosing now $K=H_0$ (the present value of the Hubble rate) we have to take
 into account that to obtain the parameter $M$ one has to impose the value of $\bar{H}=H(N)/H_0$ to be $1$ at $N=0$. 


\

\begin{figure}[H]
\begin{center}
\includegraphics[scale=0.48]{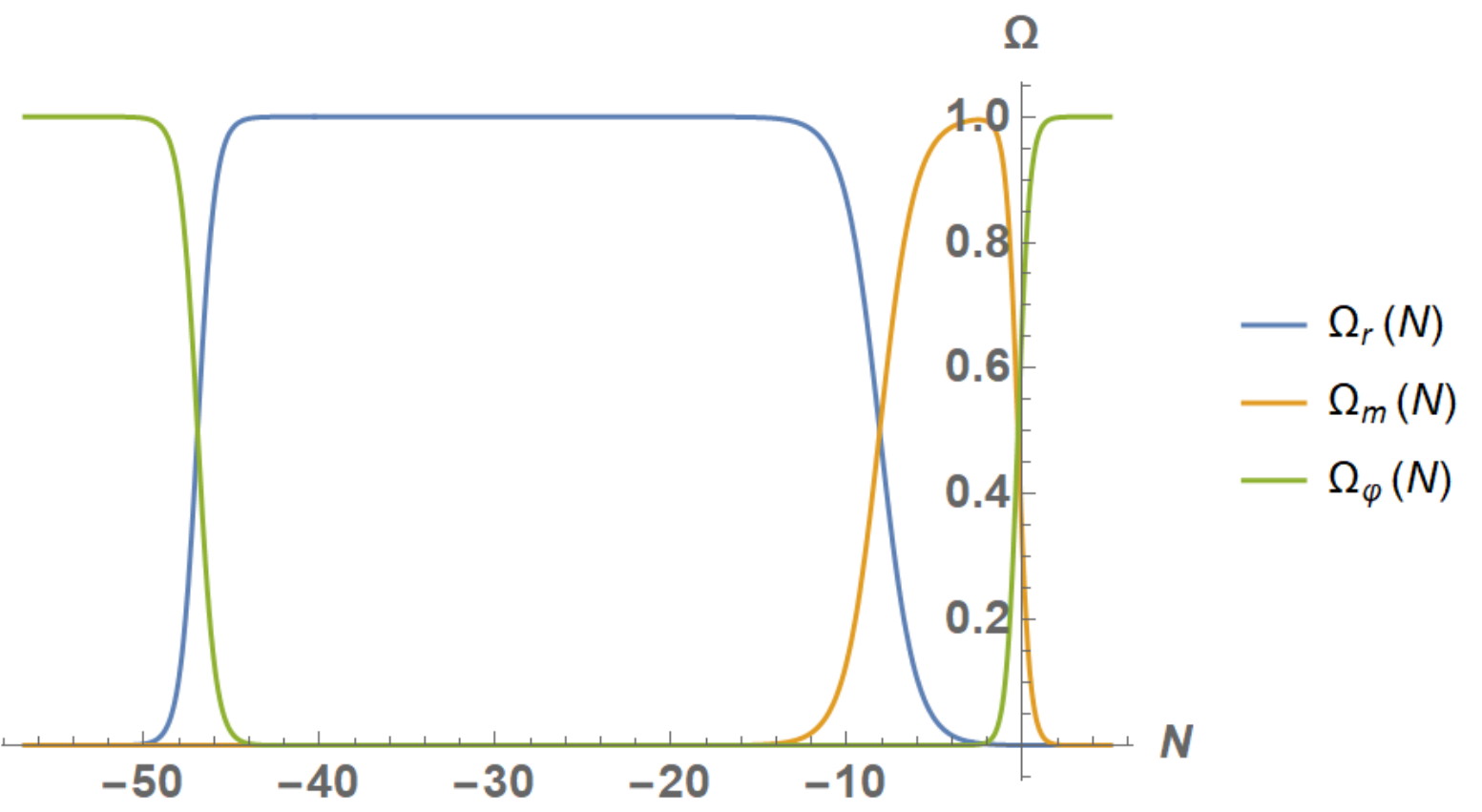}
\includegraphics[scale=0.43]{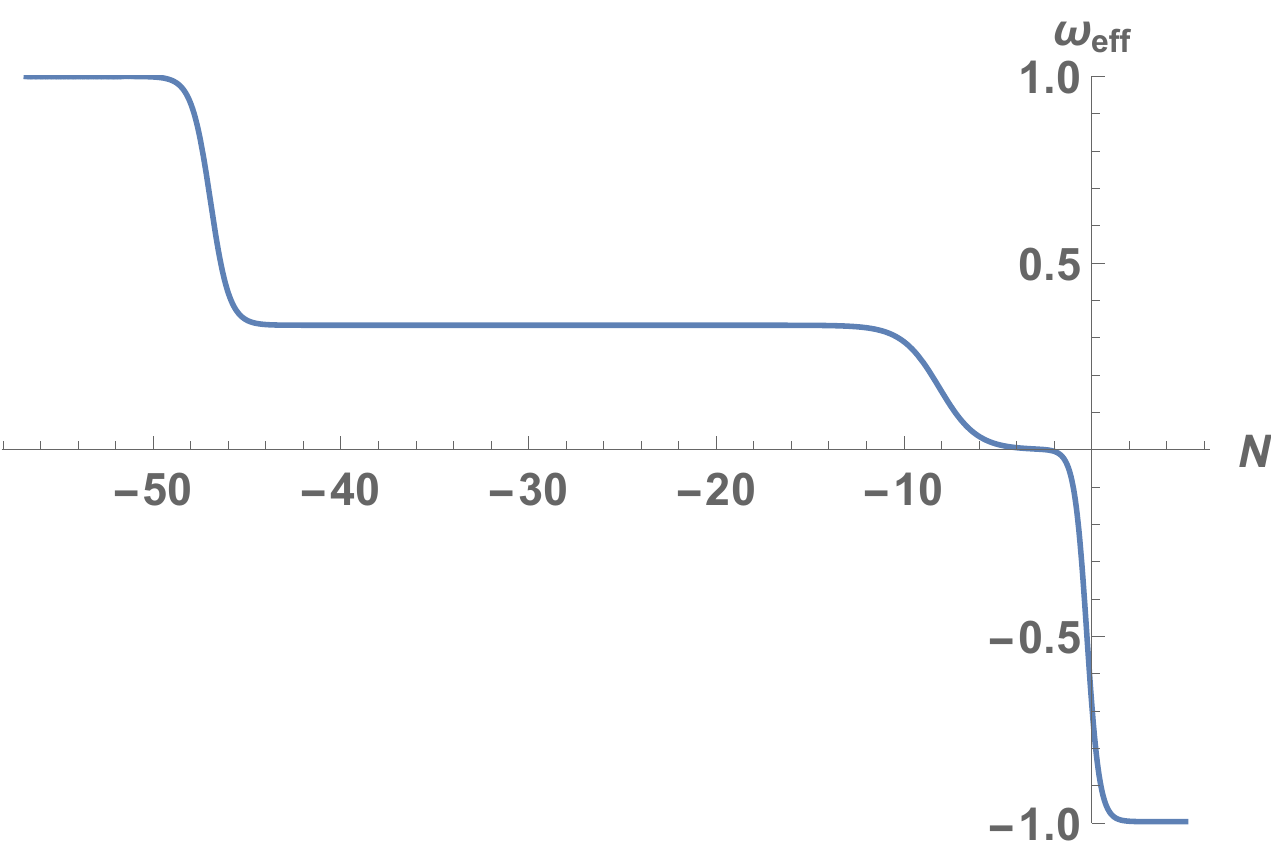}
\end{center}
\caption{
\textbf{Left:} The density parameters $\Omega_m=\frac{\rho_m}{3H^2 M_{pl}^2}$ (orange curve), $\Omega_r=\frac{\rho_r}{3H^2 M_{pl}^2}$ (blue curve) and  
$\Omega_{\varphi}=\frac{\rho_{\varphi}}{3H^2 M_{pl}^2}$ (green curve), from kination to future times, for  a reheating temperature around $10^9$ GeV. To perform numerical calculations we have taken $n=10$ and $\gamma=0.1$.  \textbf{Right:}
The effective Equation of State parameter $w_{eff}$, from kination to future times, for $n=10$ and $\gamma=0.1$. As one can see in the picture, after kination the universe enters in a large period of time
where radiation dominate. Then, after the matter-radiation equality, the universe becomes matter-dominated and,  finally,  near the present, it  enters in a new accelerated phase
where $w_{eff}$ approaches $\frac{0.1^2}{3}-1\cong -1$}, that is, it has the same effective EoS parameter as the tracker solution, meaning that the solution is in the basin of attraction of the tracker one.
\label{fig:PV_rh_eq}
\end{figure}

The numerical results are presented in Figure \ref{fig:PV_rh_eq}, where one can see that at very late times the energy density of the scalar field dominates and the universe has an eternal accelerated expansion because its evolution is the same as that of the tracker solution.

\

\section{Lorentzian Quintessential Inflation}\label{seciv}

Based on the Cauchy distribution (Lorentzian in the physics language) 
the authors of 
 \cite{Benisty:2020xqm,Benisty:2020qta} considered the following  ansatz, 
\begin{eqnarray}\label{ansatz}
\epsilon({ N})=\frac{\xi}{\pi}\frac{\Gamma/2}{{ N}^2+\Gamma^2/4},
\end{eqnarray}
where $\epsilon$ is the main slow-roll parameter, ${ N}=\ln\left(\frac{a}{a_0}\right)$ denotes the total number of e-folds (do not confuse with $\mathcal{N}$, the number of e-folds from the horizon crossing to the end of inflation),
 $\xi$ is the amplitude of the Lorentzian distribution and $\Gamma$ is its width.
 From this ansatz, we can find the exact corresponding potential of the scalar field, namely
\begin{eqnarray}\label{original}
\hspace{-0.5cm}V(\varphi)=\lambda M_{pl}^4\exp\left[-\frac{2\xi}{\pi}\arctan\left(\sinh
\left(\gamma\varphi/M_{pl} \right)  \right)\right]\boldsymbol{\cdot} 
\left(1-\frac{2\gamma^2\xi^2}{3\pi^3}\frac{1}{\cosh
\left(\gamma\varphi/M_{pl} \right) } \right),
\end{eqnarray}
where $\lambda$ is a dimensionless parameter and the parameter $\gamma$ is defined by
$$\gamma\equiv \sqrt{\frac{\pi}{\Gamma \xi}}.$$ This potential can be derived by using equations $(37)$ in \cite{Martin:2016iqo}. However, in this work we are going to use a more simplified potential,  keeping the same properties as the original potential but not coming exactly from the ansatz (\ref{ansatz}). 
However, from this potential we do recover the ansatz by using the suitable approximations, which are valid if we properly choose the parameters such that $\gamma\gg 1$ and $\frac{2\xi}{\Gamma\pi}>1$. For this reason and also because the results of the original potential happen to be exactly the same as shown in \cite{Guendelman2}, it is better to work with the simplified potential, namely

\begin{eqnarray}\label{LQI}
V(\varphi)=\lambda M_{pl}^4\exp\left[-\frac{2\xi}{\pi}\arctan\left(\sinh\left(\gamma\varphi/M_{pl} \right)  \right)\right].
\end{eqnarray}
 
\begin{figure}[H]
\begin{center}
\includegraphics[width=8cm]{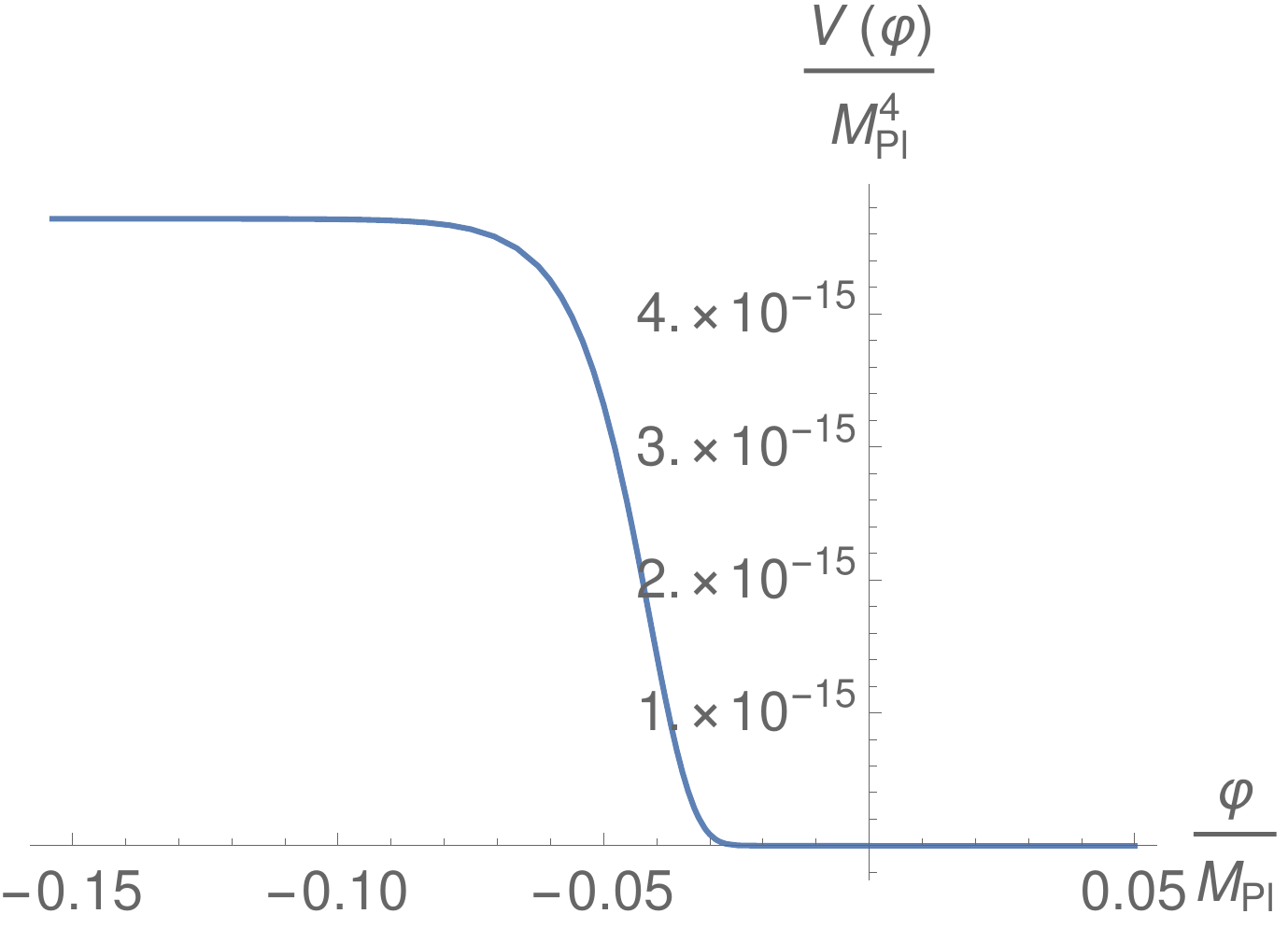}
\caption{\it{The shape of the scalar potential \eqref{exp} with  $\xi\sim 122$ and $\lambda\sim 10^{-69}$. The left side shows the inflationary energy density and the right side shows the late dark energy density. }}
\label{fig:potLQI}
\end{center}
\end{figure}

We can see the shape of the potential on Fig. \ref{fig:potLQI},  where the inflationary epoch takes place on the left-hand side of the graph, while the dark energy epoch occurs on the right-hand side.

\

\subsection{Calculation of the value of the  parameters involved in the model}

We start with 
the main slow-roll parameter, which is given by 
\begin{eqnarray}
\epsilon\equiv \frac{M_{pl}^2}{2}\left( \frac{V_{\varphi}}{V} \right)^2= \frac{2\xi/(\Gamma\pi)}{\cosh^2\left(\gamma\frac{\varphi}{M_{pl}} \right)}
=\frac{2\gamma^2\xi^2/\pi^2}{\cosh^2\left(\gamma\frac{\varphi}{M_{pl}} \right)}
\end{eqnarray}
and, since inflation ends when $\epsilon_{END}=1$, one has to assume that $\frac{2\xi}{\Gamma\pi}>1$ to guarantee the end of this period.

\

In fact, we have
\begin{eqnarray}
\varphi_{END}=\frac{M_{pl}}{\gamma}\ln\left( \sqrt{\frac{2\xi}{\Gamma\pi}} -\sqrt{\frac{2\xi}{\Gamma\pi}-1} \right)=
\frac{M_{pl}}{\gamma}\ln\left[\frac{\sqrt{2}\xi}{\pi}\left( \gamma-\sqrt{\gamma^2-\frac{\pi^2}{2\xi^2}}  \right)\right]<0,
\end{eqnarray}
and we can see that, for large values of $\gamma$, one has that $\varphi_{END}$ is close to zero. Thus, we will choose $\gamma\gg 1\Longrightarrow \Gamma\xi\ll 1$, which is completely compatible with the condition $\frac{2\xi}{\Gamma\pi}>1$.

\

On the other hand, the other important slow-roll parameter is given by
\begin{eqnarray}
\eta\equiv M_{pl}^2\frac{V_{\varphi\varphi}}{V}=\frac{2\xi\gamma^2}{\pi}\frac{\tanh\left(\gamma\frac{\varphi}{M_{pl}}  \right)}{\cosh\left(\gamma\frac{\varphi}{M_{pl}} \right)}
+\frac{4\gamma^2\xi^2/\pi^2}{\cosh^2\left(\gamma\frac{\varphi}{M_{pl}} \right)}.\end{eqnarray}

Both slow-roll parameters have to be evaluated when the pivot scale leaves the Hubble radius, which will happen for large values of 
$\cosh\left(\gamma\varphi/M_{pl} \right)$, obtaining 
\begin{equation}
\epsilon_*=\frac{2\gamma^2\xi^2/\pi^2}{\cosh^2\left(\gamma\frac{\varphi_*}{M_{pl}}  \right)}, \quad \eta_*\cong \frac{2\xi\gamma^2}{\pi}\frac{\tanh\left(\gamma\frac{\varphi_*}{M_{pl}}  \right)}{\cosh\left(\gamma\frac{\varphi_*}{M_{pl}}  \right)},
\end{equation}
with $\varphi_{*}<0$. Then, since the spectral index is given in the first approximation by $n_s\cong 1-6\epsilon_*+2\eta_*$,  one gets after some algebra
\begin{eqnarray}
n_s\cong 1+ 2\eta_*\cong 1-\gamma\sqrt{r/2},
\end{eqnarray}
where $r=16\epsilon_*$ is the ratio of tensor to scalar perturbations.

\

Now, we calculate the number of e-folds from the horizon crossing to the end of inflation, which is given by
\begin{eqnarray}
{\mathcal N}=\frac{1}{M_{pl}}\int_{\varphi_*}^{\varphi_{END}}\frac{1}{\sqrt{2\epsilon}}d\varphi= \frac{\pi}{2\gamma^2\xi}
\left[\sinh\left(\gamma\varphi_{END}/M_{pl} \right)-\sinh\left(\gamma\varphi_*/M_{pl} \right) \right]\cong \frac{\xi}{\sqrt{2\epsilon_*}}\nonumber,
\end{eqnarray}
so we have that
\begin{eqnarray}\label{n_s-r}
n_s\cong 1-\frac{2}{{\mathcal N}},\qquad r\cong\frac{8}{{\mathcal N}^2\gamma^2},
\end{eqnarray}
meaning that our model leads to the same spectral index and tensor/scalar ratio as the  $\alpha$-attractors models with $\alpha=\frac{2}{3\gamma^2}$  (see for instance \cite{Linde:1981mu}).


\begin{figure}[H]
\begin{center}
\includegraphics[width=8cm]{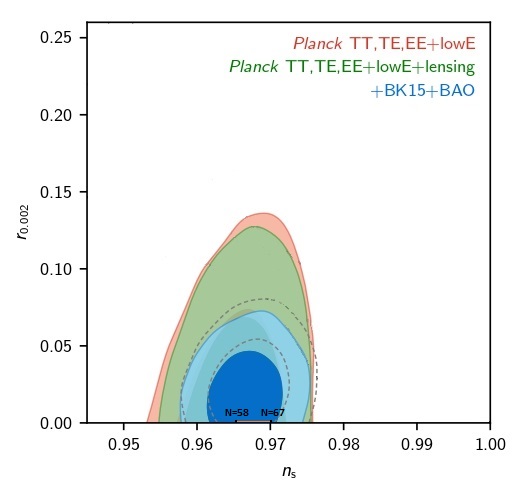}
\caption{\it{The Marginalized joint confidence contours for $(n_s,r)$ at $1\sigma$ and $2\sigma$ CL, without the presence of running of the spectral indices. We have drawn the curve for the present model for $\gamma\gg 1$ from ${\mathcal N}=58$ to ${\mathcal N}=67$} e-folds. (Figure courtesy of the Planck2018 Collaboration \cite{planck18a})}
\end{center}
\end{figure}


Finally, it is well-known that the power spectrum of scalar perturbations is given by
\begin{eqnarray}\label{power}
{\mathcal P}_{\zeta}=\frac{H_*^2}{8\pi^2\epsilon_*M_{pl}^2}\sim 2\times 10^{-9}.
\end{eqnarray}

Now, since in our case $V(\varphi_*)\cong \lambda M^4_{pl}e^{\xi}$, meaning that $H_*^2\cong \frac{\lambda M^2_{pl}}{3}e^{\xi}$, and taking into account that
$\epsilon_*\cong \frac{(1-n_s)^2}{8\gamma^2}$, one gets the constraint
\begin{eqnarray}\label{constraint}
\lambda \gamma^2 e^{\xi}\sim 7\times 10^{-11},
\end{eqnarray}
where we have chosen as a value of $n_s$ its central value $0.9649$.

\

Summing up, we will choose our parameters satisfying the condition (\ref{constraint}), with $\gamma\gg 1$ and $\xi\gg 1$, which will always fulfill the constraints $\Gamma\xi\ll 1$ and $\frac{2\xi}{\Gamma\pi}$ that we have imposed. Then, to find the values of the parameters one can perform the following heuristic argument: Taking for example $\gamma=10^2$, the constraint  (\ref{constraint}) becomes $\lambda e^{\xi}\sim 7\times 10^{-15}$. On the other hand, at the present time we will have
$\gamma\varphi_0/M_{pl}\gg 1$ where $\varphi_0$ denotes the current value of the field. Thus, we will have $V(\varphi_0)\sim \lambda M_{pl}^4e^{-\xi}$, which is the dark energy at the present time, meaning that 
\begin{eqnarray}
0.7\cong \Omega_{\varphi, 0}\cong \frac{V(\varphi_0)}{3H^2_0 M_{pl}^2}
\sim \frac{\lambda e^{-\xi}}{3}\left(\frac{M_{pl}}{H_0}\right)^2.
\end{eqnarray}

So, for  the value $H_0=67.81\; \mbox{Km/sec/Mpc}=5.94\times 10^{-61} M_{pl}$, we get the equations
\begin{eqnarray}
\lambda e^{\xi}\sim 7\times 10^{-15}
\qquad \mbox{and} \qquad \lambda e^{-\xi}\sim 10^{-120},\end{eqnarray}
whose solution is given by  $\xi\sim 122$ and $\lambda\sim 10^{-69}$.

\

If we choose $\gamma\sim 10^2$,  we see that the values of $\xi$ and $\gamma$ could be set equal in order to obtain the desired results from both the early and late inflation. From now on we will set $\xi=\gamma$, since it may help to find a successful combination of parameters because it reduces the number of effective parameters. As we will see later, numerical calculations show that, in order to have $\Omega_{\varphi,0}\cong 0.7$ (observational data show that, at the present time, the ratio of the energy density of the scalar field to the critical one is approximately $0.7$), one has to choose $\xi=\gamma\cong 121.8$. 

\


Next, we aim to find the relation between the number of e-folds and the reheating temperature $T_{reh}$.
We start once again with the formula 
\begin{eqnarray}\frac{k_*}{a_0H_0}=e^{-{\mathcal N}}\frac{H_*}{H_0}\frac{a_{END}}{a_{kin}}\frac{a_{kin}}{a_{reh}}
\frac{a_{reh}}{a_{eq}}\frac{a_{eq}}{a_{0}}
\end{eqnarray}
and, following the same steps as in the case of the Peebles-Vilenkin model, we get

\begin{eqnarray}
{\mathcal N}\cong -5+\ln\left(\frac{H_*}{H_0}  \right)+ \ln\left(\frac{a_{END}}{a_{kin}}  \right)+\frac{1}{4}\ln\left(\frac{g_{eq}}{g_{reh}}  \right)
+\frac{1}{6}\ln\left(\frac{\rho_{reh}}{\rho_{kin}}  \right)+ \ln\left(\frac{T_0}{T_{reh}}  \right).
\end{eqnarray}

\

Now, from the formula of the power spectrum of scalar perturbations \eqref{power} we infer that $H_*\sim 4\times 10^{-4}\sqrt{\epsilon_*} M_{pl}$, obtaining
\begin{eqnarray}
{\mathcal N}\cong 125+\frac{1}{2}\ln\epsilon_*+ \ln\left(\frac{a_{END}}{a_{kin}}  \right)
+\frac{1}{6}\ln\left(\frac{\rho_{rh}}{\rho_{kin}}  \right)+ \ln\left(\frac{T_0}{T_{reh}}  \right),
\end{eqnarray}
and introducing the current value of the temperature of the universe $T_0\sim 9.6\times 10^{-32}M_{pl}$ we get 
\begin{eqnarray}
{\mathcal N}\cong 54+\frac{1}{2}\ln\epsilon_*+ \ln\left(\frac{a_{END}}{a_{kin}}  \right)
-\frac{1}{3}\ln\left(\frac{H_{kin}T_{reh}}{M_{pl}^2 } \right).
\end{eqnarray}

So, since we have numerically checked that $H_{kin}\sim 4\times 10^{-8} M_{pl}$, we get
\begin{eqnarray}
{\mathcal N}+\ln {\mathcal N}\cong 55
-\frac{1}{3}\ln\left(\frac{T_{reh}}{M_{pl} } \right),
\end{eqnarray}
where we have used that $\epsilon_*=\frac{1}{2\gamma^2N^2}$ and we have also numerically computed that $\ln\left(\frac{a_{END}}{a_{kin}} \right)\cong -0.068$.

\

Finally,
taking into account that the scale of  nucleosynthesis is $1$ MeV and in order to avoid the late-time decay of gravitational relic products such as moduli fields or gravitinos  which could jeopardise  the nucleosynthesis success, one needs temperatures lower than $10^9$ GeV. So, we will assume that $1 \mbox{ MeV}\leq T_{reh}\leq 10^9 \mbox{ GeV}$, which leads to constrain the number of e-folds to $58\lesssim N\lesssim 67$. And for this number of e-folds, $0.966\lesssim n_s\lesssim 0.970$, which enters within its $2\sigma$ CL range.

\subsection{Present and future evolution: Numerical analysis}

\begin{figure}[H]
\begin{center}
\includegraphics[width=0.41\textwidth]{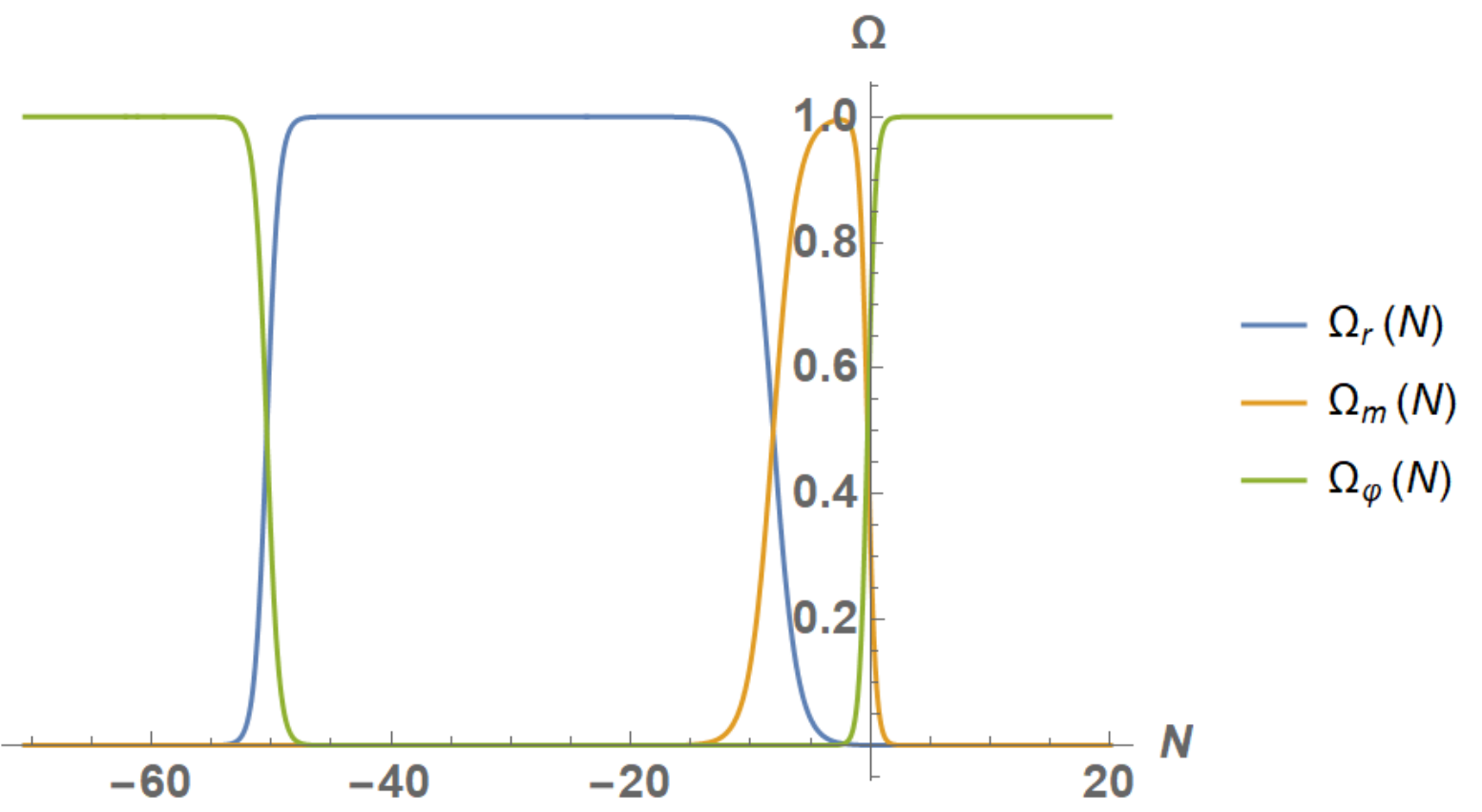}
\includegraphics[width=0.31\textwidth]{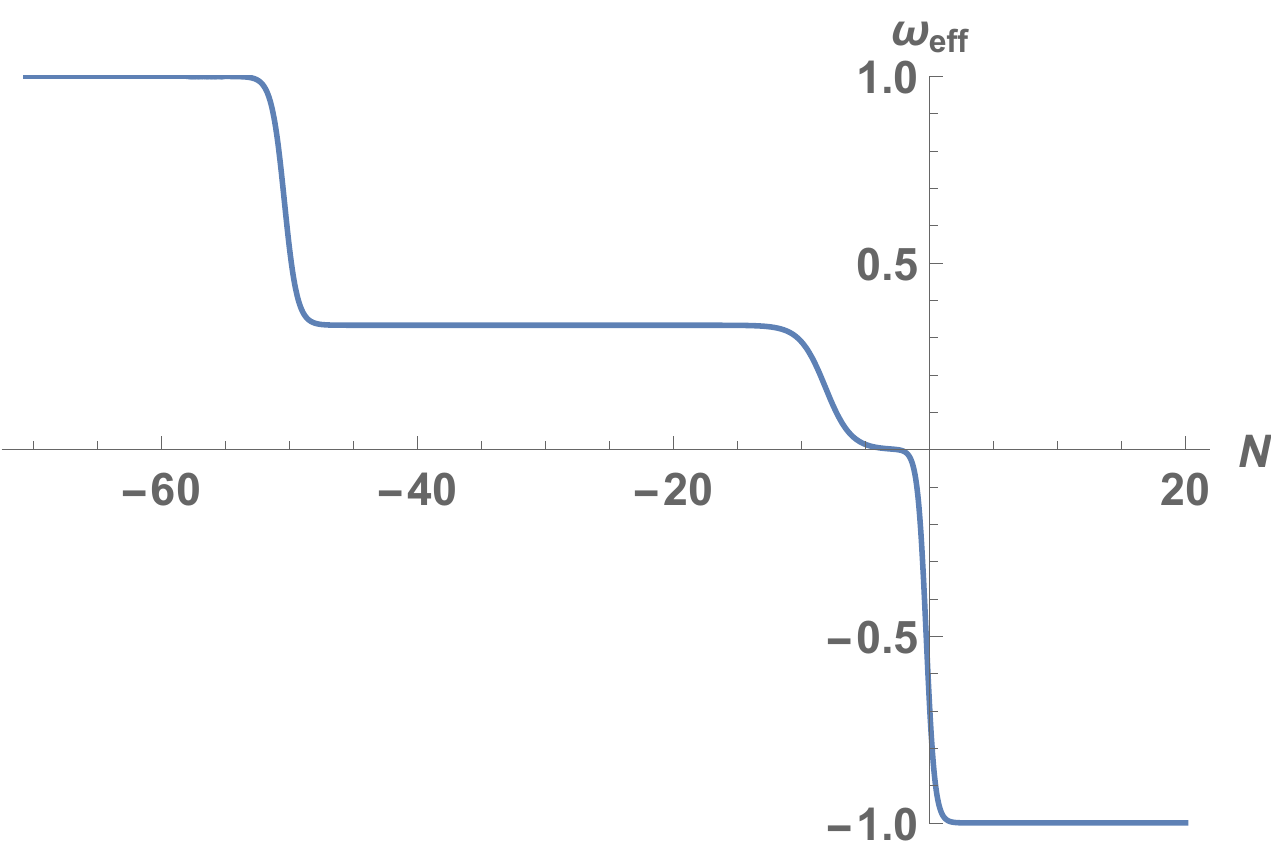}
\caption{\it{\textbf{Left:} The density parameters $\Omega_m=\frac{\rho_m}{3H^2M_{pl}^2}$ (orange curve), $\Omega_r=\frac{\rho_r}{3H^2M_{pl}^2}$ (blue curve) and $\Omega_{\varphi}=\frac{\rho_{\varphi}}{3H^2M_{pl}^2}$, from kination to future times,
{ for  a reheating temperature around $10^9$ GeV}.  \textbf{Right:} The effective Equation of State parameter $w_{eff}$, from kination to future times. As one can see in the picture, after kination the universe enters in a large period of time where radiation dominates. Then, after the matter-radiation equality, the universe becomes matter-dominated and, finally, near the present, it enters in a new accelerated phase where $w_{eff}$ approaches $-1$.}} \label{fig:Omega1}
\end{center}
\end{figure}

To understand the present and future evolution of our universe, 
we have integrated numerically the dynamical system (\ref{system}) with $K=H_0$ and initial conditions at the reheating time (obtained numerically) $\varphi_{kin}\sim -0.003 M_{pl}$ and 
$\dot{\varphi}_{kin}\sim 8\times 10^{-8} M_{pl}^2$. The results are presented in Figure 
\ref{fig:Omega1}, where one can see that in LQI the universe accelerates forever at late times with an effective EoS parameter equal to $-1$.


\

\section{$\alpha$-attractors in Quintessential Inflation}\label{secv}

The concept of $\alpha$-attractor, in the context of standard inflation, was introduced for the first time in \cite{kalloslinde}, obtaining  models which generalize  the well-known Starobinsky model (see \cite{riotto} for a detailed revision of the Starobinsky model).

In this section we consider $\alpha$-attractors in the context of Quintessential Inflation. For this reason we deal with
the following Lagrangian motivated by supergravity and corresponding to a non-trivial K\"ahler manifold (see for instance \cite{dimopoulos0} and the  references therein), combined with  a standard   exponential potential,
\begin{eqnarray}\label{lagrangian}
\mathcal{L}=\frac{1}{2}\frac{\dot{\phi}^2}{(1-\frac{\phi^2}{6\alpha}  )^2}M_{pl}^2-\lambda M_{pl}^4 e^{-\kappa \phi},
\end{eqnarray}
where $\phi$ is a dimensionless scalar field, and $\kappa$ and $\lambda$ are positive dimensionless constants.

\

In order that the kinetic term has the canonical  form, one can redefine the scalar field as follows,
\begin{eqnarray}
\phi= \sqrt{6\alpha}\tanh\left(\frac{\varphi}{\sqrt{6\alpha}M_{pl}}  \right),
\end{eqnarray}
obtaining the following potential,
\begin{eqnarray}\label{alpha}
V(\varphi)=\lambda M_{pl}^4e^{-n\tanh\left(\frac{\varphi}{\sqrt{6\alpha}M_{pl}} \right)},
\end{eqnarray}
where we have introduced the dimensionless parameter $n=\kappa\sqrt{6\alpha}$. Similarly to 
\cite{Benisty:2020qta, Benisty:2020xqm, Guendelman2}, the potential satisfies the {\it cosmological seesaw mechanism}, where the left side of the potential gives a very large energy density -the inflationary side- and the right side gives a very small energy density -the dark energy side. The asymptotic values are $V_{\pm} = \lambda \exp(\pm n)$. The parameter $n$ is the logarithm of the ratios between the energy densities, as $\xi$ in the earlier versions \cite{Guendelman2}.

\subsection{Calculation of the value of the  parameters involved in the model}

Dealing with this potential at early times, the main slow-roll parameter is given by
\begin{eqnarray}
\epsilon\equiv \frac{M_{pl}^2}{2}\left( \frac{V_{\varphi}}{V} \right)^2= \frac{n^2}{12\alpha}\frac{1}{\cosh^4\left(\frac{\varphi}{\sqrt{6\alpha}M_{pl}} \right)},
\end{eqnarray}
where we must assume that 
$ \frac{n^2}{12\alpha}>1$ because inflation ends when $\epsilon_{END}=1$, and  the other slow-roll parameter is 
\begin{eqnarray}
\eta\equiv M_{pl}^2\frac{V_{\varphi\varphi}}{V}=\frac{n}{3\alpha}\hspace{-0.1cm}\left[\hspace{-0.1cm}\frac{\tanh\left( \frac{\varphi/M_{pl}}{\sqrt{6\alpha}} \right)}{\cosh^2\left(\frac{\varphi/M_{pl}}{\sqrt{6\alpha}} \right)}
+\frac{n/2}{\cosh^4\left( \frac{\varphi/M_{pl}}{\sqrt{6\alpha}}\right)}\hspace{-0.1cm}\right].\end{eqnarray}

Both slow-roll parameters have to be evaluated at the horizon crossing, which will happen for large values of 
$\cosh\left(\frac{\varphi}{\sqrt{6\alpha}M_{pl}} \right)$, obtaining 
\begin{eqnarray}\label{parameters}
\epsilon_*\hspace{-0.1cm}=\hspace{-0.1cm}\frac{n^2}{12\alpha}\frac{1}{\cosh^4\left( \frac{\varphi_*/M_{pl}}{\sqrt{6\alpha}}\right)}, \qquad \eta_*\hspace{-0.1cm}
\cong\hspace{-0.1cm} -\frac{n}{3\alpha}\frac{1}{\cosh^2\left(\frac{\varphi_*/M_{pl}}{\sqrt{6\alpha}} \right)},
\end{eqnarray}
with $\varphi_*<0$.



\

Next, we calculate the number of e-folds from the horizon crossing to the end of inflation, which for small values of $\alpha$ is given by
\begin{eqnarray}
{\mathcal N}=\frac{1}{M_{pl}}\int_{\varphi_*}^{\varphi_{END}}\frac{1}{\sqrt{2\epsilon}}d\varphi
\cong \sqrt{\frac{3\alpha}{4\epsilon_*}},
\end{eqnarray}
so we get the standard form of the spectral index and the tensor/scalar ratio for an $\alpha$-attractor \cite{linde},
\begin{eqnarray}\label{power}
n_s\cong 
1-6\epsilon_*+2\eta_*\cong
1-\frac{2}{{\mathcal N}}, \qquad  r\cong 16\epsilon_*\cong\frac{12\alpha}{{\mathcal N}^2}.
\end{eqnarray}



Finally, it is well-known that the power spectrum of scalar perturbations is given by
\begin{eqnarray}
{\mathcal P}_{\zeta}=\frac{H_*^2}{8\pi^2\epsilon_*M_{pl}^2}\sim 2\times 10^{-9}
\end{eqnarray}
and, since in our case $V(\varphi_*)\cong \lambda M^4_{pl}e^{n}$ and, thus,  $H_*^2\cong \frac{\lambda M^2_{pl}}{3}e^{n}$, taking into account that
$\epsilon_*\cong \frac{3\alpha}{16}(1-n_s)^2$ one gets the constraint
\begin{eqnarray}\label{constraint}
\lambda  e^{n}/\alpha\sim 10^{-10},
\end{eqnarray}
where we have chosen as the value of $n_s$ its central value given by the Planck's team, i.e.,  $n_s=0.9649$ \cite{planck18}.

\


Choosing  for example $\alpha={ 10^{-2}}$, the constraint  (\ref{constraint}) becomes $\lambda e^{n}\sim {10^{-12}}$. On the other hand, at the present time we will have
$\frac{\varphi_0}{\sqrt{6\alpha}M_{pl}}\gg 1$, where $\varphi_0$ denotes the current value of the inflaton field. Hence, we will have $V(\varphi_0)\sim \lambda M_{pl}^4e^{-n}$, which is the dark energy at the present time, meaning that 
\begin{eqnarray}
0.7\cong \Omega_{\varphi, 0}\cong \frac{V(\varphi_0)}{3H^2_0 M_{pl}^2}
\sim \frac{\lambda e^{-n}}{3}\left(\frac{M_{pl}}{H_0}\right)^2.
\end{eqnarray}
Thus, taking for example the value provided by the Planck's team \cite{planck18,planck18a},  $H_0\cong 68\; \mbox{Km/sec/Mpc}\cong 6\times 10^{-61} M_{pl}$, we get the equations
\begin{eqnarray}\label{X}
\lambda e^{n}\sim  10^{-12}
\qquad \mbox{and} \qquad \lambda e^{-n}\sim 10^{-120},\end{eqnarray}
whose solution is given by  $n\sim 124$ and $\lambda\sim 10^{-66}$.

\

To end, some remarks are in order:
\begin{enumerate}
    \item We have chosen $\alpha\sim 10^{-2}$, but we can safely choose $\alpha\sim 1$. In that case the equations (\ref{X}) will become 
    \begin{eqnarray}
    \lambda e^n\sim 10^{10} \qquad \mbox{and} \qquad \lambda e^{-n}\sim 10^{-120},
    \end{eqnarray}
    whose solution is $n\sim 127$ and $\lambda\sim 10^{-65}$.
    \item If one takes $\alpha$ very small -for example of the order $10^{-2}$-, then, as has been shown in \cite{vardayan}, a much simpler model than the exponential one is possible with a linear potential. In fact, the authors of \cite{vardayan} showed that the potential
    $V(\phi)=\lambda(\phi+\sqrt{6\alpha})M_{pl}^4+\Lambda M_{pl}^4,$, which in terms of the canonically normalized field $\varphi$ has the form
    \begin{eqnarray}
    V(\varphi)=\lambda \sqrt{6\alpha}\left(\tanh\left(\frac{\varphi}{\sqrt{6\alpha}M_{pl}}\right)+1  \right)M_{pl}^4+\Lambda M_{pl}^4,
    \end{eqnarray}
    is viable for values of $\alpha\sim 10^{-2}$ and $\Lambda\sim 10^{-120}$.
    \item Another important important application of the $\alpha$-attractors is its use to alleviate the current Hubble tension \cite{finelli}. In that work, the authors, in the framework of $\alpha$-attractors, 
    include to the model  an Early Dark
Energy (EDE) component that adds energy to the Universe, because  the success in easing  the Hubble
tension crucially depends on the shape of this energy injection. In fact, in \cite{finelli} the authors use the following potential for EDE,
\begin{eqnarray}
V(\varphi)=\Lambda +V_0\frac{
\tanh^{2p}\left( \frac{\varphi}{\sqrt{6\alpha}M_{pl}}\right)}
{\left(1+\tanh\left( \frac{\varphi}{\sqrt{6\alpha}M_{pl}}\right)\right)^{2n}},
\end{eqnarray}
obtaining for $p=2$ and $n=4$ the following value of the Hubble rate at the present time: $H_0=(70.9\pm 1.1)$ $\frac{\mbox{km/s}}{\mbox{Mpc}}$
at $1\sigma$ C.L. Thus, the tension with the measurement from the SH0ES team ($H_0=(74.04\pm 1.42)$ $\frac{\mbox{km/s}}{\mbox{Mpc}}$) reduces to $1.75\sigma$, while this tension with the measurement from the Planck's team ($H_0=(67.36\pm 0.54)$ $\frac{\mbox{km/s}}{\mbox{Mpc}}$)
    was of $4.4\sigma$.
    
\end{enumerate}

\subsection{Present and future evolution: Numerical analysis}

\begin{figure}[H]
\begin{center}
\includegraphics[width=0.41\textwidth]{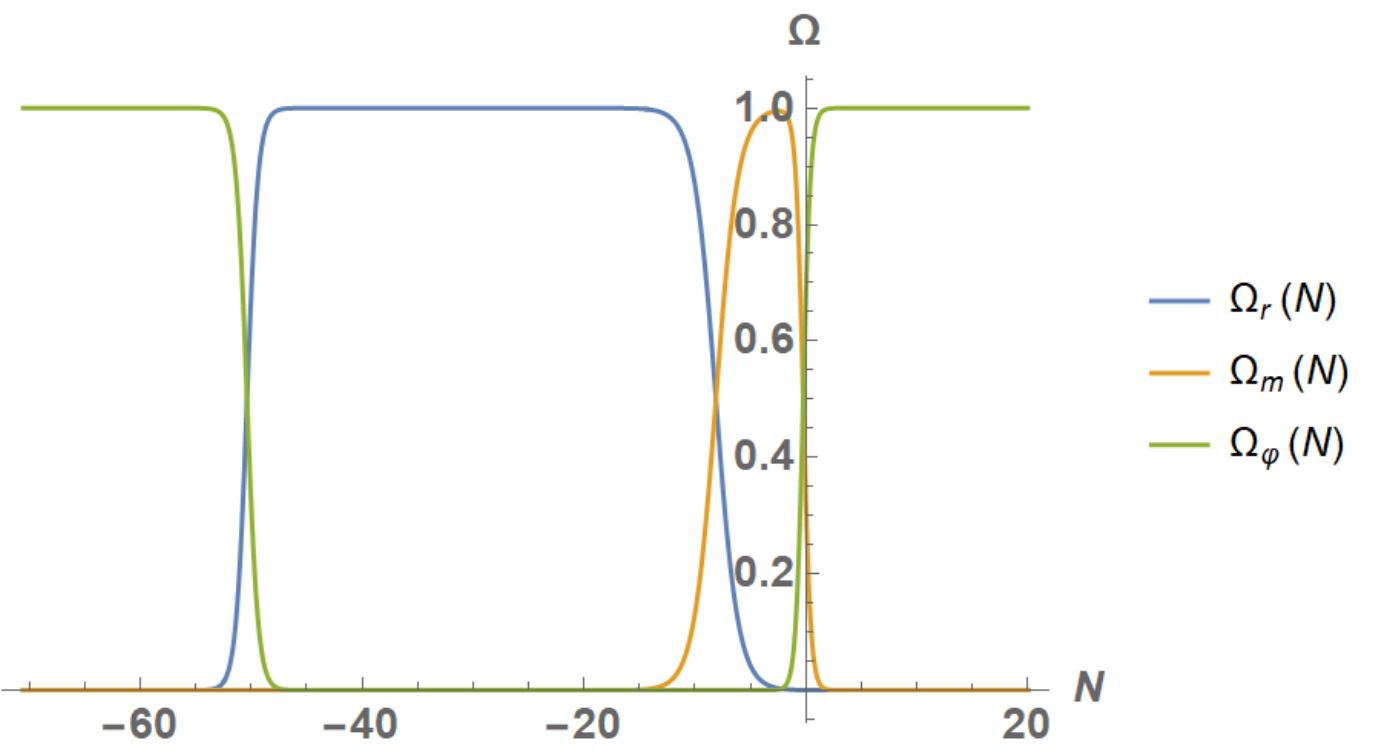}
\includegraphics[width=0.31\textwidth]{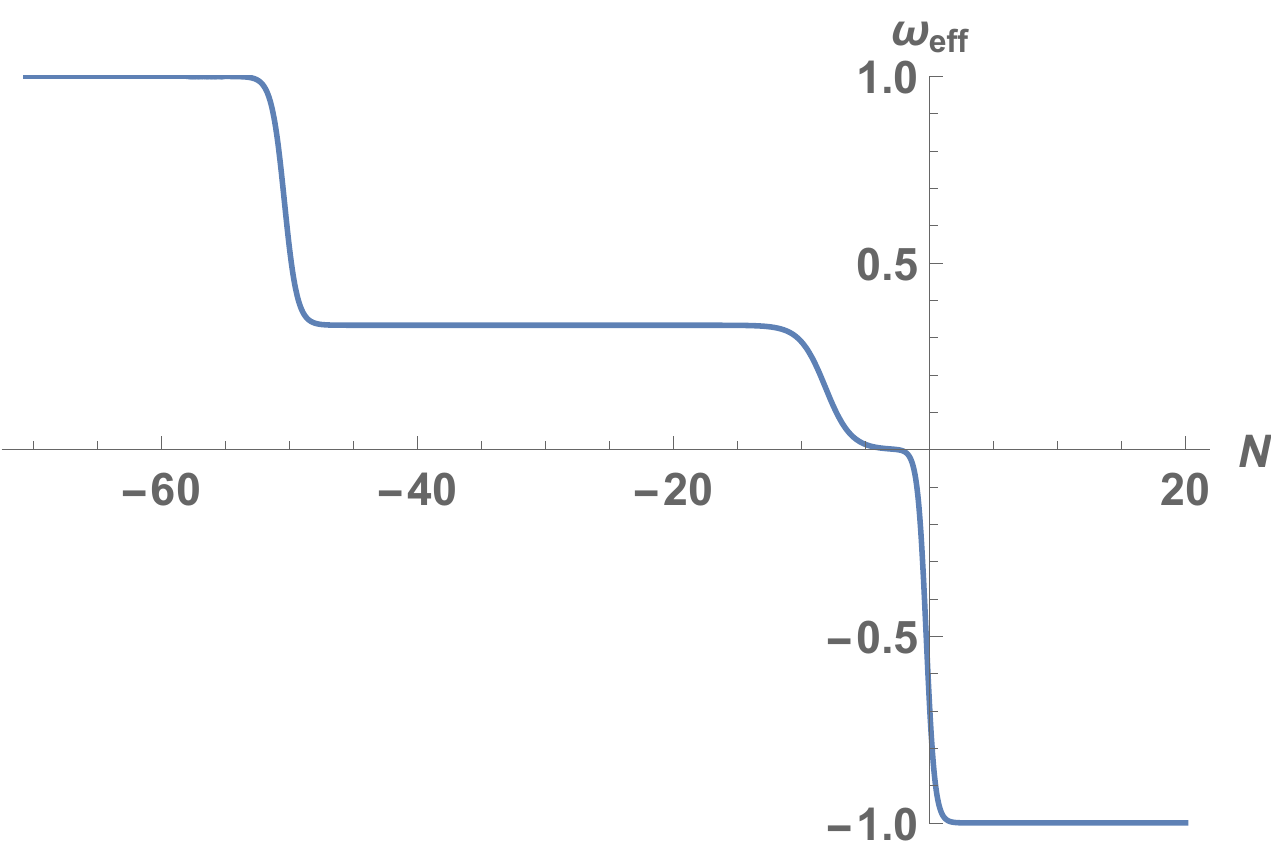}
\end{center}
\caption{\it{\textbf{Left:} The density parameters $\Omega_m=\frac{\rho_m}{3H^2M_{pl}^2}$ (orange curve), $\Omega_r=\frac{\rho_r}{3H^2M_{pl}^2}$ (blue curve) and $\Omega_{\varphi})=\frac{\rho_{\varphi}}{3H^2M_{pl}^2}$, from kination to future times
{ for  a reheating temperature around $10^9$ GeV}. \textbf{Right:} The effective Equation of State parameter $w_{eff}$, from kination to future times. As one can see in the picture, after kination the Universe enters in a large period of time where radiation dominates. Then, after the matter-radiation equality, the Universe becomes matter-dominated and, finally, near the present, it enters in a new accelerated phase where $w_{eff}$ approaches $-1$.}} \label{fig:Omega_attr}
\end{figure}

Once again, 
we have integrated numerically the dynamical system (\ref{system}) with $K=H_0$ and initial conditions at the reheating time (obtained numerically) $\varphi_{kin}\sim -0.5 M_{pl}$ and 
$\dot{\varphi}_{kin}\sim \times 10^{-6} M_{pl}^2$. The results  presented in Figure 
\ref{fig:Omega_attr} are the same as the ones obtained in the LQI scenario. So, both models predict the same
evolution of the universe.


\

\section{ Other Quintessential Inflation models}\label{secvi}

In the present section, 
some QI models, different from the ones studied in the previous sections,  are presented and  revisited.

\subsection{Dimopoulos work in Quintessential Inflation}

In this subsection we briefly  review some points of the vast contribution of K. Dimopoulos in this field, starting with the class of models introduced in \cite{Dimopoulos}, whose potential is given by
\begin{eqnarray}
V_k(\varphi)=m^4\left(1-\tanh(\varphi/M_{pl})  \right).\left(1-\sin\left(\frac{\pi\varphi/2}{\sqrt{\varphi^2+M^2}} \right)  \right)^k, 
\end{eqnarray}
where $M\ll M_{pl}$ and $m$ are two masses and $k$ is a positive integer. 

\

It is interesting to note that the potential has the following asymptotic behavior,
\begin{eqnarray}
V_k(\varphi)=\left\{\begin{array}{ccc}
   2^{k+1}m^4  & \mbox{for} & \varphi/M\ll 0, \\
    & & \\
   2^{1-k}\left(\frac{\pi}{4} \right)^{2k} e^{-2\varphi/M_{pl}}\left(\frac{m M^k}{\varphi^k}\right)^4 &  \mbox{for} & \varphi/M\gg 0,
\end{array}\right.
\end{eqnarray}
meaning that at very early times the potential approaches to a constant non-vanishing false vacuum responsible for inflation, and at late times it is approximately a quasi-exponential one which leads to quintessence.

\

The value of the parameter $m$ is obtained, as in standard inflation, using the power spectrum of scalar perturbation,
which for $k\leq 4$ leads to the value $m\sim 10^{15}$ GeV. And, in order to match with the current observational data, one has to choose 
\begin{eqnarray}
M\sim 4\times 10^2(3\times 10^{70})^{1/(4k)}10^{-30/k}M_{pl},
\end{eqnarray}
which has only sub-Planckian values for $k\leq 4$. In fact, for $k=4$ one has $M\sim 10^{18}$ GeV.

\

Another important contribution comes from {\it Modular Quintessential Inflation} \cite{Dimopoulos1}, where the author introduced the following toy potential,
\begin{eqnarray}
V_q(\varphi)=\frac{m^4}{\left[ \cosh\left(\varphi/M\right) \right]^q},
\end{eqnarray}
where $m$ and $M$ are masses and $q$ a positive integer. Once again, the asymptotic behavior is given by
\begin{eqnarray}
V_q(\varphi)=\left\{\begin{array}{ccc}
   m^4-\frac{q}{2}\left(\frac{m^2}{M} \right)^2 \varphi^2 & \mbox{for} & 0<\varphi\ll M, \\
    & & \\
   2^{q}m^4 e^{-q\varphi/M} &  \mbox{for} & \varphi\gg M,
\end{array}\right.
\end{eqnarray}
showing that the inflationary piece of the potential is like a top-hill one and the tail, which leads to quintessence, is a pure exponential one. For this modular model the minimum value of $m$ is around $10^7$ GeV and, as we have already seen, for pure exponential potentials,  in order that the tracker solution, which is an attractor, reproduces an accelerated universe, one needs that $q$ and $M$ satisfy the constraint 
$0<q\frac{M_{pl}}{M}<\sqrt{2}$. In addition, it is not difficult to show that the spectral index is given by
\begin{eqnarray}
n_s=1-2\epsilon_*-2q\left(\frac{M_{pl}}{M}\right)^2,
\end{eqnarray}
and thus, taking $n_s\cong 0.96$,  we can deduce that $q\left(\frac{M_{pl}}{M}\right)^2\leq 2\times 10^{-2}$. Finally, using the power spectrum of scalar perturbation one gets the following important relation between the parameters involved in the model:
\begin{eqnarray}
\left(\frac{m}{M_{pl}}\right)^4\sim \left( 10^{-1}- 5q\left(\frac{M_{pl}}{M}\right)^2\right)\times 10^{-7}.
\end{eqnarray}

\

Next,  we review  the treatment of $\alpha$-attractor done in \cite{dimopoulos0}. First of all, one has to introduce a negative Cosmological Constant (CC) (recall that Einstein's CC is positive) with the following form,
$\Lambda=-\lambda M_{pl}^2e^{-n}$, and thus, adding to the Lagrangian (\ref{lagrangian}) the term $\Lambda M_{pl}^2$ leads to the following effective potential,
\begin{eqnarray}\label{alpha1}
V(\varphi)=\lambda M_{pl}^4e^{-n}\left(e^{n\left(1-\tanh\left(\frac{\varphi}{\sqrt{6\alpha}M_{pl}} \right)\right)} -1\right),
\end{eqnarray}
whose main difference with the potential (\ref{alpha}) is that it vanishes at very late times.

\

During inflation $\varphi<0$ and the potential becomes as (\ref{alpha}) because $e^{-n}\ll 1$. So, as we have shown for the potential (\ref{alpha}), for small values of $\alpha$ inflation works also well when this CC is introduced in the model.

\

In the same way, for large values of the scalar field the potential will become
\begin{eqnarray}
V(\varphi)=2n\lambda e^{-n}M_{pl}^4e^{-\gamma\varphi/M_{pl}},
\end{eqnarray}
with $\gamma\equiv \sqrt{\frac{2}{3\alpha}}$. Since for an exponential potential a late-time eternal acceleration 
is  achieved when $\gamma<\sqrt{2}$, one has to choose   $\alpha>1/3$.  
However, since in this case one has to choose $\alpha>1/3$, the calculation of the spectral index and the ratio of tensor to scalar perturbations changes a little bit with respect to the case $\alpha\ll 1$, obtaining (see for details \cite{Dimopoulos:2017zvq})
\begin{eqnarray}
n_s\cong 1-\frac{2}{{\mathcal N}+\frac{\sqrt{3\alpha}}{2}} , \quad r=\frac{12\alpha}{\left({\mathcal N}+\frac{\sqrt{3\alpha}}{2}\right)^2}.\end{eqnarray}

\

Another important contribution of Dimopoulos is the introduction of the concept of {\it Warm Quintessential Inflation} in \cite{Dimopouloswarm}, which is applied to the original Peebles-Vilenkin model, obtaining the unification of early- and late-time acceleration, and providing  a very natural  mechanism to reheat our universe. In the same way, an interesting idea is to consider Quintessential Inflation in the context of Palatini $R^2$-gravity \cite{Dimopoulos3}, adapting the well-known Starobinsky model to QI.

\

About the reheating mechanism in Quintessential Inflation,  in \cite{tommi} the authors introduced a very interesting concept:   an originally subdominant, non-minimally coupled and massive scalar field with quartic self-interactions, and  without  any interaction with the inflaton field. During inflation the field is at zero vacuum, but during kination the minimum of the potential displaces from zero because the field becomes   tachyonic, and    thus,  it  is  ``kicked-off"   of the   origin  (which becomes a potential hill) due to quantum fluctuations. Finally, when it reaches the new minimum of potential it starts to oscillate coherently realising its energy and creating particles as in standard inflation.
However, as it is highlighted in \cite{rubio1,rubio2,rubio3,rubio4}, there is an  important limitation of the treatment
considered in  [54], namely the fact that reheating through
this mechanism  is far from being an homogeneous, coherent and
perturbative process. This observation has far reaching consequences,
completely absent in the original proposal, such as the production of
topological defects, the generation of a detectable gravitational wave
background or baryogenesis. Moreover, it allows the onset of radiation
domination for arbitrary potentials, and not only for quartic
interactions, as expected from the homogeneous treatment.

\

Another effective reheating can also be obtained from a combination of supergravity and string theory \cite{K1}. The model contains the inflaton and a Peccei-Quinn field, which will be the responsible for particle production. In fact, both fields are coupled as in the theory of {\it Instant Preheating}, which we will study in next section, and thus,  when the adiabatic evolution is broken the total kinetic energy of the inflaton decays into   radiation through resonant production of Peccei-Quinn particles, and the    residual  potential  density  of the inflaton field can be the responsible for  dark energy.

\

Finally, the {\it Curvaton rehating}, which we deal with in next section, is applied to QI in \cite{K2}, where bounds on the parameters of curvaton models are found. In addition,  using a minimal curvaton model, the authors showed  that the allowed parameter space is considerably larger than in the case of the usual oscillatory inflation models, i.e, than in standard inflation.

\subsection{Quintessential Inflation with  non-canonical scalar fields}

In \cite{hossain2} (see also  \cite{hossain1, Geng:2017mic}) the authors study Quintessential Inflation using the following action,
\begin{eqnarray}
S=\int d^4x \sqrt{-g} \left[\frac{M_{pl}^2}{2}R -\frac{k^2(\varphi)}{2}\partial^{\mu}\varphi \partial_{\mu}\varphi-V(\varphi)  \right]+S_m+S_r +S_{\nu}\left(\varphi, \Psi_{\nu}   \right),
\end{eqnarray}
where $S_m$ and $S_r$ are the actions for matter and radiation and $S_{\nu}$ is the action for neutrinos
(see \cite{neutrinos} and references therein for a discussion of the coupling of neutrinos and QI). Here, the Lagrangian density for massive neutrinos is given by \cite{ Geng:2017mic}
\begin{eqnarray}
{\mathcal L}_{\nu}=i\bar{\Psi}_{\nu}\gamma^{\lambda}\partial_{\lambda}\Psi_{\nu}-m_{\nu,0} e^{ \beta\varphi/M_{pl}}\bar{\Psi}_{\nu}{\Psi}_{\nu},
\end{eqnarray}
which depicts a non-minimal coupling between the inflaton and the neutrinos,
the potential is a pure exponential one,
\begin{eqnarray}
V(\varphi)= M_{pl}^4e^{-\alpha\varphi/M_{pl}},
\end{eqnarray}
and the coupling $k^2$ is given by
\begin{eqnarray}
k^2(\varphi)=\left(\frac{\alpha^2-\tilde{\alpha}^2}{\alpha^2} \right)\frac{1}{1+\beta^2e^{\alpha\varphi/M_{pl}}} +1, 
\end{eqnarray}
being $\alpha$, $\tilde{\alpha}$, $\gamma$ and $\beta$ the parameters of the model.

\

For this model the slow-roll parameters as a function of the number of e-folds ${\mathcal{N}}$ are given by \begin{eqnarray}
\epsilon_*=\frac{\tilde{\alpha}^2}{2}\frac{1}{1-e^{-\tilde{\alpha}^2\mathcal{N}}}\qquad \mbox{and}
\qquad \eta_*=\epsilon_*+\frac{\tilde{\alpha}^2}{2},
\end{eqnarray}
and thus, the spectral index and the tensor/scalar ratio are
\begin{eqnarray}
n_s\cong 1-\tilde{\alpha}^2\coth\left( \frac{\tilde{\alpha}^2\mathcal{N}}{2} \right)
\qquad \mbox{and}
\qquad r=\frac{8\tilde{\alpha}^2}{1-e^{-\tilde{\alpha}^2\mathcal{N}}}.
\end{eqnarray}

\

At first glance the model seems more complicated than the ones we have studied previously because it contains four parameters, but it simplifies very much because at late times, i.e, for large values of $\varphi$, the coupling $k^2$ goes to $1$, recovering the canonical form of the action. In addition, the effect of neutrinos, at late times,  is the modification of the potential $V(\varphi)$, becoming the following effective potential,
\begin{eqnarray}
V_{eff}(\varphi)=V(\varphi)+\rho_{\nu,0}e^{\gamma(\varphi-\varphi_0)/M_{pl}},
\end{eqnarray}
where $\varphi_0$ and  $\rho_{\nu,0}$ are the current values of the inflaton and the energy density of the massive neutrinos.

\

Therefore, since the potential $V(\varphi)$ is an exponential one, we are in the same situation studied in the subsection $3.3$, which, as we have already seen, leads to a viable model.

\subsection{Gauss-Bonnet Quintessential Inflation}

One of the problems of Quintessential Inflation is that due
to  the huge difference between the energy density scale of inflation and the current energy density  (which  is  over  a  hundred  orders  of  magnitude),  in  standard Quintessential  Inflation the  inflaton  field  typically  rolls  over  super-Planckian  distances  in  field  space, resulting  into a multitude of problems.  Firstly, the flatness of the quintessential tail may be lifted by radiative corrections.   Also,  because  the  associated  mass  is  so  small,  the  quintessence  field  may  give  rise  to  a so-called  5th  force  problem,  which  can  lead  to  a violation  of  the  equivalence  principle.   To  avoid  these problems, it is desirable to keep the field variation sub-Planckian.  In this case, however, to bridge the huge  difference  between  the  inflation  and  dark  energy  density  scales,  the  quintessential  tail  must  be steep.  But, if the quintessential tail is too steep, when the field becomes important today, it unfreezes and rolls down the steep potential not leading to accelerated expansion at all. One way to overcome this problem is to make sure the scalar field remains frozen today even though the quintessential tail is steep.  To this end,    a solution (see \cite{GB1} for a more detailed discussion) is the coupling of the field with the Gauss-Bonnet term, because such coupling impedes the variation of the field even if the potential is steep.  Thus,  a reason to use the GB scalar is that in some models the GB coupling could become important at late times making  sure that the field freezes with sub-Planckian displacement, such that it becomes the dark energy today without the aforementioned problems. Fortunately, as we have already shown,  the potential in Lorentzian Quintessential Inflation and the one of $\alpha$-attractors is steep enough and the field also freezes ($w_{eff}$ approaches to $-1$ at late times) at the present time leading to the current acceleration.

\

Therefore, here
 we will analyse the paper \cite{GB1} (see also \cite{oikonomou} for another paper about Gauss-Bonnet QI) where the authors consider the following action,
\begin{eqnarray}
S=\int d^4x \sqrt{-g} \left[\frac{M_{pl}^2}{2}[R-F(\varphi)\mathcal{G}] -\frac{1}{2}\partial^{\mu}\varphi \partial_{\mu}\varphi-V(\varphi)  \right],
\end{eqnarray}
where $F(\varphi)=F_0e^{-q\varphi/M_{pl}}$ ($q>0$)  is the coupling with the field and $\mathcal{G}=R^2-4R^{\mu\nu}R_{\mu\nu}+R^{\rho\mu\sigma\nu}
R_{\rho\mu\sigma\nu} $ is the Gauss-Bonnet scalar, which for the flat FLRW geometry has the simple form 
$\mathcal{G}=24H^2(\dot{H}+H^2)$.

\

On the other hand, the authors have chosen as a potential  a  mathematically  convenient  prototype  for  situations  in  which  an early time plateau, favoured by Planck, as well as an exponential quintessential tail are present. For example,
\begin{eqnarray}
V(\varphi)=V_0\left[ 1+\tanh\left( p\varphi/M_{pl} \right)  \right], \qquad p>0.
\end{eqnarray}

\

For this potential the number of e-folds is
$\mathcal{N}\cong \frac{1}{4p^2}e^{2p\varphi/M_{pl}}$, and the spectral index and the ratio of tensor to scalar perturbations as a function of $\mathcal{N}$ are given by
\begin{eqnarray}
n_s=1-\frac{2}{\mathcal{N}}\qquad \mbox{and} \qquad r=\frac{2}{p^2 \mathcal{N}^2},
\end{eqnarray}
which have the same form of the $\alpha$-attractors for $\alpha=\frac{1}{6p^2}$, so they clearly enter in the 2$\sigma$ C.L.

\


Finally,
since  our universe must be reheated by other means than inflaton decay, the authors employ  the instant preheating mechanism, in which, as we will see in next section,  the field is coupled to some other degrees of freedom. So, as the field is rapidly rolling down the  quintessential tail of its runaway potential, it induces massive particle production, which after the decay into lighter ones produces the  radiation bath of the hot Big Bang.  Soon afterwards, the inflaton field freezes at some value with small residual energy density, which becomes important at present, playing the role of dark energy. 

\subsection{Simple models of Quintessential Inflation}

In this subsection we will discuss very simple  models of Quintessential inflation that either do not have enough physical motivation or do not agree with the experimental data. 

\

We start with a model based on the $U(1)$ Peccei-Quinn symmetry proposed in \cite{rosenfeld}, where the real part of a complex field plays the
role of the inflaton whereas the imaginary part is the quintessence field.

\

The Lagrangian is given by 
\begin{eqnarray}
{\mathcal L}=\partial_{\mu}\Phi\partial^{\mu}\Phi^*-\lambda\left( \Phi\Phi^*-\frac{f^2}{2}\right)^2-M^4\left[\cos(\arg(\Phi)-1\right]],
\end{eqnarray}
where $\arg(\Phi)$ denotes the argument of the field $\Phi$. Writing the field as follows,
$\Phi=\frac{1}{\sqrt{2}}\phi e^{i\varphi/f}$, one finds the following equations for the real and imaginary part of the field,
\begin{eqnarray}
\ddot{\phi}+3H\dot{\phi}-\frac{\dot{\varphi}^2}{f^2}\phi+V'(\phi)=0,\nonumber\\
\ddot{\varphi}+\left(3H+\frac{\dot{g}}{g} \right)\dot{\varphi}-\frac{1}{g(\phi)}V'(\varphi)=0,
\end{eqnarray}
where $g(\phi)=\frac{\phi^2}{f^2}$ and the corresponding potentials are given by
\begin{eqnarray}
V(\phi)=\frac{\lambda}{4}\left(\phi^2-f^2 \right)^2 \qquad \mbox{and} \qquad 
V(\varphi)=M^4\left[\cos\left(\varphi/f \right)-1\right].
\end{eqnarray}

The inflaton field $\phi$ moves in a potential like a ``mexican-hat" and when it arrives at one of its minima it starts to oscillate and releases its energy producing particles as in standard inflation. On the contrary, the quintessence field $\varphi$ rolls in the potential $V(\varphi)$ producing enough dark energy to match with the current data provided that the parameters involved in the model satisfy 
\begin{eqnarray}
f>\frac{M_{pl}}{\sqrt{3}}\qquad \mbox{and} \qquad M\cong 3\times 10^{-3} \mbox{ eV}.
\end{eqnarray}

\

Another simple model was built in  \cite{bento1, bento2} in the context of brane-world context, where the modified Friedmann equation for the flat FLRW geometry becomes
\begin{eqnarray}
H^2= \frac{1}{3M_{pl, 4}^2}\left(1+\frac{\rho}{2\lambda}\right),
\end{eqnarray}
where $M_{pl,4}$ is the 4D reduced Planck's mass and the brane tension is given by
$\lambda= \frac{3}{32\pi^2}\frac{M_{pl, 5}^6}{M_{pl,4}^2}$ being $M_{pl,5}$  the 5D reduced Planck's mass.

\

In this context it has been shown that a potential composed by a sum of exponentials or a hyperbolic cosine
leads to Quintessential Inflation. For this reason the authors, adopting natural units ($M_{pl,4}=1$),  choose as a potential
\begin{eqnarray}
V(\varphi)=e^{-\alpha\varphi}\left[ A+(\varphi-\varphi_0)^2 \right].
\end{eqnarray}

\

For this potential, it is not difficult to calculate the spectral index and the tensor/scalar ratio as a function of the number of e-folds,
\begin{eqnarray}
n_s=1-\frac{4}{\mathcal{N}+1} \qquad \mbox{and} \qquad r=\frac{24}{\mathcal{N}+1},
\end{eqnarray}
which leads to the relation $6(n_s-1)=r$, which is, as in the original Peebles-Vilenkin model, clearly incompatible with the recent Planck's data (see Figure 1). 
Anyway, we continue with the model. The chosen reheating mechanism is the Instant Preheating, thus obtaining, as we will see in next section, a reheating temperature around $10^9$ GeV. Finally, the model
exhibits transient acceleration at late times for $0.96 \leq  A\alpha^2 \leq  1.26$ and $271 \leq  \alpha\varphi_0\leq  273$, while
eternal  acceleration is obtained for $2.3 \times 10^{-8}\leq A\alpha^2 \leq  0.98$ and $255 \leq \alpha\varphi_0 \leq 273$.

\

The last simple Quintessential Inflation model that we deal with was introduced in  \cite{deHaro:2016hpl}, where the idea is to find an unstable non-singular solution of the equations of General Relativity (the Friedmann and Raychaudhuri equations), as in the Starobinsky model \cite{staro} or the famous  Einstein's static model \cite{einstein},  in QI. The idea to build the model goes as follows: First of all, we look at early times for a Raychaudhuri equation of the form $\dot{H}=F(H)$ with $F(H)=-\alpha^2H^{\beta}$, and note that 
it has a finite time singularity for $\beta>1$. So, for large values of $H$,  we choose $F$ as a linear function of the Hubble rate. 
Secondly, to obtain a phase transition, one assumes that the derivative of $F$ is discontinuous at some point $H_E$ (this discontinuity enhances the particles production of heavy massive particles obtaining a finite reheating temperature compatible with the BBN success \cite{deHaro:2016hpl}), choosing for $H\lesssim H_E$
a kination regime, which corresponds to $w_{eff}=1 \Longrightarrow F(H)=-3H^2$. And finally, if the model has to take into account the current cosmic acceleration, the simplest
way is to assume that our dynamical system $\dot{H}=F(H)$ has a fixed point $H_f$, i.e., $F(H_f)=0$, meaning that $H_f$ is a de Sitter solution, which can be modeled for $H_f\lesssim H\ll H_E$ by the function $F(H)=-\alpha^2(H-H_f)^2$.

\


An example with all the properties mentioned above is 
\begin{eqnarray}
\dot{H}=\left\{\begin{array}{ccc}
  -3H_e(2H-H_e)   & \mbox{when} & H>H_E \\
  -3(H-H_f)^2   & \mbox{when} & H\leq H_E,
\end{array}\right.
\end{eqnarray}
where the  parameters involved satisfy $H_e\gg H_f$, and in order for $F$ to be continuous one has to choose $H_E=H_e+H_f+\sqrt{2H_eH_f}\cong H_e.$

\

For that model the effective EoS parameter $w_{eff}=-1-\frac{2\dot{H}}{3H^2}$ is given by 
\begin{eqnarray}
w_{eff}=\left\{\begin{array}{ccc}
  -1+2\frac{H_e}{H}\left(2- \frac{H_e}{H}  \right)   & \mbox{when} & H>H_E \\
  & &\\
 -1+2\left(1- \frac{H_f}{H}  \right)  & \mbox{when} & H\leq H_E,
\end{array}\right.
\end{eqnarray}
which shows that for $H\gg H_e$ one has an early quasi de Sitter period with $w_{eff}\cong -1$, when $H_e \gtrsim H\gg H_f $
the universe is in a kination phase ($w_{eff}=-1$), and finally,  for $H\gtrsim H_f$ one has a late quasi de Sitter regime with $w_{eff}\cong -1$.

\

On the other hand, integrating the equation $\dot{H}=F(H)$, we can see that the non-singular solution is
\begin{eqnarray}
H(t)=\left\{\begin{array}{ccc}
  \frac{H_e}{2}\left[ \left(1+2\frac{H_f}{H_e} +\sqrt{\frac{8H_f}{H_e}}\right)e^{-6H_et} +1  \right]   & \mbox{when} & t<0 \\
  \frac{H_e+\sqrt{2H_eH_f}}{3(H_e+\sqrt{2H_eH_f})t+1}  & \mbox{when} & t>0,
\end{array}\right.
\end{eqnarray}
which can be approximated by 
\begin{eqnarray}
H(t)\cong\left\{\begin{array}{ccc}
  \frac{H_e}{2}\left[ e^{-6H_et} +1  \right]   & \mbox{when} & t<0 \\
  \frac{H_e}{3H_et+1}  & \mbox{when} & t>0.
\end{array}\right.
\end{eqnarray}

\

To end, the potential can be obtained using the Raychaudhuri equation, which leads to 
\begin{eqnarray}
\varphi=M_{pl}\int \sqrt{-2\dot{H}}dt=-M_{pl}\int
\sqrt{-\frac{2}{\dot{H}}}dH.
\end{eqnarray}
Then, for the model studied here, when $H>H_E$,   one gets
\begin{eqnarray}
\varphi=-\frac{2M_{pl}}{\sqrt{3}}\sqrt{\frac{H}{H_e}-\frac{1}{2}}\Longleftrightarrow H=\frac{H_e}{2}\left( 
\frac{3\varphi^2}{2M_{pl}^2}+1\right),
\end{eqnarray}
and for $H<H_E$ a simple calculation shows that
\begin{eqnarray}
\varphi=-\sqrt{\frac{2}{3}}M_{pl}\ln\left( \frac{H-H_f}{H_e+\sqrt{2H_eH_f}} \right)+\varphi_E \nonumber \\
 {\big \Updownarrow} \qquad \qquad \qquad \qquad   \\ \nonumber
H=(H_e+\sqrt{2H_eH_f})e^{-\sqrt{\frac{3}{2}}(\varphi-
\varphi_E)/M_{pl}}+ H_f,
\end{eqnarray}
where we have introduced the notation $\varphi_E\equiv-\sqrt{\frac{2}{3}}M_{pl}
\sqrt{1+\frac{2H_f}{H_e}+\sqrt{\frac{8H_f}{H_e}}}\cong-\sqrt{\frac{2}{3}}M_{pl}$.

\

Finally, using the relation $V(H)=(3H^2+\dot{H})M_{pl}^2$, one obtains the following potential,
\begin{eqnarray}
V(\varphi)=\left\{\begin{array}{ccc}
  \frac{27H^2_eM_{pl}^2}{16}\left(\frac{\varphi^2}{M_{pl}^2}-\frac{2}{3} \right)  & \mbox{when} & \varphi<\varphi_E \\
  3H_f^2M_{pl}^2\left[ 2\left( \frac{H_e}{H_f}+\sqrt{\frac{2H_e}{H_f}} \right)e^{-\sqrt{\frac{3}{2}}(\varphi-\varphi_E)/M_{pl}} +1\right]  & \mbox{when} & \varphi>\varphi_E,
\end{array}\right.
\end{eqnarray}
which  depicts, at early times,  a $1$-dimensional Higgs potential, also called Double Well Inflationary potential, and, at late times, an exponential-quintessence potential.

\

\section{Reheating mechanisms}\label{secvii}

Different reheating mechanisms, such as gravitational particle production, instant preheating and curvaton reheating  are revisited, with all the details, in this section. 

\subsection{Gravitational particle production}

\subsubsection{Massless particle production}

Here, we consider a massless quantum field $\chi$ (intensively
considered in the early literature, see for instance \cite{ford,Zeldovich,Damour,Giovannini, pv}), which will be the responsible for particle production. This kind of field only interacts with gravity
 and we only assume that the particles are nearly conformally coupled with gravity, i.e., the coupling constant is approximately $1/6$ 
($\xi\cong \frac{1}{6}$) but not equal to $1/6$,  because free massless spinor and gauge fields are conformally-
invariant, so they do not  contribute  to the energy density of the relativistic plasma formed by the produced particles.

\

Then,  the modes, in Fourier space,  satisfy the Klein-Gordon equation 
\begin{eqnarray}\label{a2}
{\chi}''_{k}(\tau)+\left(k^2+
\left(\xi-\frac{1}{6}\right)a^2(\tau)R(\tau)\right){\chi}_{ k}(\tau)=0,
\end{eqnarray}
where $\tau$  is, once again,  the conformal time and $R(\tau)$ is the  Ricci scalar curvature.

\

To define the vacuum modes before  and after the phase transition, we 
assume that
 at early and late times the term $a^2R$ will vanish fast enough at early and late times, then its
 behavior at early and late times  is respectively
\begin{eqnarray}\label{vacuum}
 \chi_{in, k}(\tau)\simeq \frac{e^{-ik\tau}}{\sqrt{2{k}}} (\mbox{ when }\tau\rightarrow -\infty), \quad \chi_{out, k}(\tau)\simeq \frac{e^{-ik\tau}}{\sqrt{2{k}}}
 (\mbox{ when } \tau\rightarrow +\infty).
\end{eqnarray}

\

Therefore, 
 the  vacuum modes at early (``in" modes) and late times
(``out" modes) (exact solutions of (\ref{a2})) will be  given by \cite{ford}
\begin{eqnarray}\label{a36}
 \chi_{in, k}(\tau)=\frac{e^{-ik\tau}}{\sqrt{2{k}}}-\frac{\xi-1/6}{{k}}
\int_{-\infty}^{\tau}a^2(\tau')R(\tau')\sin({ k}(\tau-\tau'))\chi_{in, k}(\tau')d\tau', \nonumber\\
\chi_{out, k}(\tau)=\frac{e^{-ik\tau}}{\sqrt{2{k}}}+\frac{\xi-1/6}{{k}}
\int_{\tau}^{\infty}a^2(\tau')R(\tau')\sin({ k}(\tau-\tau'))\chi_{out, k}(\tau')d\tau',
\end{eqnarray}
and, since we are considering particles nearly conformally coupled {to gravity}, we can consider the term $(\xi-1/6)a^2(\tau)R(\tau)$ as a perturbation, and we can 
approximate the ``in''
and ``out'' modes
by the first order Picard's iteration, i.e., inserting (\ref{vacuum}) in the right-hand side of (\ref{a36}),   {as}
\begin{eqnarray}\label{a37}
 \chi_{in, k}(\tau)\cong \frac{e^{-ik\tau}}{\sqrt{2{k}}}-\frac{\xi-1/6}{{k}\sqrt{2{k}}}
\int_{-\infty}^{\tau}a^2(\tau')R(\tau')\sin({ k}(\tau-\tau')) e^{-ik\tau'} d\tau', \nonumber\\
\chi_{out, k}(\tau)\cong \frac{e^{-ik\tau}}{\sqrt{2{k}}}+\frac{\xi-1/6}{{k}\sqrt{2{k}}}
\int_{\tau}^{\infty}a^2(\tau')R(\tau')\sin({ k}(\tau-\tau'))e^{-ik\tau'}d\tau',
\end{eqnarray}
which will represent, respectively, the vacuum before and after the phase transition.

\

Thus, after the phase transition, we could write the ``in'' mode as a linear combination of the ``out'' mode and its conjugate as follows,
\begin{eqnarray}
 \chi_{in,k}(\tau)=\alpha_k\chi_{out,k}(\tau)+\beta_k\chi^*_{out,k}(\tau).
\end{eqnarray}

Imposing the continuity of $\chi$ and its first derivative at the transition time we get,
up to order $\left(\xi-1/6  \right)^2 $,
the value of these
coefficients \cite{Birrell1, Zeldovich},
\begin{eqnarray}\label{bogoliubov}
  \alpha_k\cong 1-\frac{i({\xi}-\frac{1}{6})}{2k}\int_{-\infty}^{\infty}a^2(\tau)
 R(\tau) d\tau,\quad  \beta_k\cong \frac{i({\xi}-\frac{1}{6})}{2k}\int_{-\infty}^{\infty}e^{-2ik\tau}a^2(\tau)
 R(\tau) d\tau.
\end{eqnarray}

\

Finally,
in a simple model of gravitational production of quanta with negligible rest mass 
the energy density of the produced particles due to the phase transition is given by
\cite{Birrell}
\begin{eqnarray}\label{rho}
 \rho_{\chi}=\frac{N_s}{2\pi^2a^4}\int_0^{\infty}k^3|\beta_k|^2 dk,
\end{eqnarray}
where  $N_S$ is the number of fermion and boson  fields involved in the model, which for simplicity we will assume of the order $1$, and
the integral of the $\beta$-Bogoliubov coefficient (\ref{bogoliubov}) is convergent because at early and late times the term $a^2(\tau)R(\tau)$ converges fast enough to zero.
In addition,
 if at the transition time  the first derivative of the Hubble parameter is continuous one has
$\beta_k\sim {\mathcal O}(k^{-3})$, which means that
 the energy density of the produced particles is not ultraviolet divergent. 
Then,
the energy density of the produced massless particles approximately becomes \cite{ford}
\begin{eqnarray}\label{rho1}
 \rho_{\chi}(t) \cong \bar{\mathcal N}\left({\xi}-\frac{1}{6}\right)^2H^{4}_{kin}\left(\frac{a_{kin}}{a(t)}\right)^4,
\end{eqnarray}
where $\bar{\mathcal N}$ is a dimensionless numerical factor.

\

{\bf Remark.-}
{\it The number $\bar{\mathcal N}$ is clearly model dependent. In the scenario proposed by Ford in \cite{ford} where there is
a transition from de Sitter to a matter domination
modeled by $a^2(\tau)R(\tau)\equiv\frac{12}{\tau^2+\tau_0^2}$, the number $\bar{\mathcal  N}$ can be calculated analytically giving as a result $\frac{9}{8}$.
However,
note that in this case reheating is impossible because the energy density of the produced particles decreases
faster than those of the background. To go beyond, we have calculated numerically this factor for some simple models that have a transition from a de Sitter regime to a kination one, and in all
cases $\bar{\mathcal N}$ is of the order $1$ (see for instance \cite{ha}).
}

\

On the other hand, 
it is well-known  that the thermalization of the produced particles is a very fast  but not instantaneous process  \cite{allahverdi} (see also the Section III of \cite{hyp}). In fact,  following the work of Spokoiny \cite{Spokoiny} (see also \cite{pv}), we 
can consider the  thermalization rate   $\Gamma_{th}=n_{\chi, kin}\sigma_{2\rightarrow 2}$, where the most important process for
kinetic equilibrium are $2\rightarrow 2$ scatterings with gauge boson exchange,  
whose typical energy energy is $E\sim \rho_{\chi,kin}^{1/4}$, 
in the $t$-channel, which is given by $\sigma_{2\rightarrow 2}=\frac{\alpha^3}{E^2}$ 
(see for instance the section
IV of \cite{allahverdi}), where, as usual, $\alpha^2\sim 10^{-3}$ \cite{Spokoiny}.

\

Given that
\begin{eqnarray}
  n_{\chi}(t)=\frac{1}{2\pi^2 a^3(t)}\int_0^{\infty}k^2|\beta_k|^2 dk
  =\left({\xi}-\frac{1}{6}\right)^2{\mathcal M}H^{3}_{kin}\left(\frac{a_{kin}}{a(t)}\right)^3,
\end{eqnarray}
 where for many models one finds  \cite{hap1}
\begin{eqnarray}{\mathcal M}\equiv \frac{1}{16\pi a^3_{kin}H^3_{kin}}\int_{-\infty}^{\infty} a^4(\tau)R^2(\tau) d\tau
\sim 1,
\end{eqnarray}
which is finite, because in all the considered models $R=6(\dot{H}+2H^2)$ is continuous and, as we have already explained, $a^2R$ vanishes fast enough at early and late times. So,
we get that
\begin{eqnarray}
\Gamma_{th}=\alpha^3\left({\xi}-\frac{1}{6}\right)\frac{{\mathcal M}H_{kin}}{\bar{\mathcal N}^{\frac{1}{2}}}\sim
\alpha^3\left({\xi}-\frac{1}{6}\right){H_{kin}}
.
\end{eqnarray}

\

The relativistic fluid reaches the thermal equilibrium when $H(t_{th})\sim H_{th}\sim \Gamma_{th}$ \cite{pv, Spokoiny}
and, since during kination the Hubble rate scales as $a^{-3}$, one has
\begin{eqnarray}
H_{th}=H_{kin}\left(\frac{a_{kin}}{a_{th}} \right)^3\sim \Gamma_{th}\sim \alpha^3\left({\xi}-\frac{1}{6}\right){H_{kin}}\Longrightarrow
\left(\frac{a_{kin}}{a_{th}} \right)^3\sim \alpha^3\left({\xi}-\frac{1}{6}\right),
\end{eqnarray}
 meaning that the temperature when the thermalization is achieved  is of the order
\begin{eqnarray}
T_{th}=\left(\frac{30}{\pi^2g_{th}}  \right)^{1/4} \rho_{\chi}^{1/4}(t_{th})\sim \alpha g_{th}^{-1/4}\left({\xi}-\frac{1}{6}\right)^{5/6}
H_{kin}.
\end{eqnarray}
Taking for instance $H_{kin}\sim 10^{-6} M_{pl}$ (the typical value of the Hubble rate at the beginning of kination for many models), $g_{th}=106.75$ (the effective number of degrees of freedom of the Standard Model) and ${\xi}-\frac{1}{6}\sim 10^{-2}$, one gets a thermalization temperature of the order of $10^9$ GeV.

\

 Finally, the reheating  occurs  when the energy density of the background and the one of the created particles is of the same order ($\rho_{\chi}(t_{reh})\sim \rho(t_{reh})\Longrightarrow \frac{a_{kin}}{a_{reh}}\sim \left({\xi}-\frac{1}{6}\right)
 \frac{H_{kin}}{M_{pl}}$),  which
 implies, for the typical value $H_{kin}\sim  10^{-6} M_{pl}$, a reheating temperature of the order
 \begin{eqnarray}
 T_{reh}\sim  g_{reh}^{-1/4}\rho_{\chi}^{1/4}(t_{reh})\sim  g_{reh}^{-1/4}\left({\xi}-\frac{1}{6}\right)^{3/2}
 \left(\frac{H_{kin}}{M_{pl}}\right)^2 M_{pl}\nonumber \\\sim 2\times 10^{6}g_{reh}^{-1/4}\left({\xi}-\frac{1}{6}\right)^{3/2}
 \mbox{GeV}, 
 \end{eqnarray}
which  for ${\xi}-\frac{1}{6}\sim 10^{-2}$ and $g_{reh}=106.75$ leads to a low reheating  temperature 
around $10^{3}$ GeV. 

 \subsubsection{Superheavy particle production: Calculations using the WKB approximation}

The study of the massive particle production in QI is a  bit more involved than for massless particles. Therefore,  
to warm up, we will consider, once again,  a toy model based on that of  Peebles-Vilenkin, namely
\begin{eqnarray}\label{toy}
V(\varphi)=\left\{\begin{array}{ccc}
\frac{1}{2}m^2\left(\varphi^2-M_{pl}^2+M^2\right)& \mbox{for}& \varphi\leq -M_{pl}\\
\frac{1}{2}m^2\frac{M^6}{(\varphi+M_{pl})^4+M^4}& \mbox{for}& \varphi\geq -M_{pl},
\end{array} \right.
\end{eqnarray}
where $m$ is the inflaton mass and $M$ is another very small mass whose value can be calculated in the same way as in the Peebles-Vilenkin model:
during the radiation and matter domination epochs the inflaton field is all the time of
the order $M_{pl}$ (see \cite{pv} for a detailed discussion).
Then, in the model the field will dominate at late times when
\begin{eqnarray}\frac{m^2M^6}{M^4_{pl}+M^4}\sim \frac{m^2M^5}{M^4_{pl}}\sim H_0^2M_{pl}^2\Longrightarrow M\sim \left(\frac{H_0}{m} \right)^{\frac{1}{3}}M_{pl}\sim 
10^{-18} M_{pl}\sim 
1 \mbox{ GeV},
\end{eqnarray}
where we have used that the current value of the Hubble parameter is $H_0\sim 10^{-61} M_{pl}$.

{\bf Remark}
{\it The mass of the inflaton field $m$ can be calculated using the formula of the power spectrum of scalar perturbations $P_{\zeta}=\frac{H_*^2}{8\pi^2 M_{pl}^2}\sim 2\times 10^{-9}$,
obtaining after a simple calculation
\begin{eqnarray}
m\sim \sqrt{5}\pi \epsilon_*\times 10^{-4} M_{pl}
\end{eqnarray}
and, since $\epsilon_*\sim 10^{-2}$, one finally gets $m\sim 10^{-5} M_{pl}$.
}

\

On the other hand, regarding the creation of superheavy massive particles conformally coupled with gravity that have no interaction with the inflaton field,
the time-dependent frequency of the $\chi$-particles in the $k$-mode is $\omega_k(\tau)=\sqrt{k^2+m_{\chi}^2a^2(\tau)}$, where $m_{\chi}$ denotes the mass of the quantum field $\chi$, 
 that is, the $\chi$ modes, in Fourier space, satisfy the Klein-Gordon (KG) equation
 \begin{eqnarray}\label{chi}
 \chi_k''+\omega_k^2(\tau)\chi_k=0.
 \end{eqnarray}

 Before continuing, the following observation should be made:

{\bf Remark.-}
{\it It is well-known that at temperatures of the  order of the Planck's mass quantum effects become very important  and the classical picture of the universe is not possible. However, at temperatures below $M_{pl}$, for example 
at GUT scales (i.e., when the temperature of the universe is of the order of $T\sim 10^{-3} M_{pl}\sim  10^{15}$ GeV), the beginning of the classical  Hot Big Bang (HBB) scenario is possible. Since for the  flat FLRW universe the energy density of the universe, namely $\rho$,  and the Hubble parameter $H$ 
are related through the Friedmann equation $\rho=3H^2M_{pl}^2$ and the temperature of the universe is related to the energy density via the Stefan-Boltzmann law $\rho = (\pi^2/30)g_{*} T^4$ (where  $g_*=106.75$ is the number degrees of freedom for the energy density in the Standard Model), one can conclude that a classical picture of our universe might be possible when $H\sim 10^{-5} M_{pl}\sim 10^{13}$ GeV. Then, 
if inflation starts at this scale, i.e. taking the value of the Hubble parameter at the beginning of inflation as $H_{B}\sim  10^{-5} M_{pl}$,
we will assume as a natural initial condition that the quantum $\chi$-field is in the vacuum at the beginning of inflation.
We will also choose  the mass of the  $\chi$-field two orders greater than this value of  the Hubble parameter ($m_{\chi}\sim 10^{-3} M_{pl}\sim 10^{15}$ GeV, which is a mass of the same order as those of the vector mesons responsible for transforming quarks into leptons in simple theories with SU(5)  symmetry \cite{lindebook}) because, as we will immediately see, the polarization terms will be sub-dominant and do not affect the dynamics of the inflaton field.
So, we will choose  $ 10^{13} \mbox{ GeV}\sim m\sim  H_{B}\ll m_{\chi}
\sim 10^{15} \mbox{ GeV}\ll M_{pl}\sim 10^{18} \mbox{ GeV}$.  
}

\

Thus, during the adiabatic regimes, i.e., when
 $H(\tau)\ll m_{\chi} \Longrightarrow\omega_k'(\tau)\ll \omega_k^2(\tau)
  $, one can use the WKB approximation \cite{Haro} of the equation (\ref{chi})
\begin{eqnarray}
\chi_{n,k}^{WKB}(\tau)\equiv
\sqrt{\frac{1}{2W_{n,k}(\tau)}}e^{-{i}\int^{\tau}W_{n,k}(\eta)d\eta},
\end{eqnarray}
where $n$ is the order of the approximation to calculate the $k$-vacuum mode. When some high order derivative
of the Hubble parameter is discontinuous, one has to match the $k$-vacuum modes before and after this moment, and it is precisely at this moment when one needs to use positive
frequency modes after the breakdown of the adiabatic regime in order to perform this matching, which,
following Parker's viewpoint \cite{Parker},
is the cause of the gravitational particle production. 

\

For our toy model,
note that the derivative of the potential is discontinuous at $\varphi=-M_{pl}$, which means, due to the conservation equation, that the second derivative of the 
inflaton field is discontinuous at the transition time and, consequently, from the Raychaudhuri equation
$\dot{H}=-\frac{\dot{\varphi}^2}{2M_{pl}^2}$ one can deduce that the second derivative of the Hubble parameter is also discontinuous at this time.

\

In this case one only needs the first order WKB solution to approximate  the $k$-vacuum modes before and after the phase transition
\begin{eqnarray}
\chi_{1,k}^{WKB}(\tau)\equiv
\sqrt{\frac{1}{2W_{1,k}(\tau)}}e^{-{i}\int^{\tau}W_{1,k}(\eta)d\eta},
\end{eqnarray}
where \cite{Winitzki}
\begin{eqnarray}
W_{1,k}=
\omega_k-\frac{1}{4}\frac{\omega''_{k}}{\omega^2_{k}}+\frac{3}{8}\frac{(\omega'_{k})^2}{\omega^3_{k}} .
\end{eqnarray}

Before the transition time, namely $\tau_{kin}$, which is very close to the  beginning of kination ($\varphi\cong -M_{pl}$), the  vacuum is depicted by $\chi_{1,k}^{WKB}(\tau)$, but after the phase transition this mode becomes a mix of positive and negative frequencies of the form
$\alpha_k \chi_{1,k}^{WKB}(\tau)+\beta_k (\chi_{1,k}^{WKB})^*(\tau)$. Then,
the $\beta$-Bogoliubov coefficient could be obtained matching both expressions at $\tau=\tau_{kin}$, obtaining
\begin{eqnarray}
\beta_k=\frac{{\mathcal W}[\chi_{1,k}^{WKB}(\tau_{kin}^-),\chi_{1,k}^{WKB}(\tau_{kin}^+)]}
{{\mathcal W}[(\chi_{1,k}^{WKB})^*(\tau_{kin}^+),\chi_{1,k}^{WKB}(\tau_{kin}^+)]},
\end{eqnarray}
where ${\mathcal W}[f(\tau_{kin}^-),g(\tau_{kin}^+)]=f(\tau_{kin}^+)g'(\tau_{kin}^-)
-f'(\tau_{kin}^+)g(\tau_{kin}^-)$ is the Wronskian of the functions $f$ and $g$ at the transition time, and being
$F(\tau_{kin}^{\pm})=\lim_{\tau\rightarrow \tau_{kin}^{\pm}}F(\tau)$.

\

The square modulus of the $\beta$-Bogoliubov coefficient will be given approximately by   \cite{hap}
\begin{eqnarray}\label{Beta}
 |\beta_k|^2\cong \frac{m^4_{\chi}a^{10}_{kin}\left(\ddot{H}(\tau_{kin}^+)-\ddot{H}(\tau_{kin}^-)\right)^2}{256(k^2+m_{\chi}^2a^2_{kin})^5}\cong
 \frac{m^4_{\chi}m^6a^{10}_{kin}}
 {256\omega_{kin}^{10}}
 ,
\end{eqnarray}
because
\begin{eqnarray}
\ddot{H}(\tau_{kin}^+)-\ddot{H}(\tau_{kin}^-)=
-\frac{\dot{\varphi}_{kin}}{M_{pl}^2}(\ddot{\varphi}(\tau_{kin}^+)- \ddot{\varphi}(\tau_{kin}^-))
=-\frac{\dot{\varphi}_{kin}}{M_{pl}^2}V_{\varphi}(-M_{pl}^-)= \frac{m^2\dot{\varphi}_{kin}}{M_{pl}}=
m^3,
 \end{eqnarray}
 where we have used that at the transition time all the  energy at the end of inflation,  which is
  approximately $\frac{1}{2}m^2 M_{pl}^2$ because  $\varphi_{END}=-\sqrt{2}M_{pl}$, was converted into kinetic energy.

\

Thus, for our model, the number density of the produced particles and its energy density
will be
\begin{eqnarray}
 n_{\chi}(t)\sim \frac{5m^3}{2\pi\times 16^4}\left(\frac{m}{m_{\chi}}\right)^3 \left(\frac{a_{kin}}{a(t)} \right)^3
 \sim 10^{-5}\left(\frac{m}{m_{\chi}}\right)^3 m^3\left(\frac{a_{kin}}{a(t)} \right)^3, \quad \rho_{\chi}(t)\sim m_{\chi}n_{\chi}(t).
\end{eqnarray}

Now, one has to note that there are two different situations, namely,  when the superheavy massive particles decay before and after the end of the kination regime.

\begin{enumerate}
\item {\it Decay of the superheavy particles into lighter ones before the end of kination.}

Let $\Gamma$ be the decay rate of the superheavy particles.
The decay is practically  finished when $\Gamma$ is of the same order of the Hubble rate,
i.e.,
 when ${\Gamma}\sim H(t_{dec})=H_{kin}\left(\frac{a_{kin}}{a_{dec}} \right)^3\cong
  \frac{m}{\sqrt{6}}\left(\frac{a_{kin}}{a_{dec}}\right)^3 $, 
 and thus, the corresponding energy densities will be
\begin{eqnarray}
\rho_{\varphi, dec}=3{\Gamma}^2M_{pl}^2 \quad \mbox{and} \quad \rho_{\chi, dec}\sim 2\times 10^{-5}\left( \frac{m}{m_{\chi}} \right)^2 \frac{{\Gamma}}{m}m^4.
\end{eqnarray}

\

Since the decay is before the end of kination, one has $\Gamma< H_{kin}$ and 
$\rho_{\chi, dec}<\rho_{\varphi, dec}$, which leads to the following bound,
\begin{eqnarray}
10^{-5} \left(\frac{m^2}{m_{\chi} M_{pl}}  \right)^2 m< \Gamma <H_{kin}, \end{eqnarray}
which for $m\sim 10^{13}$ GeV, $m_{\chi}\sim 10^{15}$ GeV and $H_{kin}\sim 10^{12}$ GeV becomes 
\begin{eqnarray}
10^{-6}<\Gamma/\mbox{GeV}<10^{12}.
\end{eqnarray}

Then, assuming as usual  nearly instantaneous thermalization, 
the reheating temperature, i.e., the temperature of the universe when the relativistic plasma in thermal equilibrium starts to dominate ($\rho_{\chi, reh}\sim \rho_{\varphi, reh}$), will be obtained taking into account that
\begin{eqnarray}
\rho_{\chi,reh}=\rho_{\chi, dec}\left(\frac{a_{dec}}{a_{reh}} \right)^4\sim
\rho_{\varphi, dec}\left(\frac{a_{dec}}{a_{reh}} \right)^6=\rho_{\varphi,reh}\Longrightarrow
\rho_{\chi,reh}\sim \frac{\rho_{\chi, dec}^3}{\rho_{\varphi,dec}},
\end{eqnarray}
and thus, 
\begin{eqnarray}\label{reheating1}
 T_{reh}\sim \left(\frac{\rho_{\chi,reh}}{g_{reh}} \right)^{\frac{1}{4}}\sim  \left(\frac{\rho_{\chi,reh}}{g_{reh}} \right)^{\frac{1}{4}}\sqrt{\frac{\rho_{\chi, dec}}{\rho_{\varphi, dec}}}
\sim 10^{-4}\left(\frac{m}{m_{\chi}}  \right)^{\frac{3}{2}}\left(\frac{m}{\Gamma}  \right)^{\frac{1}{4}}\left(\frac{m}{M_{pl}}  \right)^2 M_{pl},
\end{eqnarray}
which for the values of our parameters leads to a reheating temperature of the order 
\begin{eqnarray}
T_{reh}\sim 10^{-17}\left( \frac{M_{pl}}{\Gamma}\right)^{1/4}M_{pl}\sim 3\times 10^{-15}
\left( \frac{\mbox{GeV}}{{\Gamma}}\right)^{1/4}\mbox{GeV}.
\end{eqnarray}

\

Finally, using the bound of the decay rate, we deduce that the reheating temperature is bounded by
\begin{eqnarray}
3\times 10^2 \mbox{ GeV}< T_{reh}< 10^6 \mbox{ GeV}.
\end{eqnarray}

\

\item {\it Decay of the superheavy particles into lighter ones after the end of kination}

First of all, recall that we have already seen that at the end of kination one has
\begin{eqnarray}
H_{end}=\sqrt{2}H_{kin}\Theta \qquad \mbox{ with } \qquad \Theta=
\frac{\rho_{\chi, kin} }{\rho_{\varphi, kin}}\cong 2\times 10^{-5}\left(\frac{m^2}{m_{\chi} M_{pl}}  \right)^2\sim 10^{-19}.
\end{eqnarray}

Taking into account that the decay is after the end of kination we will have the bound
$\Gamma< H_{end}\sim 10^{-6}$ GeV, and thus, since the thermalization is nearly instantaneous, the reheating time coincides when the decay is completed, i.e, when $H\sim \Gamma$, and consequently, the reheating temperature will be 
\begin{eqnarray}
T_{reh}\sim g_{reh}^{-1/4}\sqrt{\Gamma M_{pl}},
\end{eqnarray}
which is bounded by 
\begin{eqnarray}
1 \mbox { MeV}< T_{reh}< 10^6 \mbox{ GeV},
\end{eqnarray}
where the lower bound is obtained taking into account that the reheating will be before the Big Bang Nucleosynthesis  which occurs around $1$ MeV. 
\end{enumerate}

\

\subsubsection{The diagonalization method}

 In the last subsection we have used the WKB method to calculate the particle production when some derivative of the potential has a discontinuity, but when the potential is smooth enough it does not work because one needs the exact solution of the equation (\ref{chi}).

So, here we will explain the so-called {\it diagonalization method}, which is essential to calculate the massive  particle creation in more realistic scenarios such as LQI and $\alpha$-QI.

\

 The idea of the method goes as follows:
 Given the quantum scalar field $\chi$ of superheavy  particles conformally coupled to gravity satisfying the KG equation (\ref{chi}),
 the modes that define the
 vacuum state, at a given initial time $\tau_i$, must
 satisfy  the condition
 \begin{eqnarray}
 \chi_{k}(\tau_i)=
 \frac{1}{\sqrt{2\omega_k(\tau_i)}}e^{-i\int^{\tau_i} \omega_k(\bar\eta)d\bar\eta}, \quad
 \chi_{ k}'(\tau_i)=
-i \omega_k(\tau_i)\chi_{ k}(\tau_i), \end{eqnarray}
and 
 the energy density of the vacuum is given by \cite{Bunch}
\begin{eqnarray}\label{vacuum-energy}
\rho_{\chi}(\tau)\equiv \langle 0| \hat{\rho}_{\chi}(\tau)|0 \rangle=
\frac{1}{4\pi^2a^4(\tau)}\int_0^{\infty} k^2dk \left(   |\chi_{ k}'(\tau)|^2+ \omega^2_k(\tau) |\chi_{ k}(\tau)|^2-  \omega_k(\tau)        \right),
\end{eqnarray}
where in order to obtain a finite energy density \cite{gmmbook} we have subtracted the energy density of the zero-point oscillations of the vacuum 
$\frac{1}{(2\pi)^3a^4(\tau)}\int d^3k  \frac{1}{2} \omega_k(\tau)$.

\

Following the method developed in \cite{zs} (see also Section $9.2$ of \cite{gmmbook}),  we will write
the modes as follows,
\begin{eqnarray}\label{zs}
\chi_{k}(\tau)= \alpha_k(\tau)\frac{e^{-i\int^{\tau} \omega_k(\bar\eta)d\bar\eta}}{\sqrt{2\omega_k(\tau)}}+
\beta_k(\tau)\frac{e^{i\int^{\tau} \omega_k(\bar\eta)d\bar\eta}}{\sqrt{2\omega_k(\tau)}},\end{eqnarray}
where $\alpha_k(\tau)$ and $\beta_k(\tau)$ are the time-dependent Bogoliubov coefficients.
Now, imposing that the modes satisfy   the conditions
\begin{eqnarray}
\chi_{k}'(\tau)= -i\omega_k(\tau)\left(\alpha_k(\tau)\frac{e^{-i\int^{\tau} \omega_k(\bar\eta)d\bar\eta}}{\sqrt{2\omega_k(\tau)}}-
\beta_k(\tau)\frac{e^{i\int^{\tau} \omega_k(\bar\eta)d\bar\eta}}{\sqrt{2\omega_k(\tau)}}\right),\end{eqnarray}
one easily  shows that   the Bogoliubov coefficients must satisfy the fundamental system 
\begin{eqnarray}\label{Bogoliubovequation}
\left\{ \begin{array}{ccc}
\alpha_k'(\tau) &=& \frac{\omega_k'(\tau)}{2\omega_k(\tau)}e^{2i\int^{\tau} \omega_k(\bar\eta)d\bar\eta}\beta_k(\tau)\\
\beta_k'(\tau) &=& \frac{\omega_k'(\tau)}{2\omega_k(\tau)}e^{-2i\int^{\tau}\omega_k(\bar\eta)d\bar\eta}\alpha_k(\tau)\end{array}\right.
\end{eqnarray}
in order for the  expression (\ref{zs}) to be a solution of the equation (\ref{chi}).

\

Finally, inserting (\ref{zs}) into the expression for vacuum energy density (\ref{vacuum-energy}), 
and taking into account that the Bogoliubov coefficients satisfy the equation $|\alpha_k(\tau)|^2- |\beta_k(\tau)|^2=1$,
one finds that
\begin{eqnarray}\label{vacuum-energy1}
\rho_{\chi}(\tau)= \frac{1}{2\pi^2a^4(\tau)}\int_0^{\infty} k^2\omega_k(\tau)|\beta_k(\tau)|^2 dk.
\end{eqnarray}

At this point,  it is very  important to notice that $|\beta_k(\tau)|^2$ encodes the vacuum polarization effects and also the particle creation, which only happens when the adiabatic evolution breaks.  In fact, the quantity 
\begin{eqnarray}N_{\chi}(\tau)=\frac{1}{2\pi^2 a^3(\tau)}\int_0^{\infty}k^2 |\beta_k(\tau)|^2 dk\end{eqnarray}
was named in the Russian literature as the number density of {\it quasi-particles}, which is very different from the number density of the produced particles because, as we will see immediately, it also contains the vacuum polarization effects.

\

As an application, we will consider our toy model (\ref{toy}) once again. Then, 
in order to obtain the value of the $\beta$-Bogoliubov coefficient
we come  back to the equation  (\ref{Bogoliubovequation}) and 
in the first approximation we take  $\alpha_k(\tau)=1$, 
getting
\begin{eqnarray}
\beta_k(\tau)=\int^{\tau}\frac{\omega_k'(\eta)}{2\omega_k(\eta)}e^{-2i\int^{\eta} \omega_k(\bar\eta)d\bar\eta}d\eta.
\end{eqnarray}
After integrating by parts before the beginning of kination, it yields
\begin{eqnarray}
\beta_k(\tau)
= \left(-\frac{\omega'_k(\tau)}{4i\omega_k^2(\tau)}+\frac{1}{8\omega_k(\tau)}\left(\frac{\omega'_k(\tau)}{\omega_k^2(\tau)}\right)'
\right.\nonumber \\\left.
+\frac{1}{16i\omega_k(\tau)}\left(\frac{1}{\omega_k(\tau)}\left(\frac{\omega'_k(\tau)}{\omega_k^2(\tau)}\right)'\right)'+....
  \right)e^{-2i\int^{\tau} \omega_k(\bar\eta)d\bar\eta}.
  \end{eqnarray}

However, after kination the $\beta$-Bogoliubov coefficient must be given by
\begin{eqnarray}\label{Bogoliubov}
\beta_k(\tau)=
\left(-\frac{\omega'_k(\tau)}{4i\omega_k^2(\tau)}+\frac{1}{8\omega_k(\tau)}\left(\frac{\omega'_k(\tau)}{\omega_k^2(\tau)}\right)'\right.\nonumber\\ \left.
+\frac{1}{16i\omega_k(\tau)}\left(\frac{1}{\omega_k(\tau)}\left(\frac{\omega'_k(\tau)}{\omega_k^2(\tau)}\right)'\right)'+....
  \right)e^{-2i\int^{\tau} \omega_k(\bar\eta)d\bar\eta}+C,\end{eqnarray}
where the constant $C$ has to be chosen in order that the $\beta$-Bogoliubov coefficient becomes continuous at $\tau_{kin}$ because the equation (\ref{Bogoliubovequation})
is a first order differential equation, and then one has to demand continuity to the solution.
Thus, after some cumbersome calculation \cite{hpa} one has
\begin{eqnarray}\label{constant}
C=\left(
\frac{m_{\chi}^2m^3a^5_{kin}}{16i\omega^5_k(\tau_{kin})}+
....
  \right)e^{-2i\int^{\tau_{kin}} \omega_k(\bar\eta)d\bar\eta},
 \end{eqnarray}
where we have denoted $a_{kin}\equiv a(\tau_{kin})$.


\

Notice that the terms that do not contain $C$ lead to sub-leading geometric quantities in the energy density. Effectively, the term $-\frac{\omega'_k(\tau)}{4i\omega_k^2(\tau)}$
leads to the following contribution to the energy density: $\frac{m_{\chi}^2 H^2}{96 \pi} \ll 3M_{pl}^2H^2$. The same happens with  $\frac{1}{8\omega_k(\tau)}\left(\omega'_k(\tau)/\omega_k^2(\tau)\right)'$, which leads to a term of order $H^4$, meaning that $\frac{H^4}{M_{pl}^2}\ll H^2$.
The product of the first and second term generates in the right-hand side of the modified semi-classical Friedmann  equation a term of the order 
$\frac{H^3m_{\chi}}{M_{pl}^2}$, which is also sub-leading compared with $H^2$. Finally, the third term of (\ref{Bogoliubov}) leads in the right-hand side of the semi-classical Friedmann equation to the sub-leading term $\frac{H^6}{m_{\chi}^2M_{pl}^2}$.

\

Fortunately, this does not happen with $C$, whose leading term  gives the main contribution  of the vacuum energy density due to the gravitational particle production. In fact, the time dependent terms, which as we have already seen are always sub-leading,  are vacuum polarization effects, and they rapidly disappear in the adiabatic regime. That is, soon after the beginning of kination  
 $|\beta_k(\tau)|^2$ approaches  to $|C|^2$, 
 whose value coincides with the one obtained using the WKB approach (see formula (\ref{Beta})),
 getting once again
\begin{eqnarray}
\rho_{\chi}(\tau)\cong \left\{\begin{array}{ccc}
0& \mbox{ when} & \tau< \tau_{kin} \\
10^{-5}\left(\frac{m}{m_{\chi}}  \right)^2m^4\left( \frac{a_{kin}}{a(\tau)} \right)^3 & \mbox{ when}    & \tau\geq \tau_{kin},
\end{array}\right.
\end{eqnarray}
which at the beginning of kination is sub-dominant with respect to the energy density of the background but it will eventually dominate because the one of the background 
decreases during kination as $a^{-6}(\tau)$.

\

In order to understand  better these results it is useful to recall, as we have already explained,  that
the authors of the diagonalization method assume that during the whole evolution of the universe quanta named {\it quasi-particles} are created and annihilated due to the interaction of the quantum field with gravity \cite{gmmbook}, i.e., this is a vacuum polarization effect. And, following this interpretation, the number density of the created {\it quasi-particles} at time $\tau$ is given by
$ N_{\chi}(\tau)=\frac{1}{2\pi^2 a^3(\tau)}\int_0^{\infty}k^2 |\beta_k(\tau)|^2 dk$. However, one has to be very careful with this interpretation and especially keep in mind that, as we have shown in our toy model (\ref{toy}), 
real particles are only created when the adiabatic regime breaks. Effectively, before the beginning of kination the main term of the $\beta$-Bogoliubov coefficient is given by $-\frac{\omega'_k(\tau)}{4i\omega_k^2(\tau)}$,
whose contribution to the energy density is $\frac{m_{\chi}^2 H^2}{96 \pi}$, and to the number density of {\it quasi-particles} is $\frac{m_{\chi}^2 H^2}{512 \pi}$, and thus,
at time $\tau$ before the beginning of kination $\rho_{\chi}(\tau)\not=m_{\chi} N_{\chi}(\tau)$, meaning that the {\it quasi-particles} do not evolve as real massive particles. On the contrary, during kination
the leading term of 
$N_{\chi}(\tau)$ is given by $ 10^{-5}\left(\frac{m}{m_{\chi}}  \right)^3m^3\left( \frac{a_{kin}}{a(\tau)} \right)^3$, so we have $\rho_{\chi}(\tau)=m_{\chi} N_{\chi}(\tau)$
and the decay follows $a^{-3}(\tau)$,
which justifies the interpretation of massive particle production.

\

So,
once we have understood the gravitational particle production,  we
review  our last work \cite{partlorentzian} about the creation of particles in the LQI scenario.
Coming back to our potential (\ref{LQI}),
first of all one has to integrate numerically the conservation equation for the inflaton field, namely
\begin{eqnarray}\label{conservation}
\ddot{\varphi}+3H\dot{\varphi}+V_{\varphi}=0,
\end{eqnarray}
where $H=\frac{1}{\sqrt{3}M_{pl}}\sqrt{\frac{\dot{\varphi}^2}{2}+V(\varphi)  }$, with initial conditions at the horizon crossing (when the pivot scale leaves the Hubble radius). Recalling that in that moment
the system is in the slow-roll phase and, since this regime is an attractor, one only has to take initial conditions in the basin of attraction of the slow-roll solution, for example,
$\varphi_*=-0.154 M_{pl}$ and $\dot{\varphi}_*=0$, where the ``star" denotes, as usual,  that the quantities are evaluated at the horizon crossing.

\

Once  the evolution of the background is obtained, and in particular the evolution of the Hubble rate, one can compute the evolution of the scale factor, which is 
given by 
\begin{eqnarray}
a(t)=a_*e^{\int_{t_*}^t H(s)ds},
\end{eqnarray}
where  one can choose  $a_*=1$ as the value of the scale factor at the horizon crossing.

\

From the evolution of the scale factor, 
 one can see in Figure  9 that a spike appears during the phase transition from the end of inflation to the beginning of kination, i.e., when the adiabatic evolution is broken and particles are gravitationally produced.
 
 \begin{figure}[H]
 \begin{center}
\includegraphics[width=0.4\textwidth]{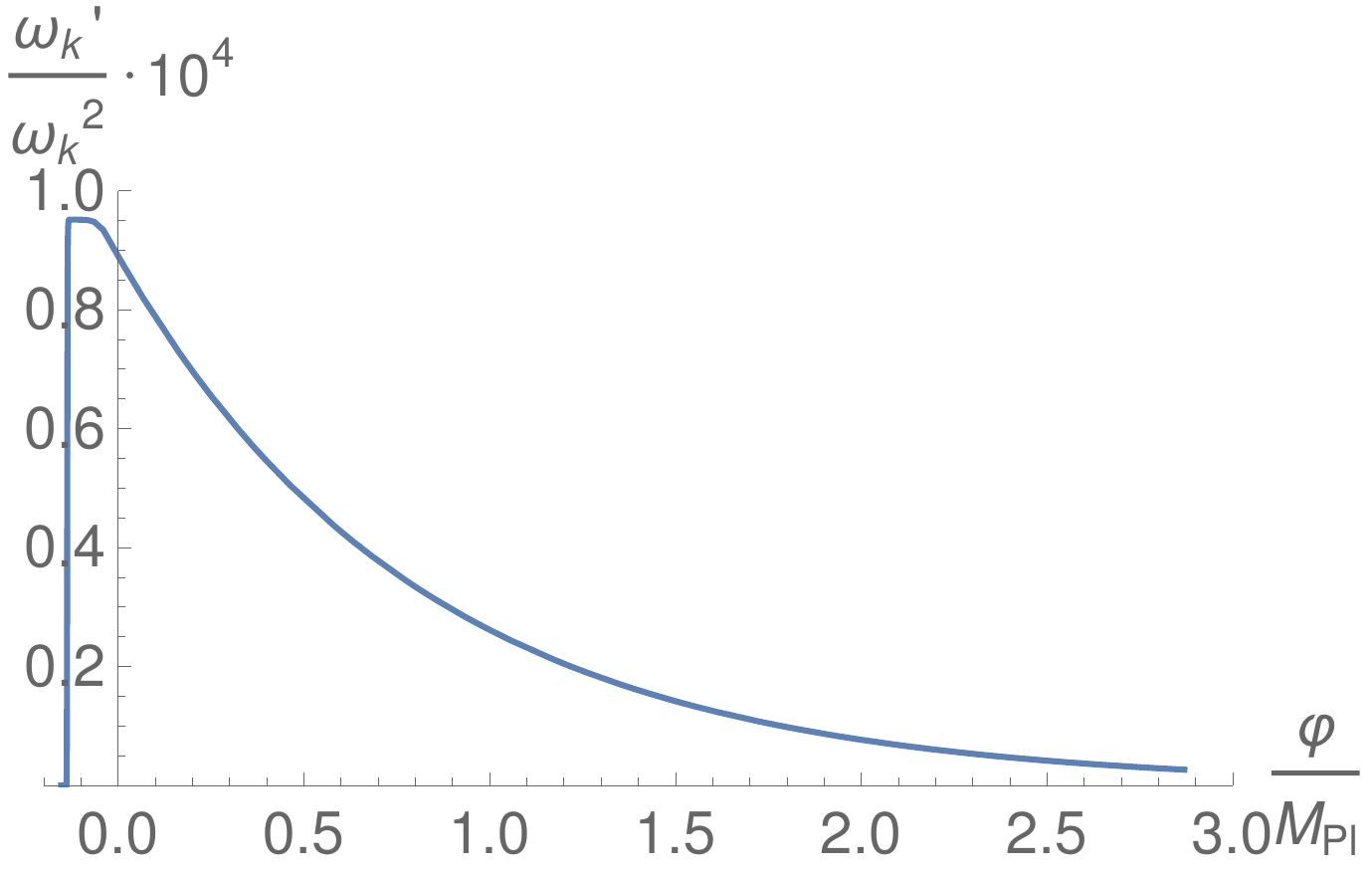}
\caption{Plot of the adiabatic evolution for a heavy field with mass $m_{\chi}\cong 10^{15}$ GeV, when the background is given by the Lorentzian Quintessential Inflation potential. The value $k=a_{kin}H_{kin}$ has been used for the quantities $a_{kin}$, $H_{kin}$ of this model.}
\end{center}
\label{adi}
\end{figure}

Next, taking into account what we have learned about   our  toy model (\ref{toy}), one can  numerically solve the equation (\ref{Bogoliubovequation}), 
with initial conditions  $\alpha_k(\tau_*)=1$ and $\beta_k(\tau_*)=0$ at the horizon crossing (there were neither particles nor polarization effects at that moment because during the slow-roll regime the derivatives of the Hubble rate are negligible compared with the powers of $H$, i.e.,  the system is in the adiabatic regime).

\

\begin{figure}[H]
    \centering
    \includegraphics[width=0.5\textwidth]{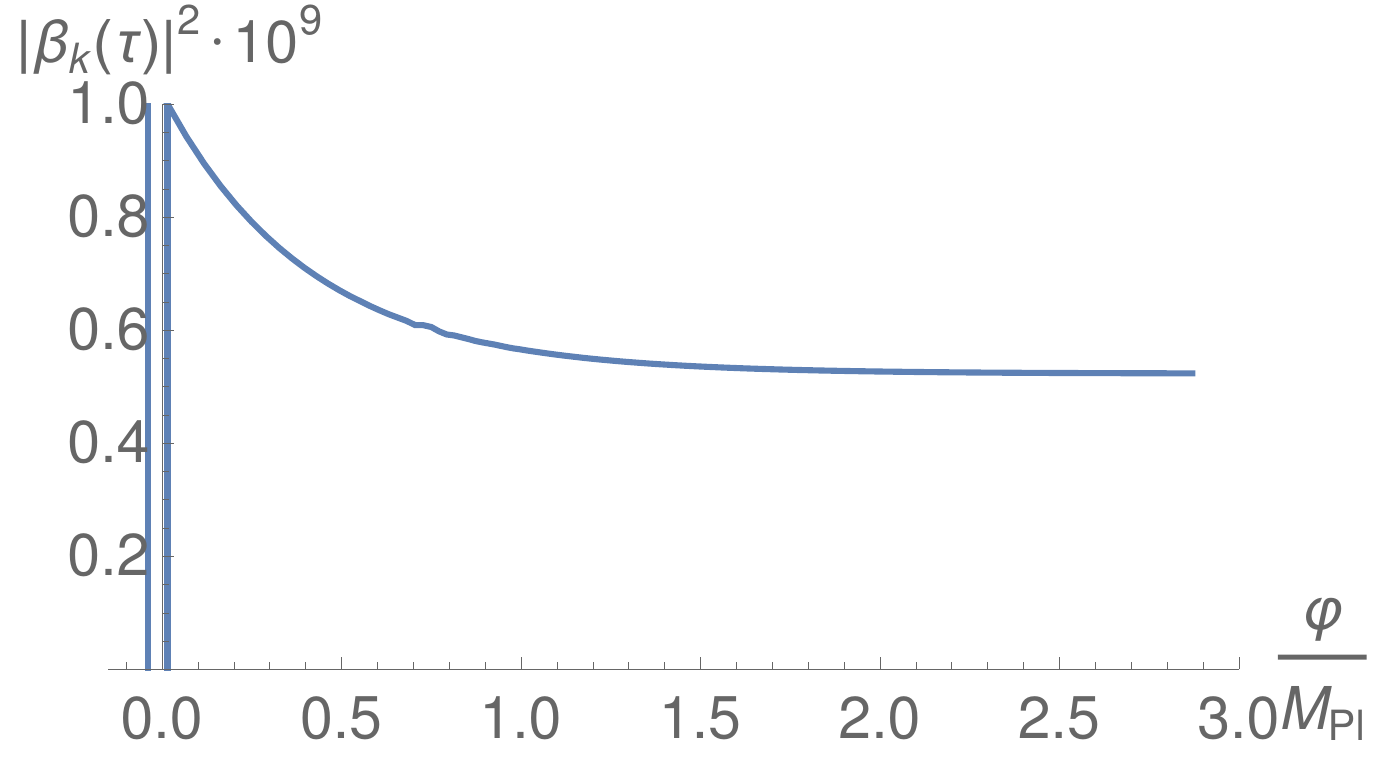}
    \caption{Evolution of $|\beta_k(\tau)|^2$}
    \label{fig:bogoliubov}
\end{figure}

 In order to get rid of complex exponentials it is useful to transform the equation \eqref{Bogoliubovequation} into a second order differential equation, namely
\begin{eqnarray}
\left\{ \begin{array}{cc}
     &  \alpha_k''(\tau)=\alpha_k'(\tau)\left(\frac{\omega_k''(\tau)}{\omega_k'(\tau)} - \frac{\omega_k'(\tau)}{\omega_k(\tau)}+2i\omega_k(\tau) \right)+ \left(\frac{\omega_k'(\tau)}{2\omega_k(\tau)} \right)^2\alpha_k(\tau)\\
     & \beta_k''(\tau)=\beta_k'(\tau)\left(\frac{\omega_k''(\tau)}{\omega_k'(\tau)} - \frac{\omega_k'(\tau)}{\omega_k(\tau)}-2i\omega_k(\tau) \right)+ \left(\frac{\omega_k'(\tau)}{2\omega_k(\tau)} \right)^2\beta_k(\tau)
\end{array}\right..
\end{eqnarray}
Given that $\alpha_k(\tau_*)=1$ and $\beta_k(\tau_*)=0$ leads to $\alpha_k'(\tau_*)=0$, 
one is interested in solving the equation for $\alpha_k(\tau)$, which can be split into the real and imaginary form in the following way,
\begin{eqnarray}
\left\{ \begin{array}{cc}
     &  \alpha_{k,Re}''(\tau)=\alpha_{k,Re}'(\tau)\left(\frac{\omega_k''(\tau)}{\omega_k'(\tau)} - \frac{\omega_k'(\tau)}{\omega_k(\tau)}\right)-2\omega_k(\tau)\alpha_{k,Im}''(\tau) + \left(\frac{\omega_k'(\tau)}{2\omega_k(\tau)} \right)^2\alpha_{k,Re}(\tau)\\
     & \alpha_{k,Im}''(\tau)=\alpha_{k,Im}'(\tau)\left(\frac{\omega_k''(\tau)}{\omega_k'(\tau)} - \frac{\omega_k'(\tau)}{\omega_k(\tau)}\right)-2\omega_k(\tau)\alpha_{k,Re}''(\tau) + \left(\frac{\omega_k'(\tau)}{2\omega_k(\tau)} \right)^2\alpha_{k,Im}(\tau)
\end{array}\right.,
\end{eqnarray}
and then $|\beta_k(\tau)|^2=|\alpha_k(\tau)|^2-1$ because of the well-known conservation property of the Wronskian. For the value $k=a_{kin}H_{kin}$, one can see in Figure \ref{fig:bogoliubov} that $|\beta_k(\tau)|^2$ stabilizes soon to a non-zero value after the beginning of kination, containing only particle production effects. One can numerically check that   this happens for the range $0.05\lesssim\frac{k}{a_{kin}H_{kin}}\lesssim7\times 10^4$, which leads to values of $|\beta_k|^2$ of the order of $10^{-10}$ and $10^{-11}$.

Then, introducing these values of the $\beta$-Bogoliubov coefficient in the equation (\ref{vacuum-energy1}), one obtains a vacuum energy density of the order of $10^{44} \mbox{GeV}^4$. So, the energy density of the produced particles evolves as 
\begin{eqnarray}
\rho_{\chi}(\tau)=\bar{\rho}_{\chi}\left(\frac{\bar{a}}{a(\tau)}\right)^3,
\end{eqnarray}
where $\bar{\rho}_{\chi}\sim 10^{44} \mbox{ GeV}^4\sim 10^{-30} M_{pl}^4$ and $\bar{a}$ are, respectively, the energy density of the produced particles 
and the value of the scale factor at the end of the  non-adiabatic phase. Finally, the energy density of the background at this moment is given by
$\bar{\rho}_{\varphi}=3\bar{H}^2M_{pl}^2\sim   10^{57}\mbox{ GeV}^4\sim  10^{-17} M_{pl}^4$, showing that the energy density of the produced particles is sub-leading at the beginning of kination, but will eventually be dominant because, during the kination regime, the energy density of the inflaton field decreases as $a^{-6}$.

\

Once again, two different situations arise:
\begin{enumerate}
    \item Decay of the superheavy particles into lighter ones before the end of kination.
    
    In this case,
 the energy density of the  the inflaton field and the one of the relativistic plasma, when the decay is finished, 
i.e., 
 when ${\Gamma}\sim H_{dec}=\bar{H}\left(\frac{\bar{a}}{a_{dec}} \right)^3\sim 
  10^{10}\left(\frac{\bar{a}}{a_{dec}}\right)^3 $ GeV $\sim  10^{-8}\left(\frac{\bar{a}}{a_{dec}}\right)^3 M_{pl}$ , is given by
\begin{eqnarray}\label{rhochiphidec}
\rho_{\varphi, dec}=3{\Gamma}^2M_{pl}^2 \quad \mbox{and} \quad \rho_{\chi, dec}\sim  10^{-29} 
\left(\frac{\bar{a}}{a_{dec}}\right)^3M_{pl}^4 \sim  10^{-21} \Gamma M_{pl}^3.
\end{eqnarray}

Imposing that the end of the decay precedes the end of kination, that means, $ \rho_{\chi, dec}\leq \rho_{\varphi, dec}$, one gets
$\Gamma\geq 10^{-21} M_{pl}$,
and, since the decay is after the beginning of the kination and for our LQI  model 
$H_{kin}\sim 10^{-8} M_{pl}$ one gets $\Gamma\leq H_{kin}\cong 4\times 10^{-8} M_{pl}$. So,  the following bound for the decay rate is obtained,
\begin{eqnarray}\label{bound}
 10^{-21} M_{pl}\leq \Gamma \leq  10^{-8} M_{pl}.\end{eqnarray}

\

Finally,  the reheating temperature, i.e., the temperature of the universe when the relativistic plasma in thermal equilibrium starts to dominate, 
which happens when $\rho_{\varphi, reh}\sim \rho_{\chi, reh}$, can be calculated as follows:
Since after the decay the evolution of the respective energy densities is given by
\begin{eqnarray}
\rho_{\varphi, reh}=\rho_{\varphi, dec}\left(\frac{a_{dec}}{a_{reh}}\right)^6,\qquad 
 \rho_{\chi, reh}= \rho_{\chi, dec} \left(\frac{a_{dec}}{a_{reh}}\right)^4,\end{eqnarray}
one has
$\frac{ \rho_{\chi, dec}}{\rho_{\varphi,dec}}=\left(\frac{a_{dec}}{a_{reh}}\right)^2$
and, thus, the reheating temperature will be
\begin{eqnarray}\label{reheating1}
 T_{reh}=  \left(\frac{30\rho_{\chi, reh}}{\pi^2g_{reh}} \right)^{1/4}= 
 \left(\frac{30\rho_{\chi,dec}}{\pi^2g_{reh}} \right)^{1/4}
\sqrt{\frac{\rho_{\chi,dec}}{\rho_{\varphi,dec}}} 
\cong  10^{-16} \left( \frac{M_{pl}}{\Gamma}\right)^{1/4} M_{pl},
\end{eqnarray}
where $g_{reh}=106.75$ is the effective number of degrees of freedom for the Standard Model. So, taking into account the bound (\ref{bound})
the reheating temperature ranges between  $10^3$ GeV and $10^7$ GeV.

    \item Decay of the superheavy particles into lighter ones after the end of kination.
    
    In the case  that the decay of the $\chi$-field is after the end of kination,
one  has to impose ${\Gamma}\leq  H_{end}$. Now, taking  into account that at the end of kination 
\begin{eqnarray}\label{31}
H^2_{end}=\frac{2\rho_{\varphi, end}}{3M_{pl}^2}\quad \mbox{and} \quad \rho_{\varphi, end}=\bar{\rho}_{\varphi}\left( \frac{\bar{a}}{a_{end}} \right)^6=
\frac{ \bar{\rho}_{\chi}^2}{\bar{\rho}_{\varphi}},
\end{eqnarray}
where it is used that the kination ends when ${ \rho}_{\chi, end}\sim{{\rho}_{\varphi, end}}$, meaning 
$\left(\frac{\bar{a}}{a_{end}} \right)^3=
\frac{\bar{\rho}_{\chi}}{\bar{\rho}_{\varphi}}$. So, the condition ${\Gamma}\leq H_{end}$ leads to the bound 
\begin{eqnarray}\label{bound1}
\Gamma\leq  10^{-22} M_{pl}.
\end{eqnarray}

\

 On the other hand, assuming once again instantaneous thermalization, the reheating temperature (i.e., the temperature of the universe when the thermalized plasma starts to dominate) will be obtained when all the superheavy particles decay, i.e., when $H\sim \Gamma$, obtaining
\begin{eqnarray}
T_{reh}=\left( \frac{30}{\pi^2 g_{reh}} \right)^{1/4}\rho_{\chi,dec}^{1/4}= \left( \frac{90}{\pi^2 g_{reh}} \right)^{1/4}\sqrt{{\Gamma}M_{pl}}~,
\end{eqnarray}
where one has to take into account that, after  the end of the kination regime, the energy density of the produced particles dominates the one  of the inflaton field. 

\

Consequently,  since the BBN epoch occurs at the $1$ MeV regime 
and taking once again  $g_{reh}=106.75$, one can find that, in that case,  the reheating temperature is bounded by
\begin{eqnarray}
1 \mbox{ GeV}\leq T_{reh} \leq  10^7 \mbox{ GeV }.
\end{eqnarray}

\end{enumerate}

\

\subsection{Instant preheating}

In this subsection we review the work of Felder,  Kofman and Linde \cite{fkl} (see also \cite{fkl0}), where {\it Instant Preheating} for  QI scenarios was introduced and discussed in detail.

\

Essentially,
instant reheating is based on the interacting part of the Lagrangian density, namely $-\frac{1}{2}g^2\varphi^2\chi^2$, where 
g is a coupling constant,
 which  due to the phase transition between the end of inflation and the beginning of kination ceases to be in the vacuum state to produce superheavy massive particles. These created particles will
interact decaying into light ones, becoming a relativistic plasma whose energy density eventually
dominates those of the background and as a consequence the universe will thermalize and will reheat.

\

To deal with particle production, note that, in Fourier space,  the 
$k$-mode of the quantum field $\chi$ satisfies the equation of a time dependent harmonic oscillator
\begin{eqnarray}
{\chi}_k''+\omega_k^2(\tau){\chi}_k=0,
\end{eqnarray}
where the derivative is with respect to the conformal time $\tau$ and the square of the frequency is given by
\begin{eqnarray}
\omega_k^2(\tau)=k^2+a^2(\tau)\left[m_{\chi}^2+g^2\varphi^2(\tau)+\left(\xi-\frac{1}{6}\right)R(\tau)\right],
\end{eqnarray}
being once again $m_{\chi}$  the bare mass of the quantum field,  $R(\tau)$  the scalar curvature and $\xi$  the coupling constant with gravity.

\

To clarify ideas about preheating (i.e., particle production) we consider the simplest toy model
\begin{eqnarray}\label{pot}
V(\varphi)=\left\{\begin{array}{ccc}
\frac{1}{2}m^2 \varphi^2 & \mbox{for}& \varphi<0\\
0 &\mbox{for}& \varphi\geq 0,
\end{array}\right.
\end{eqnarray}
although the reasoning will serve for a general class of Quintessential  Inflation potentials.

\

Near the phase transition, which actually corresponds to the beginning of kination ($\varphi_{kin}=0$), one has $\varphi(\tau)\cong \varphi'_{kin}\tau$. Then, if we simplify choosing    a conformal coupling 
$\xi=\frac{1}{6}$, 
we can approximate the frequency $\omega_k(\tau)$ by
$\sqrt{k^2+a^2_{kin}(m_{\chi}^2+g^2(\varphi'_{kin})^2\tau^2)}$, where
we have disregarded  the expansion of the
universe taking  $a(\tau)=a_{kin}$ in order to obtain the well-known overbarrier problem in scattering theory \cite{nikishov}, 
which can be analytically solved.
Effectively, the $\beta_k$-Bogoliubov coefficient is related with the reflection 
coefficient via the formula (see \cite{popov, marinov, haro} and \cite{meyer, fedoryuk} for a mathematical explanation)
\begin{eqnarray}
|\beta_k|^2=e^{-Im\int_{\gamma}\omega_{k}(\tau) d\tau},
\end{eqnarray}
where $\gamma$ is a closed path in the complex plan that contains the two turning points
$\tau_{\pm}=\pm i\frac{\sqrt{k^2+ a^2_{kin}m_{\chi}^2}}{ga_{kin}\varphi'_{kin}}.$

A simple calculation yields that the number of particles in the $k$-mode is given by \cite{haro}
\begin{eqnarray}\label{particle}
n_k\equiv |\beta_k|^2=e^{-\frac{\pi(k^2+a^2_{kin}m_{\chi}^2)}{ga_{kin}\varphi'_{kin}}}.
\end{eqnarray}

\

Then, the number density of produced particles is given by in terms of the cosmic time $t$ \cite{Birrell}, namely
\begin{eqnarray}
n_{\chi}(t)\equiv \frac{1}{2\pi^2 a^3(t)}\int_0^{\infty} k^2|\beta_k|^2dk=
\frac{(g\dot{\varphi}_{kin})^{\frac{3}{2}}}{8\pi^3}\left(\frac{a_{kin}}{a(t)}  \right)^3
e^{-\frac{\pi m_{\chi}^2}{g\dot{\varphi}_{kin}}}.
\end{eqnarray}

Analogously, the energy density of the produced particles is given by \cite{Birrell}
\begin{eqnarray}
\rho_{\chi}(t)\equiv \frac{1}{2\pi^2 a^4(t)}\int_0^{\infty}\omega_k(t) k^2|\beta_k|^2dk.
\end{eqnarray}

Then, at the phase transition one has
\begin{eqnarray}
\rho_{\chi, kin}= \frac{g^2\dot{\varphi}^2_{kin}}{4\pi^4}
e^{-\frac{\pi m_{\chi}^2}{g\dot{\varphi}_{kin}}}                ,
\end{eqnarray}
and at late times  $\rho_{\chi}(t)\cong \sqrt{m_{\chi}^2+g^2\varphi^2(t)}n_{\chi}(t)$, which means that at late times the $\chi$-particles acquire an effective mass equal to
\begin{eqnarray}
m_{eff}(t)\equiv \sqrt{m_{\chi}^2+ g^2\varphi^2(t)}.
\end{eqnarray}

\

Now, we will see that 
 three  constraints must be imposed \cite{fkl} in order to have a viable model:

\begin{enumerate}\item
  If we do not want an exponential suppression of the energy density, one has to choose a bare mass 
 satisfying $m_{\chi}\leq \sqrt{g\dot{\varphi}_{kin}}$. In fact, for the sake of simplicity,  we will take $m_{\chi}=0$, and thus, the effective mass is given by 
 $m_{eff}=g|\varphi|$.

 \item For masses -the effective mass of the field $\chi$ is $g|\varphi|$- greater than the Hubble parameter the  vacuum polarization energy density 
  due to the field $\chi$, which for $H\ll g|\varphi|$ can be calculated using the WKB approximation, is of the order $\frac{H^6}{g^2\varphi^2} $  \cite{kaya}. This
  quantity is   smaller than the energy density  of the background ($\sim H^2M_{pl}^2$) when the pivot scale leaves the Hubble radius, meaning that
 the polarization effects  will not affect the last stages of inflation.  
 Therefore, since
$ H\leq \sqrt{\frac{V}{3M_{pl}^2}}$, for a quadratic potential the condition  $H\ll g|\varphi|$ is accomplished  imposing $g\gg \frac{m}{M_{pl}}$.
 

 \item The energy density of the produced $\chi$-particles cannot dominate before its decay into light particles, which will form the   relativistic plasma,
 because, if so,  the  force driving the inflation
back to $\varphi=0$ 
 will not disappear and the inflaton field would not continue its movement forward up to $\infty$.
Effectively, the  interaction term $ \frac{1}{2}g^2\varphi^2\chi^2$ entails that  after the phase transition the inflaton field satisfies the equation
\begin{eqnarray}\label{inflaton}
\ddot{\varphi}+3H\dot{\varphi}=-g^2\chi^2\varphi.
\end{eqnarray}

When the energy density of the $\chi$-particles is sub-dominant the right-hand side of (\ref{inflaton}) is negligible and the field rolls towards $\infty$, but when 
it is dominant the right-hand side ceases to be negligible, meaning that the inflaton is under the action of  the quadratic potential $V(\varphi)=\frac{1}{2}g^2\varphi^2\chi^2$,
so the inflaton field will roll down to zero, which may produce a new inflationary phase. 

Therefore,  to avoid this situation, we have to calculate 
  when the energy density of the background and the field $\chi$ are of the same order, that is,
\begin{eqnarray}
\rho_{\varphi}(t)\sim \rho_{\chi}(t)\Longleftrightarrow 3H^2(t)M_{pl}^2\sim g\varphi(t)n_{\chi}(t).
\end{eqnarray}

\

To obtain these quantities we use that for  the model presented here, after the phase transition the universe enters in a kination regime and  one has
\begin{eqnarray}
\dot{H}=-3H^2 \Longleftrightarrow H(t)=\frac{1}{3t} \Longleftrightarrow a(t)=a_{kin}\left(\frac{t}{t_{kin}}\right)^{\frac{1}{3}}=a_{kin}
\left({3H_{kin}t}\right)^{\frac{1}{3}},
\end{eqnarray}
and taking into account the Raychaudhuri equation $\dot{H}=-\frac{\dot{\varphi}^2}{2M_{pl}^2}$, we get
\begin{eqnarray}
\varphi(t)=M_{pl}\int_{t_{kin}}^t \sqrt{-2\dot{H}(s)}ds=M_{pl}\int_{H(t)}^{H_{kin}}\sqrt{\frac{-2}{\dot{H}(H)}}dH=
\sqrt{\frac{2}{3}}M_{pl}\ln(3H_{kin}t).
\end{eqnarray}

\

Now,  using that $\rho_{\chi}(t)\cong g\varphi (t) n_{\chi}(t)$, one  gets
\begin{eqnarray}
\rho_{\chi}(t)\cong 3\times 10^{-3}g^{5/2}M_{pl}^{5/2}H^{1/2}_{kin} \frac{1}{t}\ln(3H_{kin}t),\quad
\mbox{and} \quad \rho_{\varphi}(t)\cong 3\times 10^{-1} M_{pl}^2 \frac{1}{t^2}.
\end{eqnarray}

Then, both quantities are of the same order
when
\begin{eqnarray}
t\sim \bar{t}\equiv \frac{10^2}{g^{5/2}\sqrt{H_{kin}M_{pl}}}.
\end{eqnarray}

And we obtain an important constraint for this theory: the decaying time, i.e., when the particles have decayed into a relativistic plasma,  must be smaller than $\bar{t}$, 
in order for the back-reaction to be subdominant so that the inflaton field rolls monotonically towards  $\infty$.

\end{enumerate}

On the other hand,
assuming as usual that  there is no substantial drop of energy between the end of inflation and the phase transition time, 
and using that the value of the power spectrum of the curvature perturbation when the pivot scale leaves the Hubble radius is given by \cite{btw}
${\mathcal P}_{\zeta}\cong \frac{H_*^2}{8\pi^2 M_{pl}^2\epsilon_*}\sim 2\times 10^{-9}$,
one obtains 
\begin{eqnarray}
m^2\sim 3\times 10^{-9} \pi^2(1-n_s)^2   M_{pl}^2,
\end{eqnarray}
 where we have used that for our model one has   $\epsilon_*=\frac{2M_{pl}^2}{\varphi^2_*}\cong \frac{1-n_s}{4}$, being  $n_s$  the spectral index. 
Then,  
  since the recent observational data constrain the value of the spectral index to be $n_s=0.968\pm 0.006$ \cite{Planck}, taking its central value  one gets 
 $m\sim 5\times 10^{-6} M_{pl}$ and, as a consequence, $H_{kin}\sim H_{end}\sim \frac{m\varphi_{end}}{\sqrt{6}M_{pl}}\sim 3\times10^{-6} M_{pl}$. 
 Thus, for values of $g \leq 10^{-2}$, one gets  $\bar{t}\geq 10^{10} M_{pl}^{-1}$ and   $\varphi(\bar{t})\sim M_{pl}$, meaning that  for times  $t\in [10^{6} M_{pl}^{-1}, \bar{t}]$ the value of the inflaton field remains close to $M_{pl}$ and the effective mass of the $\chi$-field will approximately  be 
   $gM_{pl}$.

\

Since the decay must be before than $\bar{t}$, 
the condition $t_{dec}<\bar{t}$ , where $t_{dec}\cong \frac{1}{3\Gamma}$ is the time when the field $\chi$ decayed, i.e.,  $H_{dec}\sim {\Gamma}$, leads 
 to the relation 
$g<  10^{2}\left(\frac{\Gamma}{M_{pl}}  \right)^{2/5}$, which together with the condition $g\gg \frac{m}{M_{pl}}\sim 10^{-6}$ constrains the value of the decay rate to satisfy
$\Gamma\gg  10^{-20} M_{pl}\sim 10^{-2}$ GeV.

\

Let's now calculate the  temperature at the reheating time, 
which as we have already seen is given by
\begin{eqnarray}
T_{reh}\sim g_{reh}^{-1/4}\rho_{\chi,reh}^{1/4}\sim g_{reh}^{-1/4}
\rho_{\chi, dec}^{1/4}\sqrt{ \frac{\rho_{\chi, dec}}{\rho_{\varphi, dec}} }.
\end{eqnarray}

\

Taking into account that, if $t_{dec}>10^6 M_{pl}^{-1}\Longrightarrow \Gamma<10^{-6} M_{pl}$, the effective mass is approximately $gM_{pl}$, one has 
\begin{eqnarray}
\rho_{\varphi, dec}= 3{\Gamma}^2M_{pl}^2 \quad\mbox{and}\quad \rho_{\chi, dec}\sim gM_{pl}n_{\chi, dec}\sim 
10^{-2}g^{5/2}\sqrt{H_{kin}M_{pl}}M_{pl}^2{\Gamma},
\end{eqnarray}
getting that
\begin{eqnarray}
T_{reh}\sim  10^{-2} g^{15/8} H^{3/8}_{kin} M_{pl}^{7/8}{\Gamma}^{-1/4}
\sim 10^{-4} g^{15/8}\left( \frac{M_{pl}}{\Gamma} \right)^{1/4} M_{pl}.
\end{eqnarray}

\

The values of the parameters $g$ and $\Gamma$ must satisfy the constraint $10^{-6}\ll g <10^2 \left(\frac{\Gamma}{M_{pl}} \right)^{2/5}$, and choosing for instance $g\sim 10^{-4}$ and $\Gamma \sim 10^{-12} M_{pl}$
one obtains a reheating temperature around $10^9$ GeV.



 
 \

 Finally, one can conclude that the   value of the parameters
 involved in the theory must
  accomplish all the following requirements:
 \begin{enumerate}
 \item  $10^{-6}\ll g <10^2 \left(\frac{\Gamma}{M_{pl}} \right)^{2/5}$. (The back-reaction is not important at the last stages of inflation and the $\chi$-field decays before the end of kination)
 \item $10^{-20} M_{pl}\ll \Gamma< 10^{-6}M_{pl}$. (The decay ends when the value of the inflaton field is of the order of the reduced Planck mass).
 \item $ 10^{-19}<g^{15/8}\left( \frac{M_{pl}}{\Gamma} \right)^{1/4}<10^{-7}$. (Reheating temperatures guaranteeing the BBN success).
 \end{enumerate}


\

\subsection{ Curvaton reheating in Quintessential Inflation}
In this section we review the so-called {\it curvature reheating} mechanism in Quintessential Inflation. To do that we follow \cite{FL}, but taking into account that for the authors of that paper, contrary to our convention, the reheating temperature is the temperature of the universe when the curvaton field has totally decayed in relativistic particles. Recall that the universe is really reheated when the energy density of inflaton is of the same order than the one of the decay products,
provided that  the curvaton field had previously decayed. 

\

We assume that the potential of the curvaton field, namely $\sigma$,  is quadratic, $V(\sigma)=\frac{1}{2}m_{\sigma}^2\sigma^2$, where the mass of the curvaton is chosen to be  smaller  than the value of the Hubble parameter at the end of inflation $m_{\sigma}\ll H_{end}$. At this moment the curvaton field is in a slow-roll regime because the 
condition $m_{\sigma}\ll H_{end}\Longleftrightarrow V_{\sigma\sigma}\ll H_{end}^2$  means that the curvaton potential is flat enough at the end of  inflation \cite{LW}. Then, in order to avoid a second inflationary stage now driven by the curvaton, one has to impose that its energy density is subdominant when the curvaton starts to oscillate, which happens when $m_{\sigma}\cong H$  \cite{LW} (see also the section $5.4.1$ of \cite{mukhanovbook} for a detailed discussion of the quadratic potential).  Then,
\begin{eqnarray}
\rho_{\sigma}(t_{osc})\equiv \rho_{\sigma, osc}<\rho_{\varphi, osc}=3H^2_{osc}M_{pl}^2,
\end{eqnarray}
where $t_{osc}$ is the time when the curvaton starts to oscillate. 
Taking into account that $H_{osc}\cong m_{\sigma}$ and 
since at the beginning of the oscillations, i.e., at the end of the slow-roll period for the curvaton, $H\dot{\sigma}\sim V_{\sigma} \Longrightarrow \dot{\sigma}_{osc}\sim m_{\sigma}\sigma_{osc}$, 
one has
$\rho_{\sigma,osc}\cong m_{\sigma}^2\sigma^2_{osc}$ obtaining the bound
$\sigma^2_{osc}<3 M_{pl}^2$.

\

Here we consider a Quintessential Inflation potential, as the Peebles-Vilenkin one or similar, according to which the universe enters in a kination regime immediately after the end of inflation. Then, after the phase transition, the energy density of the background evolves as $a^{-6}$, and those of the curvaton as $a^{-3}$ because during the oscillation regime the effective Equation of State parameter for a power law potential 
$V(\sigma)=V_0\left(\frac{\sigma}{M_{pl}}\right)^{2n}$ 
is given by $w_{eff}\cong \frac{n-1}{n+1}$ \cite{turner}.

\

Now, let ${\Gamma}$ be the decay rate of the curvaton. Then, there are two different situations: 
\begin{enumerate}
\item
The curvaton decays when it was subdominant.

In this first case, the curvaton decays into radiation (recall that  the thermalization is nearly instantaneous)  at a time ${t_{dec}}$ satisfying $H_{dec}\sim  {\Gamma}$, and thus, one has 
 \begin{eqnarray}
 \rho_{\sigma, dec}<\rho_{\varphi, dec}\Longrightarrow \rho_{\sigma,osc}\frac{{\Gamma}}{m_{\sigma}}<3{\Gamma}^2M_{pl}^2,
\end{eqnarray}
where we have used that the energy density of the curvaton decays as $a^{-3}$, that the universe is in the kination phase  (the Hubble parameter also decays as $a^{-3}$)  and $H_{osc}\cong m_{\sigma}$. Then, since 
$  \rho_{\sigma,osc}\cong  m_{\sigma}^2\sigma^2_{osc}$ and ${\Gamma}\cong H_{dec}\leq H_{osc}\cong m_{\sigma}$, one gets the constraint
\begin{eqnarray}\label{bound}
\frac{\sigma^2_{osc}}{3M_{pl}^2}\leq \frac{{\Gamma}}{m_{\sigma}}\leq 1.
\end{eqnarray}

Now, to obtain the reheating temperature, that is, when  $\rho_{\varphi, reh}\sim \rho_{\sigma, reh}$, 
one gets
\begin{eqnarray}
T_{reh}\sim g_{reh}^{-1/4}\rho_{\sigma, reh}^{1/4}\sim g_{reh}^{-1/4}\rho_{\sigma, dec}^{1/4}\sqrt{\frac{\rho_{\sigma, dec}}{\rho_{\varphi, dec}}}
\sim \frac{\rho_{\sigma, dec}^{3/4}}{M_{pl}{\Gamma}}
\sim \frac{m_{\sigma}^{3/4}|\sigma(t_{osc})|^{3/2}}{M_{pl}{\Gamma}^{1/4}},
\end{eqnarray}
where we have used that $\rho_{\sigma, dec}=\rho_{\sigma,osc}\left( \frac{a_{osc}}{a_{dec}}\right)^3
=\rho_{\sigma, osc}\frac{H_{dec}}{H_{osc}}\cong m \sigma^2_{osc}{\Gamma}$.

Then, using the bound (\ref{bound}) we can see that the reheating temperature is constrained to be in the range
\begin{eqnarray}\label{temperature}
\frac{m_{\sigma}^{1/2}|\sigma_{osc}|^{3/2}}{\sqrt{3}M_{pl}}\leq T_{reh}\leq \frac{m_{\sigma}^{1/2}|\sigma_{osc}|}{{3}^{1/4}M^{1/2}_{pl}}.
\end{eqnarray}

\

On the other hand, when the decay of the curvaton takes place when it is subdominant, the power spectrum of the curvature perturbation is given by \cite{LU, ABM}     
${\mathcal P}_{\zeta}= \frac{1}{1296\pi^2}\frac{m_{\sigma}^2}{{\Gamma}^2} \frac{H_*^2\sigma_*^2}{M^4_{pl}}$, which from the bound (\ref{bound}) leads to 
\begin{eqnarray}
\frac{1}{1296\pi^2} \frac{H_*^2\sigma_*^2}{M^4_{pl}}\leq {\mathcal P_{\zeta}} \leq \frac{1}{144\pi^2} \frac{H_*^2}{\sigma_*^2},
\end{eqnarray}
where we have used that before the oscillations the curvaton rolls slowly and, thus, $\sigma_{osc}\sim \sigma_*$.
Now, taking into account that 
${\mathcal P_{\zeta}}\sim 2\times 10^{-9}$, one gets the bounds
\begin{eqnarray}
\frac{H_*}{|\sigma_*|}\geq  2 \times 10^{-3} \quad \mbox{ and } \quad {H_*}{|\sigma_*|}\leq  5 \times 10^{-3} M_{pl}^2. 
\end{eqnarray}

\

 Then, choosing $H_*\sim 4\times 10^{-5} M_{pl}$, which for 
 $n_s\cong 0.96$  is the value of the Hubble parameter when the pivot scale leaves the Hubble radius
if the 
curvaton field is not present \cite{ha},  and taking into account that the condition $H_*\ll |\sigma_*|$ must be satisfied in order to guarantee the Gaussianity of the curvature perturbation \cite{LW}, 
one can safely  take $|\sigma_*|\sim 10^{-2} M_{pl}$, which agrees with the bound $\sigma_* ^2\sim \sigma^2_{osc}<3M_{pl}^2$.

For these values, the equation  (\ref{temperature}) becomes 
\begin{eqnarray}
5\times 10^{-4} \sqrt{m_{\sigma}M_{pl}} \leq T_R \leq 7\times 10^{-3} \sqrt{m_{\sigma}M_{pl}}, \end{eqnarray}
which means that in order to get the nucleosynthesis bounds one has to choose  light masses of the curvaton, namely $m_{\sigma}\leq 10^{-14} M_{pl}\sim 2\times 10^4$ GeV.

\item 
 The curvaton decays when the curvaton field dominated the universe.

 Assuming once again instantaneous thermalization, since the curvaton decays when it dominates, that is, when  $\rho_{\varphi, dec}\leq \rho_{\sigma, dec}$, 
 the reheating time will occur at the decay time ($H_{reh}=H_{dec}\sim \Gamma$).
\

In this case the  condition 
\begin{eqnarray}3M_{pl}^2\Gamma^2=
\rho_{\sigma , reh}< \rho_{\sigma, osc} =m_{\sigma}^2\sigma_{osc}^2,
\end{eqnarray}
together  with $H_{reh}< H_{osc}$, leads to the constraints 
\begin{eqnarray}\label{bound1}
\frac{{\Gamma}^2}{m_{\sigma}^2}<\frac{\sigma^2_{osc}}{3 M_{pl}^2} \qquad \mbox{and}
\qquad \frac{{\Gamma}}{m_{\sigma}}< 1.
\end{eqnarray}

\

Next, in this case,  the reheating temperature is $T_{reh}\sim g_{reh}^{-1/4}\rho_{\sigma, reh}^{1/4}\sim \sqrt{M_{pl}{\Gamma}}$ and the constraint leads to the bound
\begin{eqnarray}\label{temperature1}
T_{reh}\leq \sqrt{\frac{m_{\sigma}}{ M_{pl}}}|\sigma_{osc}|.
\end{eqnarray}

\

On the other hand, when the curvaton decays after its domination, the power spectrum of the curvature perturbation is given by \cite{LW}
\begin{eqnarray}
{\mathcal P}_{\zeta}\cong \frac{1}{9\pi^2}
\frac{H_*^2}{\sigma_*^2}
\sim 2\times 10^{-9}\Longrightarrow \frac{H_*}{|\sigma_*|}\sim 4 \times 10^{-4} M_{pl}.
\end{eqnarray}

 Then, choosing as in the previous case $H_*\sim 4\times 10^{-5} M_{pl}$,  one has $|\sigma_*|\sim 10^{-1} M_{pl}$, and,  
 since the curvaton  rolls slowly before the oscillations, one can safely take $|\sigma_{osc}|\sim |\sigma_*|\sim 10^{-1} M_{pl}$, which satisfies the bound
(\ref{bound1}).

\

Finally, 
from these values and the equation (\ref{temperature1}), one can conclude that only for curvaton light masses satisfying  
$m_{\sigma}\leq 5\times 10^{-16} M_{pl}\sim 10^3$ GeV
a reheating temperature compatible with the BBN success is obtained.

 \end{enumerate}

\

\section{Impact of the production of Gravitational Waves in the BBN success}\label{secviii}

\subsection{Overproduction of GWs}

It is well-known that 
 a reheating due to the gravitational production of light particles is incompatible with the overproduction of Gravitational Waves  because both satisfy the same equation, and thus, they scale in the same way. For this reason it is impossible that at the reheating time the ratio of the energy density of the GWs to the energy density of the relativistic plasma was smaller that $10^{-2}$, which is essential to prevent any complication in the BBN success \cite{pv}. 
In fact, the minimum value of this ratio, which is obtained when the light particles are minimally coupled with gravity ($\xi=0$), is of the order $1/N_s$, where $N_s$ is the number of light fields. As has been explained in \cite{pv} in a minimal GUT theory $N_s=4$,
 which corresponds to the electroweak  Higgs doublet, and only in supersymmetric theories  $N_s$ is of the order $10^2$ (see the explanation given by  Peebles-Vilenkin in \cite{pv}).

 However, as we will see in this subsection,  for our LQI model,  when the reheating is due to the gravitational creation of heavy particles,  the overproduction of GW's do not modify the BBN.

So, to show this, first of all we recall that
 the production of Gravitational Waves during the phase transition from the end of inflation to the beginning of kination is \cite{Giovannini99}
 \begin{eqnarray}
\rho_{GW}(\tau)\cong \frac{H_{kin}^4}{2\pi^3}\left( \frac{a_{kin}}{a(\tau)}\right)^4\cong 
10^{-2} H_{kin}^4\left( \frac{a_{kin}}{a(\tau)}\right)^4.
\end{eqnarray}

The success of the BBN demands that the ratio of the energy density of GWs to the one of the produced particles at the reheating time satisfies
\cite{attractor1}
\begin{eqnarray}\label{bbnconstraint}
\frac{\rho_{GW, reh}}{\rho_{\chi, reh}}\leq 10^{-2}.
\end{eqnarray}

First of all we see that the constraint (\ref{bbnconstraint}) is  overpassed when the decay of the superheavy massive particles is previous to the end of  kination. Effectively,
if the decay occurs before  the end of kination one has 
\begin{eqnarray}
\frac{\rho_{GW, reh}}{\rho_{\chi, reh}}=
\frac{\rho_{GW, dec}}{\rho_{\chi, dec}}
\end{eqnarray} 
because the light relativistic particles evolve as the GW's.
Now, taking into account that during kination the energy density of the background scales as $a^{-6}$, it yields that 
\begin{eqnarray}
\left(\frac{a_{kin}}{a_{dec}} \right)^4=
\left(\frac{\rho_{\varphi,dec}}{\rho_{\varphi,kin}} \right)^{2/3},
\end{eqnarray} 
and then, since in LQI one has $H_{kin}\sim 4\times 10^{-8}M_{pl}$ and using the results obtained previously (see the formula \eqref{rhochiphidec}, we get 
\begin{eqnarray}
\frac{\rho_{GW, reh}}{\rho_{\chi, reh}}\cong 
 10^{-1} \left(\frac{\Gamma}{M_{pl}} \right)^{1/3}.
\end{eqnarray} 

Therefore, the bound (\ref{bbnconstraint}) is overpassed when
\begin{eqnarray}
\Gamma\leq 7\times 10^{-4} M_{pl},
\end{eqnarray}
which is completely compatible within the bound (\ref{bound}).

\

On the contrary, when  the decay of superheavy particles is produced after the end of kination, and assuming once again instantaneous thermalization,  the reheating time will coincide with the decay one. Then, since
$\rho_{\chi, dec}=3{\Gamma}^2M_{pl}^2$ and 
\begin{eqnarray}
H_{dec}=H_{end}\left( \frac{a_{end}}{a_{dec}} \right)^{3/2}\Longrightarrow \left( \frac{a_{end}}{a_{dec}} \right)^{3/2}=\sqrt{\frac{3}{2}}\frac{\Gamma M_{pl}\sqrt{\bar{\rho}_{\varphi}}}
{ \bar{\rho}_{\chi}},
\end{eqnarray}
where we have used that $H_{end}= \sqrt{\frac{2}{3}}
\frac{\bar{\chi, \rho}}{M_{pl}\sqrt{\bar{\rho}_{\varphi}}}$,  we  will have 
\begin{align}
\rho_{GW, dec}=\rho_{GW, end}\left( \frac{a_{end}}{a_{dec}} \right)^4= 
\rho_{GW, end}\left(  \sqrt{\frac{3}{2}}\frac{\Gamma M_{pl}\sqrt{\bar{\rho}_{\varphi}}}
{\bar{\rho}_{\chi}}   \right)^{8/3}
\end{align}
and, using that 
\begin{eqnarray}
\left(\frac{a_{kin}}{a_{end}}  \right)^4=  \frac{\rho_{\varphi, end}}{\rho_{\varphi, kin}} \quad \mbox{and} \quad 
\rho_{\varphi, end}=\frac{ \bar{\rho}_{\chi}^2}
{\bar{\rho}_{\varphi}},
\end{eqnarray}
we get
\begin{eqnarray}
\rho_{GW, dec}=\left(\frac{2}{16}\right)^{1/3}\left( \frac{H_{kin}}{M_{pl}} \right)^2
\left(\frac{\bar{\rho}_{\varphi}}{\bar{\rho}_{\chi}^2}\right)^{1/3}(\Gamma M_{pl})^{8/3}\cong 10^{-2}\Gamma^{8/3} M_{pl}^{4/3},
\end{eqnarray}
and thus,
\begin{eqnarray}
\frac{\rho_{GW, reh}}{\rho_{\chi, reh}}=  \frac{\rho_{GW, dec}}{\rho_{\chi,dec}} 
\cong 3\times 10^{-3} \left(\frac{\Gamma}{M_{pl}} \right)^{2/3}\leq  10^{-17},
\end{eqnarray}
where we have used the bound (\ref{bound1}), so the constraint (\ref{bbnconstraint}) is clearly overpassed.

\subsection{BBN constraints from the logarithmic spectrum of GWs}
\label{subsec-gw1}

It is well-known that during inflation GWs  are produced (known as primordial GWs, in short PGWs) and in the post-inflationary period, i.e., during kination, the logarithmic spectrum of GWs, namely
$\Omega_{GW}$ defined as $\Omega_{GW}\equiv \frac{1}{\rho_c}\frac{d\rho_{GW}(k)}{d\ln k }$ (where $\rho_{GW}(k)$ is the energy density spectrum of the produced GWs; $\rho_c=3H_0^2M_{pl}^2$, where $H_0$ is the present value of the Hubble parameter, is the so-called {\it critical density}) scales as $k^2$ \cite{rubio}, producing a spike in the spectrum of GWs at high frequencies. Then, so that GWs do not destabilize the BBN, the following bound must be imposed  (see Section 7.1 of \cite{maggiore}),
\begin{eqnarray}\label{integral}
I\equiv h_0^2\int_{k_{BBN}}^{k_{end}} \Omega_{GW}(k) d \ln k \leq 10^{-5},
\end{eqnarray}
where $h_0\cong 0.678$ parametrizes the experimental uncertainty to determine the current value of the Hubble constant and $k_{BBN}$, $k_{end}$ are the momenta associated to the horizon scale at the BBN and at the end of inflation respectively. As has been shown in \cite{Giovannini1}, the main contribution of the integral \eqref{integral} comes from the modes that leave the Hubble radius before the end of the inflationary epoch and finally re-enter during  kination, that means, for $k_{end}\leq k\leq k_{kin}$, where
$k_{end}=a_{end}H_{end}$ and $k_{kin}=a_{kin}H_{kin}$.  For these modes one can calculate the  logarithmic spectrum of GWs as in \cite{Giovannini} (see also \cite{rubio, Giovannini2, Giovannini3,Giovannini:2016vkr} where the graviton spectra
in quintessential models have been reassessed, in a model-independent way, using numerical techniques),

\begin{eqnarray}\label{Omega}
\Omega_{GW}(k)=\tilde{\epsilon}\Omega_{r}h^2_{GW} \left(\frac{k}{k_{end}}  \right)\ln^2\left(\frac{k}{k_{kin}}  \right),
\end{eqnarray}
where $h^2_{GW}=\frac{1}{8\pi}\left(\frac{H_{kin}}{M_{pl}}  \right)^2$
is the amplitude of the GWs; $\Omega_{r}\cong 2.6\times 10^{-5} h_0^{-2}$ is the present density fraction of radiation, and the quantity $\tilde{\epsilon}$, which is approximately equal to $0.05$ for the Standard Model of particle physics,  takes into account the variation of massless degrees of freedom between decoupling and thermalization (see \cite{rubio, Giovannini1} for more details). As has been derived in \cite{Giovannini1}, the specific form of the expression above comes from the behavior of the Hankel functions for small arguments. Now, plugging expression (\ref{Omega}) into (\ref{integral}) and disregarding the sub-leading logarithmic terms, one finds  
\begin{eqnarray}\label{constraintx}
 2\tilde{\epsilon}h_0^2\Omega_{r}h^2_{GW}\left( \frac{k_{kin}}{k_{end}} \right)\leq 10^{-5}
 \Longrightarrow 
 10^{-2}\left( \frac{H_{kin}}{M_{pl}} \right)^2\left( \frac{k_{kin}}{k_{end}} \right)\leq 1
 \Longrightarrow 
 10^{-17}\left( \frac{k_{kin}}{k_{end}} \right)\leq 1  \end{eqnarray}
  because in our LQI model we have $H_{kin}\sim 4\times 10^{-8} M_{pl}$.

  \

To calculate the ratio $k_{kin}/k_{end} $,  we  will have to study the following
 two different situations:

\begin{enumerate}

\item When the decay of superheavy particles occurs before the end of kination.

In this case the reheating  time coincides with the end of kination, so
a simple calculation leads to 
\begin{eqnarray}
\frac{k_{kin}}{k_{end}} =\frac{k_{kin}}{k_{reh}}.
\end{eqnarray}

Using  the formulas
\begin{eqnarray}
\rho_{\varphi, reh}=\rho_{\varphi, kin}\left(\frac{a_{kin}}{a_{reh}}\right)^6\qquad \mbox{and}\qquad
\rho_{\chi, reh}=\rho_{\chi, dec}\left(\frac{a_{dec}}{a_{reh}}\right)^4,
\end{eqnarray}
we get 
\begin{eqnarray}
\frac{a_{kin}}{a_{reh}}=\left( \frac{\rho_{\chi, dec}}{\rho_{\varphi, kin}} \right)^{1/6}
\left(\frac{a_{dec}}{a_{reh}}\right)^{2/3}=\left( \frac{\rho_{\chi, dec}}{\rho_{\varphi, kin}} \right)^{1/6}\left( \frac{\rho_{\chi, dec}}{\rho_{\varphi, dec}} \right)^{1/3},
\end{eqnarray}
where we have used the relation $\left(\frac{a_{dec}}{a_{reh}}\right)^2=\frac{\rho_{\chi, dec}}{\rho_{\varphi, dec}}$. Then, taking into account that $H_{reh}=\Gamma \left(\frac{a_{dec}}{a_{reh}}\right)^3$, we obtain
\begin{eqnarray}
\frac{k_{kin}}{k_{end}} =\frac{k_{kin}}{k_{reh}}=
\frac{H_{kin}a_{kin}}{H_{reh}a_{reh}}=\frac{H_{kin}}{\Gamma}\frac{\rho_{\varphi, dec}}{\rho_{\chi, dec}}
\left( \frac{\rho_{\varphi, dec}}{\rho_{\varphi, kin}} \right)^{1/6}.
\end{eqnarray}

\

 Finally, using our previous results
  \begin{eqnarray}
  \rho_{\varphi, dec}=3\Gamma^2M_{pl}^2, \quad \rho_{\chi, dec}\sim 10^{-21}\Gamma M_{pl}^3 \quad \mbox{and} \quad  \rho_{\varphi, kin}=3H_{kin}^2M_{pl}^2,
  \end{eqnarray}
we arrive at
\begin{eqnarray}
\frac{k_{kin}}{k_{end}}\sim 10^6\left(\frac{\Gamma}{M_{pl}} \right)^{1/3},
\end{eqnarray}
and consequently the constraint becomes 
\begin{eqnarray}
\left(\frac{\Gamma}{M_{pl}} \right)^{1/3}\leq 10,
\end{eqnarray}
which is obviously overpassed for all the viable values of $\Gamma$, i.e, for all values between $10^{-21} M_{pl}$ and $10^{-8} M_{pl}$.

\item When the  decay of superheavy particles occurs  after the end of kination.

Since  during the kination period  we have $H_{end}=H_{kin}\left(\frac{a_{kin}}{a_{end}}\right)^3$, one gets
\begin{eqnarray}
\frac{k_{kin}}{k_{end}}=\left(\frac{H_{kin}}{H_{end}} \right)^{2/3}
\end{eqnarray}
and, taking into account that
\begin{eqnarray}
H_{end}=\sqrt{\frac{2\bar{\rho}_{\chi}^2}{3M_{pl}^2\bar{\rho}_{\varphi}}}\sim 3\times 10^{-22}M_{pl}\qquad \mbox{and} \qquad H_{kin}\sim 4\times 10^{-8} M_{pl}, 
\end{eqnarray}
the bound is completely overpassed.

\end{enumerate}

\

\section{Conclusions}\label{secix}

In the present  review we have dealt with some important Quintessential Inflation scenarios, namely the original Peebles-Vilenkin one -the first Quintessential Inflation model introduced at the end of the 90's- and some of its improved versions, a kind of exponential scenarios, the Lorentzian Quintessential Inflation and also Quintessential Inflation in the context of $\alpha$-attractors. Studying these models, first of all we have checked  that at early times they provide results -spectral index of scalar perturbations, ratio of tensor to scalar perturbations, number of e-folds from the horizon crossing to the end of inflation- agreeing with those of the standard inflation and matching with the current observational data. Next, the models should contain a phase transition from the end of inflation to the beginning of kination where the adiabatic evolution of the universe must be broken in order to have enough particle production to reheat the universe after its cooling produced during inflation. Once we have checked that the models contain this phase transition we have chosen the reheating mechanism, in our case the gravitational particle production of superheavy particles, although in this review we have also studied other mechanisms such as the Instant Preheating or the Curvaton reheating, and we have calculated numerically the energy density of the produced particles and
the corresponding
reheating temperature,  whose maximum value is for the Lorentzian Quintessential Inflation  model around $10^7$ GeV. 

\

We have also  dealt with the overproduction of Gravitational Waves, produced as well during the phase transition from the end of inflation to the beginning of kination,  showing that in the case of Lorentzian Quintessential Inflation -and also for the $\alpha$-attractors in the context of QI- all the bounds imposed are overpassed, so 
they  do not disturb the Big Bang Nucleosynthesis success. 

\

Finally,  
we have found the numerical value of the other parameter (these models normally only depend on two parameters, the first one is determined using the power spectrum of scalar perturbations when the pivot scale  leaves the Hubble radius), imposing that the model matches with the current observational data at the present time.  To do it, we have solved numerically the corresponding dynamical system imposing initial conditions, which are determined from the values of the inflaton field and its first derivative during the slow-roll period,  at the beginning of the radiation era. In this way, we have  checked numerically that, at late times, the models enter in a  dark energy regime (via quintessence) able to explain the current observational data.

\vspace{6pt} 




\section*{Acknowledgments}

{The investigation of J. de H. has been supported by MINECO (Spain) grant  MTM2017-84214-C2-1-P, and  in part by the Catalan Government 2017-SGR-247. L.A.S. thanks the School of Mathematical Sciences (Queen Mary University of London) for the support provided.}

\end{document}